\documentclass[fleqn,usenatbib]{mnras}

\usepackage{newtxtext,newtxmath}
\usepackage[T1]{fontenc}

\DeclareRobustCommand{\VAN}[3]{#2}
\let\VANthebibliography\thebibliography
\def\thebibliography{\DeclareRobustCommand{\VAN}[3]{##3}\VANthebibliography}

\usepackage{graphicx}	% Including figure files
\usepackage{subcaption} % subfigures
\usepackage{amsmath}	% Advanced maths commands
\usepackage{siunitx}

\newcommand{\SYCL}{\textsc{SYCL}}
\newcommand{\ROCM}{\textsc{ROCm}}
\newcommand{\CUDA}{\textsc{CUDA}}
\newcommand{\OpenMP}{\textsc{OpenMP}}
\newcommand{\MPI}{\textsc{MPI}}
\newcommand{\SHAMROCK}{\textsc{Shamrock}}
\newcommand{\AABB}{\textit{AABB}}

\newcommand{\gl}[1]{{#1}}
\newcommand{\tdc}[1]{{#1}}

\newcommand{\TODO}[1]{\textcolor{red}{\textbf{TODO {#1}} }}

\newcommand{\CPP}{\textsc{C++}}
\newcommand{\Python}{\textsc{Python}}

\newcommand{\unitau}{\si{\astronomicalunit}}

\usepackage{pgfplots}
\pgfplotsset{compat=1.6}

\usepackage{fontawesome}

\usepackage{tikz}
\usetikzlibrary{shapes.symbols}
\usetikzlibrary{shapes.geometric, arrows.meta}
\usetikzlibrary{calc}
\usetikzlibrary{backgrounds} 
\usetikzlibrary{shadows}

\usepackage[linesnumbered,ruled,vlined]{algorithm2e}

\SetCommentSty{mycommfont}
\SetKwRepeat{Do}{do}{while}%
\SetKwFor{Rtreefor}{rtree\char`_for}{do}{endfor}

\usepackage{tcolorbox}

\usepackage{fancyvrb}
\fvset{fontsize=\smaller}

\input{figures/sources/gtexcpp.sty}

\title[Shamrock SPH solver]{The \SHAMROCK{} code: I- Smoothed Particle Hydrodynamics on GPUs.}

\author[T. David-$\,\!$-Cléris et al.]{
T. David-$\,\!$-Cléris,$^{1,2}$\thanks{E-mail: davidclt@univ-grenoble-alpes.fr}
G. Laibe,$^{1,3}$
and Y. Lapeyre$^{1}$
\\
$^{1}$ENS de Lyon, CRAL UMR5574, Universite Claude Bernard Lyon 1, CNRS, Lyon, F-69007, France.\\
$^{2}$Univ. Grenoble Alpes, CNRS, IPAG, 38000 Grenoble, France.\\
$^{3}$Institut Universitaire de France
}

\date{Accepted XXX. Received YYY; in original form ZZZ}

\pubyear{\the\year{}}

\begin{document}
\label{firstpage}
\pagerange{\pageref{firstpage}--\pageref{lastpage}}
\maketitle

\begin{abstract}
We present \SHAMROCK{}, a performance portable framework developed in \textsc{C++17} with the \SYCL{} programming standard, tailored for numerical astrophysics on Exascale architectures.
The core of \SHAMROCK{} is an accelerated parallel tree with negligible construction time, whose efficiency is based on binary algebra. The Smoothed Particle Hydrodynamics algorithm of the \textsc{Phantom} code is implemented in \SHAMROCK{}. On-the-fly tree construction circumvents the necessity for extensive data communications.
%In a Sedov blast test performed with 65 billion of particles, \SHAMROCK{} achieves a single time step in just 7 seconds using the 1024 MI250X GPUs of the \textsc{Adastra} Cluster. This equates to processing 9 billion particles per second, with 64 million particles per MI250X. 
In tests displaying a uniform density with global timesteping with tens of billions of particles, \SHAMROCK{} completes a single time step in a few seconds using over the thousand of GPUs of a super-computer. This corresponds to processing billions of particles per second, with tens of millions of particles per GPU. The parallel efficiency across the entire cluster is larger than $\sim 90\%$.
\end{abstract}

\begin{keywords}
Methods: numerical
\end{keywords}

%%%%%%%%%%%%%%%%%%%%%%%%%%%%%%%%%%%%%%%%%%%%%%%%%%

%%%%%%%%%%%%%%%%% BODY OF PAPER %%%%%%%%%%%%%%%%%%

\section{Introduction}

The study of the formation of structures in the Universe is a field in which non-linear, non-equilibrium physical processes interact at many different scales, requiring ever greater computing resources to simulate them, right up to Exascale (one quintillion operations per second). To increase energy efficiency with acceptable CO$_{2}$ emissions, recent super computers have been designed with specialised hardware such as ARM central processing units (CPUs) or graphics processing units (GPUs). GPUs involve multiple computational units that perform the same operation on multiple data simultaneously through specific instructions (Single Instruction Multiple Data, or SIMD parallel processing). This type of hardware differs radically from standard x86 CPUs, requiring a complete rewrite of CPU-based codes.

Considerable efforts have recently been invested into developing codes adapted to the new hybrid architectures aimed at Exascale (e.g. \textsc{Idefix}: \citealt{Idefix} ; \textsc{Parthenon}: \citealt{Parthenon} ; \textsc{Quokka}: \citealt{QUOKKA}). Most of the codes in the community targeting exascale are grid-based. However, these methods can be inadequate for systems with complex geometries or misaligned flows, highlighting the need for SPH and, more broadly, Lagrangian-based approaches on exascale architectures. Currently, problems involving complex geometries, such as multiple distorted systems, are primarily studied using SPH methods. However, resolution is often limited by a lack of MPI scalability and GPU support, preventing efficient execution on exascale architectures. In particular, simulating small-scale instabilities embedded within global structures requires at least a billion SPH particles, which is an impractical number for codes without exascale capabilities.
The performance of hydrodynamical codes is conditioned by the rate at which data involved in the solver can be prepared, explaining the efficiency of grid-based Eulerian codes developed to date. For example, the multiphysics Godunov code \textsc{Idefix} uses a fixed grid, so no overhead is required when executing the numerical scheme. On the other hand, simulating moving disordered particles on Exascale architectures is a tremendous challenge, regardless of whether they are tracers for Eulerian methods, super particles for Lagrangian methods or interpolation points for Fast Multiple Moments. The rule of thumb is that performance decreases when the number of neighbours increases and when they are unevenly distributed.
Our code \textsc{Shamrock} is a performance portable framework aiming at hybrid CPU-GPU multi-node Exascale architectures. The design of \SHAMROCK{} makes it appealing for with particle-based methods such as Smoothed Particle Hydrodynamics (e.g. \citealt{gizmo,phantom,gadget}), while remaining inherently compatible with any distribution of numerical objects (grids, particles) and numerical schemes (grid-based or Lagrangian). Our strategy in \SHAMROCK{} is that the tree used for neighbour search is never updated. Instead, we are aiming for a highly efficient fully parallel tree algorithm that allows on-the-fly building and traversal, for any distribution of cells or particles. The specific nature of GPU architectures calls for a different design from the state-of-the-art methods developed for CPUs (e.g. \citealt{2011Gafton}).
The simulation domain undergoes an initial partitioning into sub-domains, fostering communication and interface exchange through an coarse layer of MPI parallelism. The core of \SHAMROCK{} is its fine layer of parallelism, which consists in processing each sub-domains on GPUs using the \textsc{SYCL} standard. This includes a tree algorithm that delivers the performance required for handling any number of objects compatible with current GPU capabilities. It combines state-of-the-art algorithms based on Morton codes \citep{morton1966, WarrenSalmon1993, Lauterbach2009FastBC} with specific optimizations, including the Karras algorithm \citep{karras2012maximizing}. The resulting tree-building time is negligible. The overall performance of \SHAMROCK{} hinges on the performance on neighbour finding \textit{on a single GPU}. Hence the need for a tree building and traversal procedure that doesn't bottleneck the hydrodynamical time step. 
In Sect.~\ref{sec:sph-muti-gpu}, we begin by discussing SPH and its specific challenges on Multi-GPU architectures. Next, in Sect.~\ref{sec:SHAMROCK}, we introduce the \SHAMROCK{} framework. In Sect.~\ref{sec:tests}, we present results from standard hydrodynamical tests, as well as the application of \SHAMROCK{} to simulating a circumbinary astrophysical disc. Our implementation is nearly identical to that of the \textsc{Phantom} code, enabling performance assessments and comparisons on both single CPU, GPU and multiple GPUs (Sect.~\ref{sec:perf}). Finally, in Sect.~\ref{sec:prospects}, we discuss potential future directions for \SHAMROCK{}.

\iffalse
In Sect.~\ref{sec:num}, we first present a tree algorithm that has the required level of performance for any number of objects compatible with current GPU capabilities. It combines state-of-the-art algorithms on Morton codes \citep{morton1966,WarrenSalmon1993,Lauterbach2009FastBC} with specific optimisations, a key feature being the Karras algorithm \citep{karras2012maximizing}. The resulting tree building time is negligible. We detail the subsequent implementation of a Smoothed Particle Hydrodynamics solver (SPH) in \SHAMROCK{} in Sect.~\ref{sec:SHAMROCK}, and present the results obtained on standard astrophysical tests in Sect.~\ref{sec:tests}. Our implementation is almost identical to that of the \textsc{Phantom} code, facilitating performance assessments and comparisons on both one and multiple GPUs (Sect.~\ref{sec:perf}). We discuss potential future directions for \SHAMROCK{} in Sect.~\ref{sec:prospects}.
\TODO{Check refs}
\fi
\section{SPH on Multi-GPU architectures}
\label{sec:sph-muti-gpu}

\subsection{SPH}
\subsubsection{Equation of motions}

Smoothed Particle Hydrodynamics (SPH), initially introduced by 
\citet{1977Lucy, 1977Gingold}, is a Lagrangian approach widely 
employed in astrophysics. It is used for its capacity to handle 
complex geometries, adapt resolution to follow mass, address free 
boundary conditions, and offer an alternative approach to grid-based 
methods for validating nonlinear solutions.
In \SHAMROCK{} we have implemented the SPH hydrodynamical scheme  
of the \textsc{Phantom} code \cite{phantom}. 
Density estimates for each particle $a$ are obtained following
\begin{align}
\rho_a & = \sum_b m_b W_{ab}(h_a), \label{eq:SPHmass} \\
W_{ab}(h_a)&  = \frac{C_{\mathrm{norm}}}{h_{a}^{3}} f 
\left( \frac{\lvert \mathbf{r}_a-\mathbf{r}_b\rvert }{ h_a}\right),
\label{eq:rhosphsum}
\end{align}
$\mathbf{r}_{a}$, $m_{a}$, and $h_{a}$ represent the position, 
fixed mass, and smoothing length of particle $a$, respectively. 
When $h_{a}$ is itself a function of the density, 
Eqs.~\ref{eq:SPHmass} -- \ref{eq:rhosphsum} should be solved 
consistently (see Sect.~\ref{sec:adaptsmoothinglenght}). 
In practice, 
particles are semi-regularly arranged. The interpolation kernel $W$ is 
a bell-shaped function that converges weakly towards a delta Dirac 
distribution when the smoothing length $h$ goes to zero. 
$C_{\mathrm{norm}}$ is a normalisation constant for the kernel, 
calculated for a three dimensional domain of simulation.
In \SHAMROCK{}, typical functions $f$ with finite compact supports, 
such as \cite{Schoenberg1946} B-splines like $M_4$, $M_5$, $M_6$ , 
or Wendland functions like $C_2$, $C_4$, $C_6$ 
(see e.g., \citealt{Wendland1995}), are implemented. 

The as detailed in \cite{2012Price, phantom} the equation of motions 
are :
\begin{align}
{\mathrm{d} \mathbf{v}_a \over \mathrm{d} t} & = \sum_b m_b \left( \frac{P_a + q^a_{ab}}{\rho_a^2 \Omega_a} \nabla_a W_{ab}(h_a) + \frac{P_b + q^b_{ab}}{\rho_b^2 \Omega_b} \nabla_a W_{ab}(h_b)  \right) , \label{eq:sphvel}\\
{\mathrm{d} {u}_a \over \mathrm{d} t} &= {P_a + q^a_{ab} \over \rho_a^2 \Omega_a} \sum_b m_b \mathbf{v}_{ab} \cdot \nabla_a W_{ab}(h_a) + \Lambda_{\rm cond} , \label{eq:sphuint}
\end{align}
where
\begin{align}
    q_{a b}^a & = \begin{cases}-\frac{1}{2} \rho_a v_{\mathrm{sig}, a} \boldsymbol{v}_{a b} \cdot \hat{\boldsymbol{r}}_{a b}, & \boldsymbol{v}_{a b} \cdot \hat{\boldsymbol{r}}_{a b}<0 \\ 0 & \text { otherwise, }\end{cases} ,\\
    v_{\mathrm{sig}, a} & = \alpha_a^{\rm AV} c_{s,a} + \beta |\mathbf{v}_{ab} \cdot \hat{\boldsymbol{r}}_{a b}|~, \, \alpha_a^{\rm AV} \in [0,1] .\\
    \Omega_a & =  1 - \frac{\partial h_a}{\partial \rho_a} \sum_b m_b {\partial W_{ab} (h_a) \over \partial h_a}, \label{eq:sphomega}\\
    \Lambda_{\rm cond} & = \sum_b m_b \beta_u v_{\rm sig}^u (u_a - u_b) {1 \over 2} \left[
       {F_{ab}(h_a) \over \Omega_a \rho_a} + 
       {F_{ab}(h_b) \over \Omega_b \rho_b}
    \right] ,\\
    v_{\rm sig}^u & = \sqrt{ |P_a - P_b| \over (\rho_a + \rho_b) /2},\\
    F_{ab}(h_a) &= \hat{\mathbf{r}}_{ab} \cdot \nabla_a W_{ab}(h_a)
\end{align}

\subsubsection{Neighbouring criterion}

The conservative SPH formalism, derived from the variational principle, involves the kernel terms $\nabla W_{ab}(h_a)$ and $\nabla W_{ab}(h_b)$.The SPH forces can be nonzero for any distance $r_{ab}$ between two particles $a$ and $b$ that is smaller than the maximum smoothing length of either particle, formally expressed as $a$ and $b$, formally expressed as $r_{ab} < \max(h_a, h_b)$. This implies that the list of neighbours of a particle $a$ does not depend solely on its own properties, but also on those of its neighbours Hence, the interaction criterion depends also on the properties of the particle $b$.

In order to look for neighbours, one must have a criterion to check if a group of particles (such as an intermediate node in the search tree) interact with the particle of interest. This is achieved by generalizing the pairwise interaction criterion into a particle-to-group interaction criterion (see App.~\ref{sec:sph-interact-crit}). Lastly, the ghost zones between subdomains introduce a final type of interaction criterion, known as the group-group criterion, which identifies whether two distinct particle groups contain a pair of particles with a nonzero interaction (see App.~\ref{sec:patch-interact-crit}).

\subsection{Multi-GPU architectures}

Modern computer hardware harnesses graphics processing units (GPUs) as computing accelerators. A typical compute node configuration consists of several GPUs connected to a CPU via a PCI express or other proprietary interconnect. Each GPU is equipped with its network card, enabling direct communications from one GPU to another with the Message Passing Interface (MPI) protocol, using so called IPC (Interprocess communications) handles. 
The GPUs themselves are 
specialised hardware capable of exploiting the full bandwidth of their high speed memory in tandem with a high compute throughput using SIMT (Single instruction, Multiple Threads) and SIMD (Single Instruction Multiple Data). This design renders GPUs more potent and energy-efficient than CPUs, especially for processing parallelized tasks involving simple, identical operations.
Performing simulations on architectures comprising thousands of GPUs introduces several challenges: evenly distributing the workload among the available GPUs (load balancing problem), communicating data between domains to perform the computation while moving the communication directly to the GPU if possible (communication problem), structuring and organising the workload on the GPU into GPU-executed functions called \textit{compute kernels} to make the best use of hardware capabilities (algorithmic problem). The first two points are are common issues associated with MPI, while the third is specific to GPU architecture, raising the question of choosing an appropriate backend.

\subsection{Challenges}

\subsubsection{Large number of particles}

The number of particles in a given simulation may exceed the available memory of a single GPU. This imposes a decomposition strategy to distribute the computational workload across the available hardware.
To perform such decomposition we decompose the simulation into subdomains called \textit{patches} and distribute them across the available GPUs or CPUs.
As SPH particles move, they must be able to transfer between GPUs when crossing simulation subdomains. Moreover, if particles concentrate locally, the domain decomposition must be adapted to maintain efficient load balancing.
To address this, \SHAMROCK{} employs an adaptive patch-based decomposition strategy, where subdomains can split or merge, similar to an AMR grid. These patches operations are performed based on an estimated load value (see Sec.~\ref{sec:coarse-par-mpi-decomp}).

\subsubsection{Uneven and dynamical particle distribution}

The distribution of SPH particles is uneven and evolves dynamically. Their positions, as well of the corresponding number of neighbours, are not known in advance. Therefore, a tree algorithm is used to search for neighbours on the fly. Several efficient algorithm has been developed for CPUs (e.g. \citealt{2011Gafton}). On GPUs, a key challenge for tree algorithms consists in ensuring an even distribution of computational load across particles to maintain the internal load balancing of the GPU, and prevent performance degradation. In designing the kernels of \SHAMROCK{} kernels, we apply the strategy of reducing task unevenness whenever possible.

\subsubsection{Tree traversal on GPUs}

On GPUs, dynamic allocation is generally not allowed. Therefore, when performing a tree traversal, the current paths being tested must be stored in an array with a predefined size. Additionally, since this array is frequently accessed, it should reside in the innermost level of memory, namely the registers.  However, register space is limited to 128 integers or floats. Exceeding this limit forces the GPU to reduce the number of parallel threads to fit within the available register bank, potentially impacting performance. Using a complete binary tree with a known depth satisfies these constraints, since the stack required to store the currently tested path can be sized according to the tree depth (see App.~\ref{sec:treetrav}).
While an octree would better align with the geometry, it would require a stack size approximately seven times larger, exceeding register capacity. This would lead to register spilling and, consequently, reduced performance.

\subsubsection{Unknown interaction radius}

From Eq.~\ref{eq:sphvel}, the interaction criterion of particle $a$ depends not only on its smoothing length $h_{a}$, but also on the smoothing lengths of its neighbours $h_{b}$. To handle this in practice, we compute the maximum smoothing length possible within each tree cells during the tree traversal. The details of this interaction criterion are discussed in details in App.~\ref{sec:sph-interact-crit}.

\subsubsection{Inefficiency of the single loop approach}

One of the leading approaches for an effective implementation of an SPH code, referred to as the single-loop approach, consists in merging the various solver steps into a single updating loop \citep{phantom}.  However, when decomposing into subdomains, the number of SPH particles required in the ghost zones of each subdomain depends on their smoothing lengths. This implies that updating the smoothing length of the particles changes the size of the ghost zones. This motivates an alternative approach where the smoothing lengths are updated first, the ghost zones are then communicated based on the new smoothing lengths, and finally, the derivative update is performed (see App.~\ref{sec:sph-timestep}).
This approach is also motivated by the limited sizes of the registers on GPUs. Employing a single-loop approach would necessitate a large computational kernel that demands too many registers, potentially leading to register spilling or reduced occupancy, both of which negatively affect performance.  In general, we favour many smaller kernels over a few larger ones.

\subsubsection{Large number of neighbours}

In a typical 3D SPH simulation, the average number of neighbours is around 60 (the value obtained for a $M_4$ kernel with $h_{\rm fact} = 1.2$). This number can increase when using larger kernels such as $M_{5}$ and $M_{6}$, or even more for Wendland kernels. This is much larger than typical values encountered in finite volume or finite element methods, where only about 6 neighbouring cells are involved. Such large number of neighbours has to be taken in consideration in the development of the code, since this leads to larger ghost zones and larger neighbour index tables. In practice, we have chosen to store the list of the indexes of the neighbours since the performance gains outweigh the additional memory required for these larger tables.

Moreover, when using a tree to search for neighbours, we aim to have them distributed over as few tree nodes as possible to avoid additional search steps. This key point is central for the efficiency of \SHAMROCK{}. To achieve this, we have developed an original reduction algorithm, which compresses the deepest leafs of the tree (this algorithm is presented in details in App.\ref{sec:reduc-alg}).

%\textcolor{green}{ distrib aleat part, part move,  sym h,  omega ghost zones,  60 neighbors (teaser reduc)  }

\section{The \SHAMROCK{} framework}
\label{sec:SHAMROCK}

\subsection{Choice of languages and standards}

GPU vendors have developed various standards, languages and libraries to handle GPU programming, the most widely used to date for scientific applications being \CUDA{} and \ROCM{}, which are vendor-specific. To address the issue of portability, libraries and standard have been created to enable the same code to be used on any hardware from any vendors. Current options include \textsc{Kokkos} \citep{Kokkos}, \textsc{Openacc} and \OpenMP{} (target). 
The \SYCL{} standard, released by Khronos in 2016, is a domain-specific embedded language compliant with \textsc{C++17}, which is compiled to the native \CUDA{}, \ROCM{} or \OpenMP{} backend. With a single codebase, one can directly target directly any GPUs or CPUs from any vendors, eliminating the need for separate code paths for each supported hardware.
To date, the two main \SYCL{} open source compilers are \textsc{AdaptiveCpp} \citep{HipSYCL1,HipSYCL2}, and \textsc{OneAPI/DPC++}, which is maintained by Intel. 
Among other heterogeneous parallelisation libraries, we use the \SYCL{} standard to develop \SHAMROCK{}, since it offers robustness, performance \citep{LumiSYCL}, portability \citep{CloverLeafSYCL,PerfSYCLCUDA} and potential for durability. \SYCL compilers can also generally compile directly to a native language without significant overhead, delivering near-native performance on Nvidia, AMD and Intel platforms (e.g. tests with \textsc{Gromacs}, \citealt{GromacsSycl,abraham2015gromacs, alekseenko2024gromacs}).
Since \textsc{C++} code written using \SYCL{} is compiled directly to the underlying backend (\CUDA{} or \ROCM{} or others), we harness direct GPU communication and use vendor libraries directly in the code.

\subsection{Two level parallelism}

\begin{figure*}
    \centering
    %\captionsetup{format=sanslabel}
    \begin{subfigure}{.30\textwidth}
       \centering
       %\fbox{
       \includegraphics[height = 0.24\textheight]{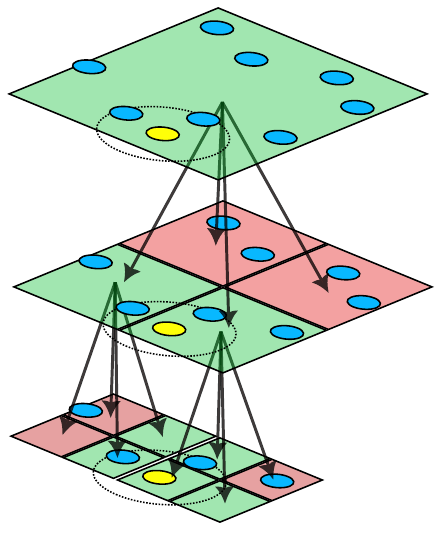}
       %}
       \caption{}
   \end{subfigure}
   \begin{subfigure}{.30\textwidth}
       \centering
       %\fbox{
       \includegraphics[height = 0.24\textheight]{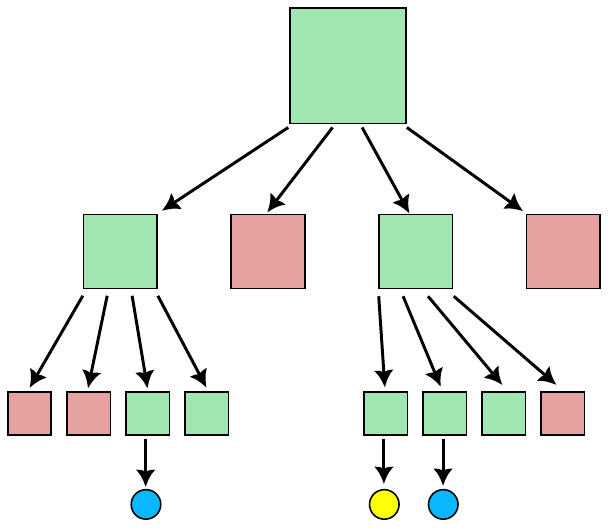}
       %}
       \caption{}
   \end{subfigure}

    \caption{An example of a tree structure for particle sets and its use in neighbour search: (a) shows the top layer with the particles in the simulation domain, recursively subdivided into 4 subdomains. The particle of interest is highlighted in yellow, domains intersecting the interaction radius in green, and non-intersecting domains in red. (b) illustrates the hierarchical structure of the same tree.}
    \label{fig:tree-schem}
 \end{figure*}

Most modern supercomputers adopt a fat-tree topology, where computing nodes are grouped into islands, with all nodes within an island directly connected to each other. The supercomputer consists of multiple islands, each connected to the others. This structure introduces locality effects, as peak communication performance (in terms of latency and bandwidth) is only achieved for communications within the same island.
For this reason, we aim for close subdomains of the simulation to ideally be within the same island, such that the communication topology deviates as little as possible from the communicator topology.
To do so, we use a Hilbert curve ordering for load balancing, which provides strong locality in how subdomains are distributed across computing nodes. The details of the load balancing procedure are provided in App.~\ref{sec:load-balancing}.

In heterogeneous supercomputers, an additional layer of decomposition is introduced as every computing nodes posses multiple accelerators (typically GPUs). A simple way to account for the hierarchy of the communicator topology is to assign a process per accelerators and treat communications as communications between accelerators rather than between nodes. This strategy aligns with the implementation of accelerator-aware \MPI{} routines and allows for direct communication of data between two accelerators without going through the host CPU (see CUDA-aware \MPI{} for example), since most heterogeneous supercomputers have network cards directly attached to the accelerators.

In the following, we use \textit{coarse parallelism} to refer to the decomposition of the problem across available accelerators, and \textit{fine parallelism} to describe the highly parallel tasks executed within each accelerator.

\subsubsection{Coarse parallelism : MPI decomposition}
\label{sec:coarse-par-mpi-decomp}

We start by discussing coarse parallelism, since its design influences the implementation of fine parallelism.
Coarse parallelism in \SHAMROCK{} involves dividing the simulation box, which contains potentially inhomogeneously distributed objects, into smaller subdomains that are distributed across the nodes of the computing cluster.
For convenience, we shall further refer to these subdomains as \textit{patches}.

In \SHAMROCK{}, patches are constructed following a procedure of recursive refinement. Starting from the simulation box, patches are divided into eight patches of equal sizes by splitting in two equal parts the original patch on each axis. 
The resulting structure is an \textit{octree}, where each node is either a leaf or an internal node with eight children.
The patches managed by \SHAMROCK{} are the leaves of this octree. We call this structure the \textit{\textit{patch octree}} of \SHAMROCK{}.
The {patch octree} is similar to the structure of a three-dimensional grid that has been adaptively refined (AMR grid). The cells of this AMR grid would correspond to the patches of \SHAMROCK{}.
Similar to an AMR grid, patches can be dynamically subdivided or merged.

To each patch $p$, we associate an \textit{estimated load} $\mathcal{W}_p$, which is an estimate of the time required to perform the update of the patch. The load depends \textit{a priori} on the type of simulation chosen by the user (e.g. fixed or refined grids, particles).
If the estimated load of a patch exceeds a maximum threshold, i.e. $\mathcal{W}_p > \mathcal{W}_{\rm max}$, the patch is subdivided. If the estimated load is below a minimum threshold $\mathcal{W}_{\rm min}$, the patch is flagged for a merge operation. In \SHAMROCK{}, patch merging is performed when all eight patches corresponding to the same node in the patch octree are flagged.
To avoid cycling between subdivisions and merges, we enforce
$\mathcal{W}_{\rm max} > 4 \mathcal{W}_{\rm min}$.
Hence, the decomposition of the simulation box into patches is only controlled by the values of $\mathcal{W}_{\rm min}$ and $\mathcal{W}_{\rm max}$ and is independent of the number of \MPI{} ranks.

For a given patch decomposition, we distribute the patches across the \MPI{} ranks, such that every rank has a similar sum of the patches' estimated load. For that purpose we employ an Hilbert curve, which is efficient as each patch has a similar load on average, which is provided by the decomposition based on the estimated load. 
Hence, we distribute the patches across the GPUs according to an Hilbert curve. To avoid confusion, We do not decompose the domain along the Hilbert curve, but distribute the patches across the GPUs according to it.

For the user, this approach allows for a simulation to be restarted even when the number of accelerators is changed without modification to the behaviour of the code.
Additionally, since the domain decomposition is written in term of patches the result should be independent of the number of accelerators available.
\SHAMROCK{} maps several patches to a given \MPI{} rank in a dynamical manner (see \ref{sec:load-balancing}). We call this decomposition an \textit{abstract domain decomposition}. 
In practice, we find that 10 patches per \MPI{} rank provides a compromise between the level of granularity required for effective load balancing and the overheads associated with patch management.
The details of the updating step for the domain decomposition are given in App.~\ref{sec:schedstep}.

\subsubsection{Coarse parallelism : \MPI{} communications}

In hydrodynamical simulations, interactions among objects are predominantly local, resulting in each patch being connected to only a limited number of neighbouring patches (see the treatment of \textit{ghost-zones} detailed in App.\ref{sec:patch-interact-crit}). A crucial element of communication management in \SHAMROCK{} consists in upholding this sparsity, thus avoiding unnecessary \MPI{} communications. Additionally, we aim at having abstracted communications in the code, in order for the communications routines to be hidden from the user as well as allowing us to change the implementation of the communications without the need to change any subsequent calls to the protocol.
As such, we have developed in \SHAMROCK{} a sparse communication protocol where inputs and outputs are a the combination of a buffer on the GPU, as we employ direct GPU-GPU communications, and a pair of sender receiver \MPI{} rank (details of the implementation of the sparse communication are described in App.~\ref{sec:sparse-mpi-comm}).
Lastly, as particles move, they need to be exchanged between patches. This operation is similar to ghost-zone communication and can therefore use the same communication protocols.

\subsubsection{Fine parallelism : Tree traversal}

Having now decomposed the problem across the available accelerators, we need to be able to search efficiently for neighbours in a parallel fashion.
To do so, a common technique is to make use of a tree, or more specifically a \textit{Bounding Volume Hierarchy} (BVH) \citep{RubinWhitted1980}.
In this approach, particles are put in boxes which span a range of coordinates that encompass all the contained objects (hence the name \textit{Bounding Volume}).
Those bounding volume can be defined hierarchically, with each box bounding volume enclosing the volume of its children, forming a \textit{tree} structure. A leaf in this structure refers to a box that contains no child boxes but only the aforementioned objects. An example of such a tree structure in two dimensions is sketched in Fig. \ref{fig:tree-schem}).

Various types of tree exist. In \SHAMROCK{}, we will use a complete binary tree rather than octrees. A complete binary tree is defined as a tree where each node is either a leaf or an internal node with exactly two children.
The patch structure of \SHAMROCK{} forms an octree. However, for fine parallelism, a binary tree is used instead since the register size of a GPU is limited, and a tree traversal procedure using a stack with an octree would require too many registers leading to register spilling, and therefore reduced occupancy and performance.

In \SHAMROCK{}, tree traversal is performed in two steps (details of the algorithm implementation are provided in App.~\ref{sec:twostageneighcache}). The first stage involves selecting a leaf of the tree and identifying the indices of other leaves that may contain neighbours of the objects within the chosen leaf. This step can be executed in parallel for all leaves. Once the neighbouring leaves for each leaf are determined, the next step is to identify the parent leaf of each object. We then search for neighbouring objects within the neighbouring leaves of the parent leaf. The corresponding algorithms, along with a rigorous interaction criteria for defining neighbours, are comprehensively detailed in Appendix \ref{sec:sph-interact-crit}. This two-stages search contrasts with a one-stage approach, which would involve direct search for neighbours using the tree.

Lastly, since the neighbour searching operation can be quite costly, for each particle, we aim to reduce the number of leaves containing neighbours of the target particle. In practice, this is achieved using a novel \textit{reduction algorithm} (detailed in Appendix \ref{sec:reduc-alg}), which, in the best-case scenario for SPH, targets around 10 particles per leaf. This approach, combined with the two-stage search, yields a tenfold speedup compared to a dual- or single-stage approach without reduction (see Fig.~\ref{fig:neighcache}), where the test is outlined in Section \ref{sec:cachebuild}). As such, reduction is one of the key algorithm of \SHAMROCK{}.

\subsubsection{Fine parallelism : Tree building}

To build the binary tree without compromising the performance of the code or unnecessarily over-communicating the trees within ghost zones, we have implemented an advanced tree-building method originally developed for ray-tracing trees \citep{karras2012maximizing}, which ensures that the tree building time is negligible compared to the time of the entire hydrodynamical step. This algorithm, which relies on Morton codes and binary algebra, is one of the key of the performance of \SHAMROCK{}. This tree-building efficiency allows us to discard trees at the end of each time step and simply recompute them when necessary. Furthermore, for ghost zones, we append them to the subdomain that is currently updated, and build the tree over the subdomain extended by its ghost zones.
As a result, the tree can therefore be updated during the simulation to account for changes in the distribution of objects. In practice, the current tree building time accounts for about 5\% of the total timestep cost. Detail of the tree building procedure are provided in Appendix \ref{sec:treebuild}.
We note here that the advantage of the \cite{karras2012maximizing} algorithm relies on its portability to GPUs as well as its single pass approach as opposed to a bottom-up approach with multiple passes (e.g. \citealt{TokuueIshiyama2024}).

\subsubsection{Fine parallelism: neighbours cache}

In \SHAMROCK{}, in order to perform operations involving neighbours, we rely on a neighbour list.
We use a list of neighbours for 2 reasons. Firstly, without a list, one should perform in SHAMROCK at least 2 neighbour search (one for the smoothing length iteration, one for derivatives). Secondly, as mention in Sect. 2.3.2, tasks performed on GPUs should be as homogeneous as possible to avoid warp divergence.  As such, nesting physical operation within the result of a neighbour search would be detrimental for performance. As a result, one should first search for neighbours and then, perform operations with them. Thirdly, we have noticed that such approach is always beneficial to performance in our case, even though this increase significantly memory consumption in the code.

\subsection{Universality of the framework}

\begin{figure}
    \centering
    \includegraphics[width=1\linewidth]{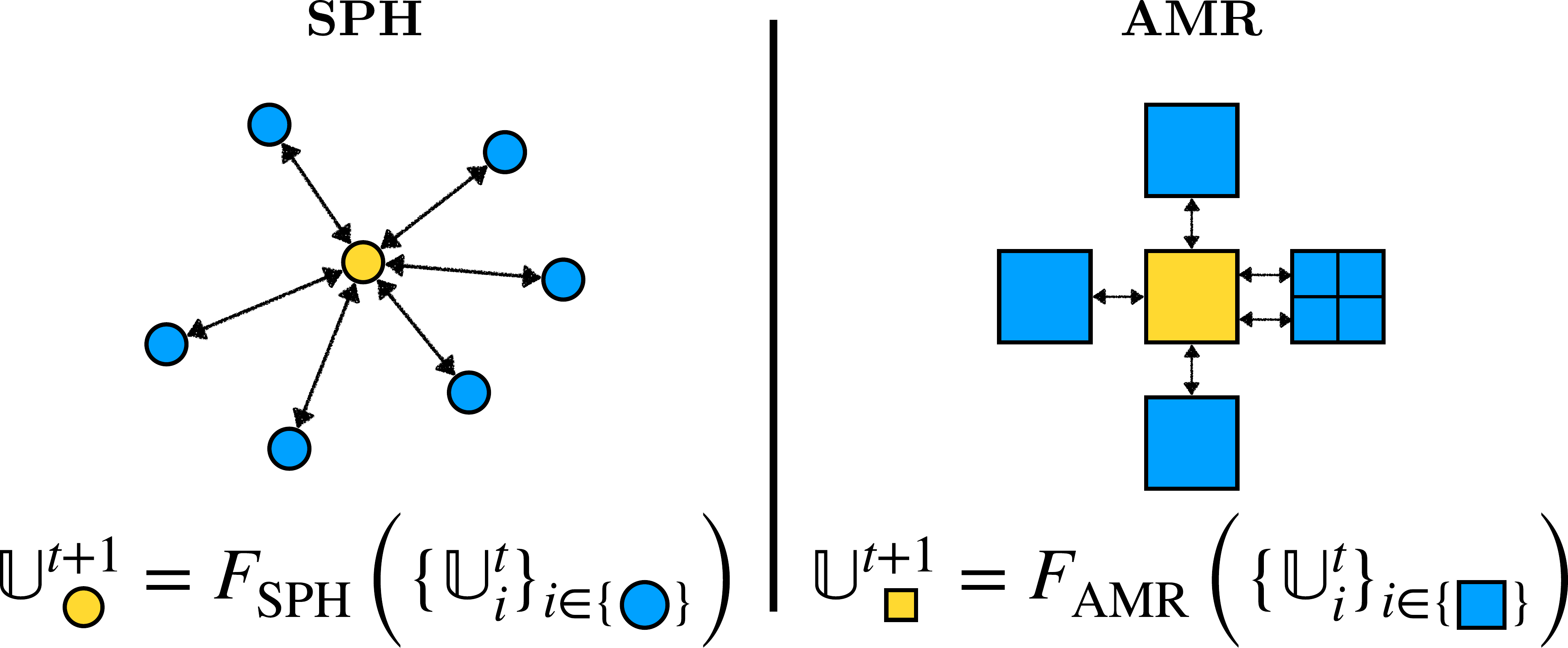}
    \caption{Numerical integration of an hydrodynamic quantity $\mathbb{U}$ involves finding neighbours $i$ (particles, cells), then adding their contributions according to the chosen solver $F$.}
    \label{fig:illust_amr_sph}
\end{figure}
\begin{figure}
    \centering
    \includegraphics[width=0.7\linewidth]{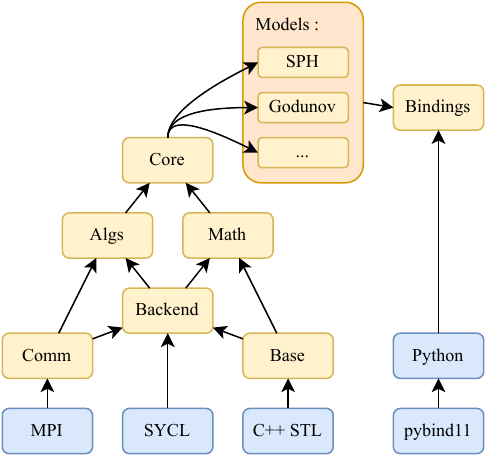}
    \caption{Internal structure of \SHAMROCK{}: functionalities for calculating neighbour finding are organised in different layers of abstraction, enabling the independent treatment of any numerical scheme (Models).}
    \label{fig:framework_org}
\end{figure}

% hydro: PDES -> discrete derivatives, local inyterpolations
% non-local such as gravity : hierchical sums over neighbours

% ref a la figure

%ICI-----

Computational fluid dynamics consists in discretising a physical system of partial differential equations, alongside specifying initial and boundary conditions. Deterministic numerical schemes can be viewed as being the combination of neighbour finding and specified arithmetic (see Fig.~\ref{fig:illust_amr_sph}). As such, an algorithm capable of operating on neighbours, as well as handling the distribution of the problem across multiple computing nodes can provide a generic framework for implementing schemes frequently used in astrophysics, such as Lagrangian Smoothed Particle Hydrodynamics (SPH) and Eulerian Adaptive Mesh Refinement (AMR) grid-based approaches, within a unified structure.  This is the very purpose of \SHAMROCK{}: to abstract optimised neighbour search as well as domain decomposition and load balancing, in a way that allows users to implement new schemes with minimal modifications. Fig.\ref{fig:framework_org} sketches the \SHAMROCK{} framework: a collection of libraries connected by standardised interfaces, where models (CFD solvers or analysis modules) are implemented atop these libraries.

\subsection{SPH scheme in \SHAMROCK{}}

We base the implementation of the SPH scheme from the one of \cite{phantom}, with the following key differences:
\begin{itemize}
    \item We replace the KD-tree used in \textsc{Phantom} by our radix tree and neighbour finding procedure, enabling on-the-fly tree reconstruction without requiring updates or storage between timesteps. This also simplifies the management of ghost zone management and the problem distribution, which is crucial for \textsc{Shamrock}'s scalability. (see Appendices \ref{sec:num}). 
    \item We do not use the single loop approach, since the result from the smoothing length update sets the value of $\Omega_a$ which must be communicated to the ghost zones before computing the new derivatives. For that reason, we communicate ghost zones extended by a tolerance to account for possible evolutions of the smoothing length. We then update the smoothing length and compute the new value of $\Omega_a$.  Finally, we communicate the remaining fields and $\Omega_a$ before computing the new derivatives and completing the timestep integration.The full timestep loop is detailed in Appendix \ref{sec:sph-timestep}.    
    \item Rather than using a local cache for neighbour indices, we compute a global index cache. This approach involves more memory, but provides the best overall performance in our case (see Sect. \ref{sec:single-gpu-cpu-perf}). 
\end{itemize}

\subsection{Element of software design}

The software design of \SHAMROCK{} relies on :
\begin{itemize}
\item A modular organisation of the code structured around interconnected \textsc{cmake} projects.
\item Python bindings implemented using \textsc{pybind} \citep{jakobpybind11}. \SHAMROCK{} runs are set via Python scripts that interact with the code through these bindings, allowing \SHAMROCK{} to function as both a Python library and a Python interpreter (typically on supercomputers) without modifying the run script.
\item A comprehensive, automated library of test that supports multiple configurations of compilers, targets, and versions.
\item Version control development using \textsc{Git}, with CI/CD pipelines running on \textsc{Github}. In this workflow, pull requests are merged only after passing all tests and receiving approval from a contributor. Issues and documentation are also managed within the same repository.
\item Automated code deployment across machines using environment scripts.
\end{itemize}
Further details are provided in App.~\ref{sec:sd}

\section{Physical tests}
\label{sec:tests}

\subsection{Generalities}

\gl{Firstly, we validate the SPH solver by performing convergence tests against classical problems having analytical solutions, such as the Sod tube and the Sedov-Taylor blast test. The hydrodynamic tests presented in this Section are performed using the $M_6$ kernel with $h_{fact} = 1.0$, with an average number of neighbours of $113$ neighbours for each SPH particle (almost no difference is expected in the results when using other spline kernels, e.g. \citealt{phantom}). In all tests, momentum is conserved to machine precision. The choice of the CFL condition result in energy deviations that do not exceed $10^{-6}$ relative error with respect to the initial value.}

\gl{Secondly, we examine the residuals obtained from comparing the results generated by \SHAMROCK{} and \textsc{Phantom}. Since both codes use the same SPH algorithm, such an analysis is required for conducting further performance comparisons. L2 errors are estimated using the norm
\begin{align}
\mathcal{L}_2 (A_{\rm sim}, A_{\rm ref}) = \sqrt{\frac{1}{N_{\rm part}} \sum_i \vert A_{i,\rm sim}- A_{i,\rm ref}\vert^2} ,
\end{align}
where $A_{i,\rm ref}$ represents the reference quantities, while $A_{i,\rm sim}$ denotes the quantities computed in the simulation.}

\subsection{Advection}

\begin{figure}
    \centering
    \includegraphics[width=0.95\linewidth]{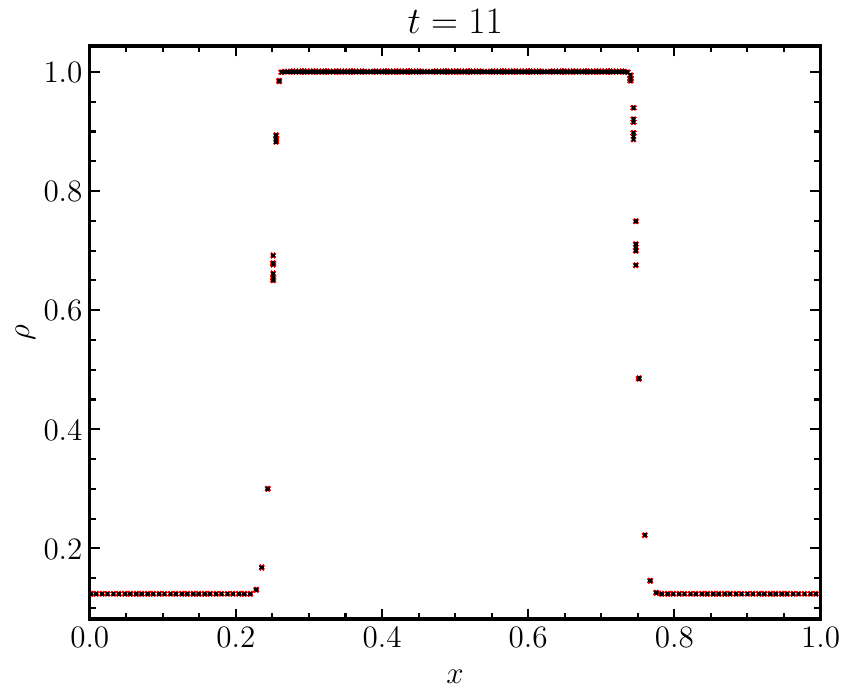}
    \caption{\gl{Advection of a density step across several traversal of a periodic box, in code units. SPH being Galilean invariant, the results (black dots) precisely match the initial setup (red crosses) down to machine precision, thus validating the boundary treatment in \SHAMROCK{}.}
    }
    \label{fig:advecttest}
\end{figure}

\gl{We first perform an advection test in a periodic box of length unity to verify the correct treatment of the periodic boundaries by \SHAMROCK{}. Three lattices of $(16\times 12 \times 12)$, $(64\times24\times24)$ and $(16\times12\times12)$ particles having velocities $v_{x} = 1.0$ are initially juxtaposed, such that $\rho = 1.0$ if $0.25 \le x \le 0.75$, and $\rho = 0.1$ elsewhere. We let the simulation evolve until the step has crossed several times the boundaries of the box (we choose $t = 11$, although any other time, even very large, yields the same outcome since SPH is Galilean invariant). The result obtained at the end of the simulation is identical to the initial setup to machine precision.}

\subsection{Sod tube}

\begin{figure*}
    \centering
    \includegraphics[width=0.73\linewidth]{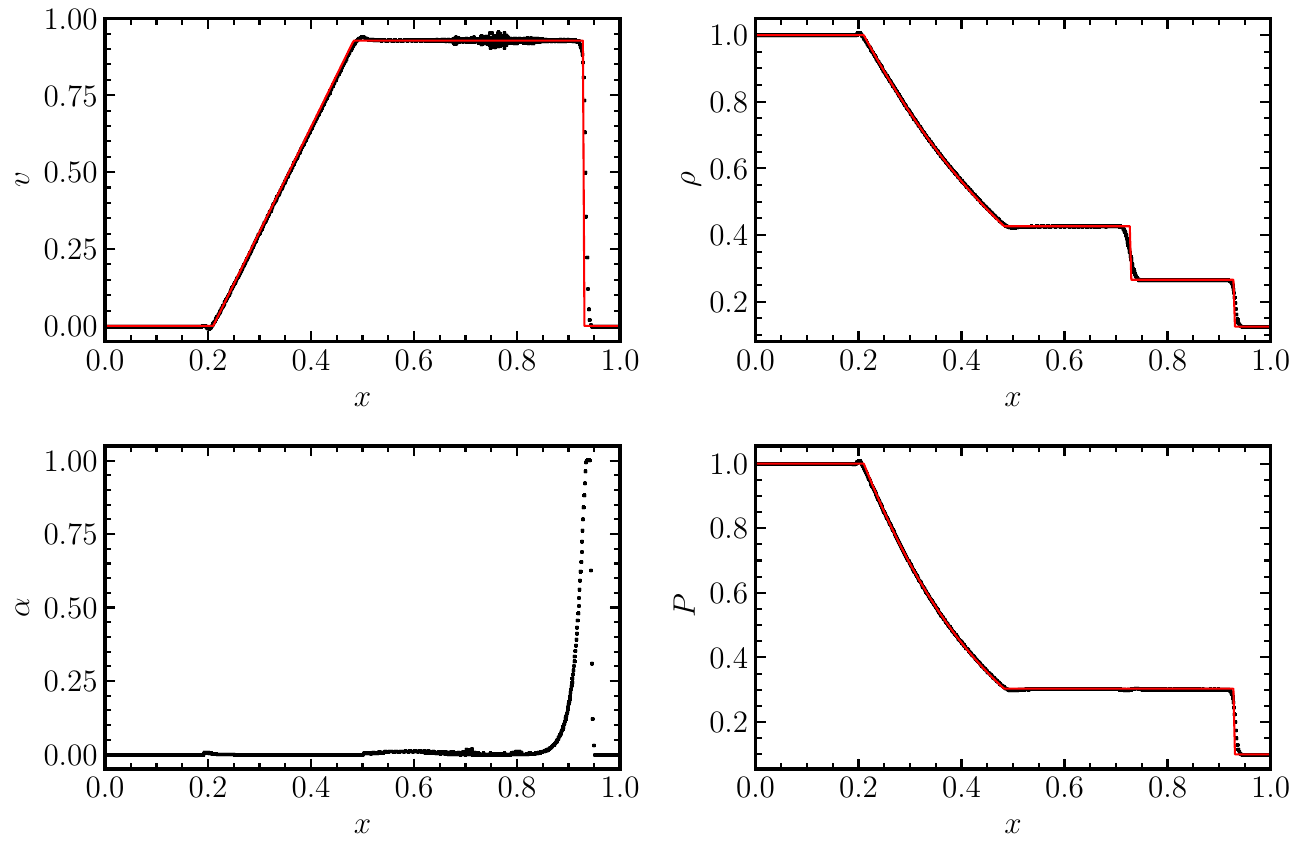}
    \caption{
        Result obtained for the Sod-tube test by juxtaposing two tubes of $24\times24\times512$ particles
        in $x\in[-0.5,0.5]$ and $12\times12\times256$ particles in $x\in[0.5,1.5]$ organised in hexagonal compact packing lattices. The density is set to $\rho=1$ in $x\in[-0.5,0.5]$ and $\rho=0.125$ in $x \in [0.5,1.5]$. Initial pressure is $P=1$ for $x\in[-0.5,0.5]$ and $P=0.1$ for
        $x\in[0.5,1.5]$, with zero initial velocities. An adiabatic
        equation of states with $\gamma=1.4$ is used. Boundaries are periodic, and only half of the simulation is displayed. 
        The simulation is performed until $t=0.245$, and numerical results are compared against the
        analytic solution. \tdc{We additionally show the values of the shock viscosity parameter $\alpha$.}
    }
    \label{fig:sodtubetest}
\end{figure*}

\begin{figure}
    \centering
    \includegraphics[width=0.9\linewidth]{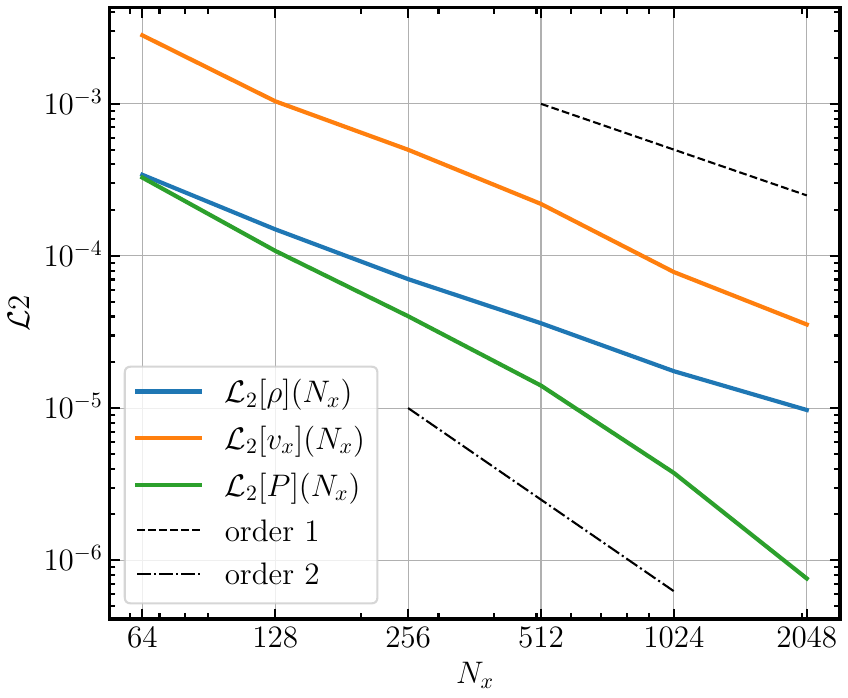}
    \caption{\gl{$\mathcal{L}_2$ errors obtained for the Sod shock tube test presented on Fig.\ref{fig:sodtubetest} as a function of the number $N_x$ of particles used on the $x$ axis for $x\in[-0.5,0.5]$.  We observe second-order convergence on the pressure, first-order convergence in density, and in-between convergence in velocity, as found in other SPH codes.}}
    \label{fig:sodtubetestL2err}
\end{figure}

\gl{We perform a Sod-tube test \citep{1978Sod} by setting up a box with a discontinuity between a left state and a right state initially positioned at $x=0.5$. In the left state $x < 0.5$, the density and the pressure are set to $\rho_{\rm l} = 1$ and $P_{\rm l} = 1 $, while in the right state $x > 0.5$, they are set to $ \rho_{\rm r} = 0.125$ and $P_{\rm r}=0.1$ respectively. 
To initialise the density profile, we use a periodic box in which we setup a $24\times 24\times 256$ hexagonal close packed lattice in $x\in[-0.5,0.5]$ and $12\times 12\times 128$ in $x\in[0.5,1.5]$.
The initial velocity is uniformly set to zero throughout the simulation. The size of the simulation box size is adjusted such that we ensure periodicity across the $y$ and $z$ boundaries.
We use $\gamma=1.4$ to align our test with the Sod tube test commonly performed in grid codes. No particle relaxation step is used in this test, since the initial distribution of SPH particles closely resembles a relaxed distribution akin to a crystal lattice. We use periodic boundaries in the $x$ direction.
Shock viscosity is setup with the default parameters of \SHAMROCK{}, namely $\sigma_d = 0.1,~\beta = 2,~\beta_u = 1$.
The setup presented above is then evolved until $t=0.245$. Fig.~\ref{fig:sodtubetest} shows results obtained for velocity, density, and pressure, displaying additionally the shock-capturing parameter $\alpha$. For $N_{x} = 128$ particles $\mathcal{L}_2$ errors are $\sim 10^{-3}$ in $v$ and $\sim 10^{-4}$ in $\rho$ and $P$, similarly to what is obtained with other SPH codes.
Similar setups are used to perform convergence analysis, except for the lattice, for which we use $24\times 24\times N_x$, and $12\times 12\times (N_x/2)$ particles instead respectively. Results obtained when varying the value of $N_x$ are reported in Fig.~\ref{fig:sodtubetestL2err}. We observe second-order convergence on the pressure, first-order convergence in density, and in-between convergence in velocity.
The scattering observed in the velocity field behind the shock corresponds to  particle having to reorganise the crystal lattice, a typical feature of SPH (e.g. \citealt{phantom}).
Letting the shock evolve further and interact with the periodic boundary, we verify that we obtain a second symmetric shock, as expected.}

\subsection{Sedov-Taylor blast}
\label{sec:sedovtaylor}

\begin{figure*}
    \centering
    \includegraphics[width=0.95\linewidth]{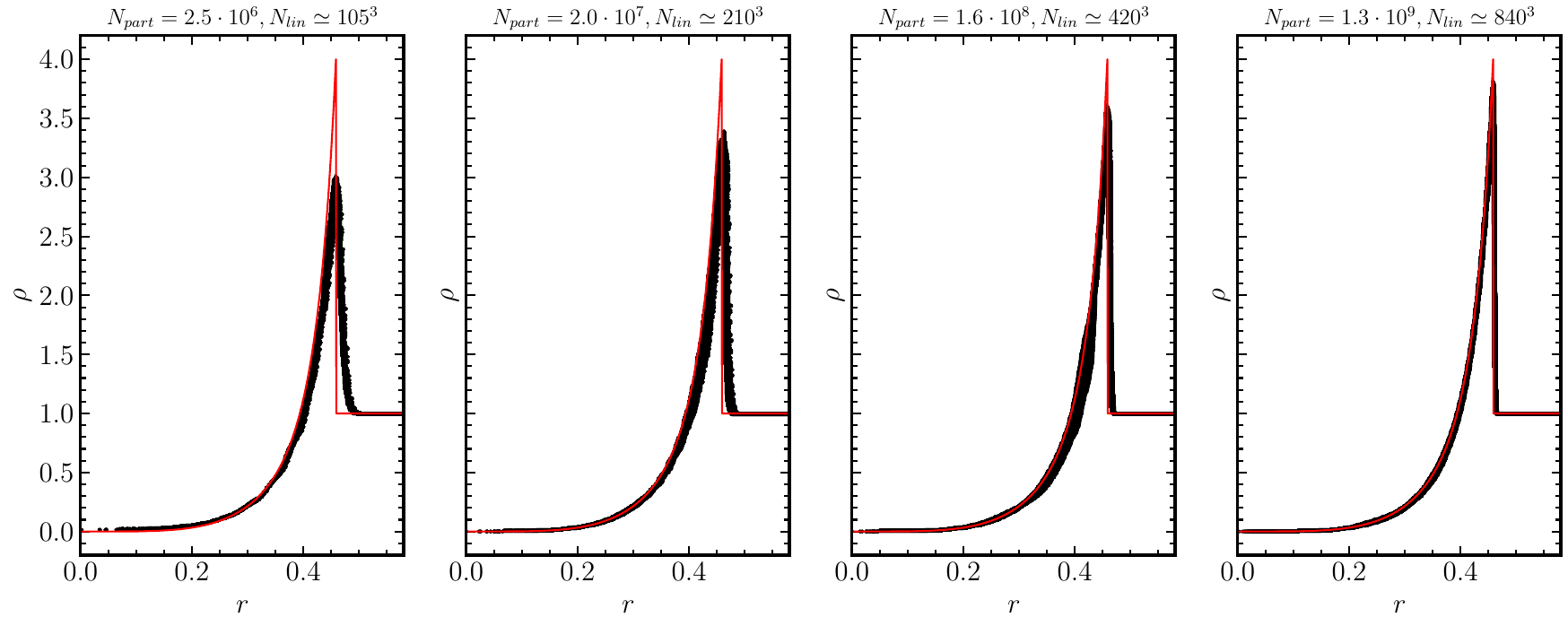}
    \caption{\tdc{Result of the densities (black dots) obtained for the Sedov-Taylor blast test described in Sect.~\ref{sec:sedovtaylor} at $t = 0.1$ for $N_{\rm part}$ particles, with $N_{\rm part} = 2.5 \cdot 10^6, 2.0 \cdot 10^7, 1.6 \cdot 10^8, 1.3 \cdot 10^9$ SPH particles (global time-stepping), corresponding to an inter particle spacing on the HCP lattice of respectively $10^{-2}/4, 10^{-2}/8, 10^{-2}/16, 10^{-2}/32$. Results are represented against the analytical solution (solid red line). The legend provides the effective linear resolution $N_{\rm lin}$ corresponding the cubic root of the number of particle displayed on each graphs which are truncated at $r=0.58$.}}
    \label{fig:hydrotestsedov}
\end{figure*}

\begin{figure}
    \centering
    \includegraphics[width=0.9\linewidth]{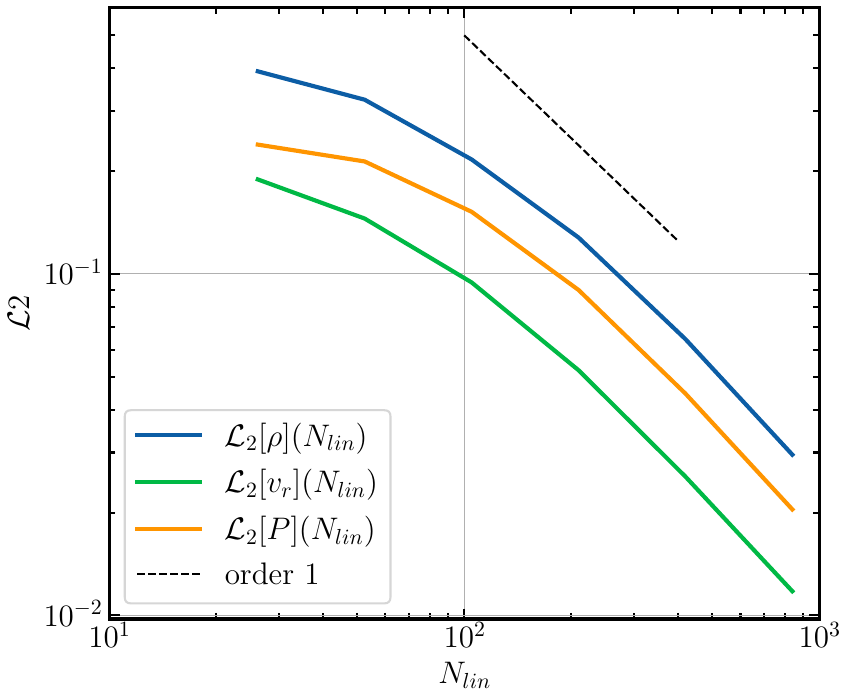}
    \caption{\gl{Convergence study of the 
    Sedov-Taylor blast test presented in Fig.~\ref{fig:hydrotestsedov}. Order one convergence is achieved for $v$, $\rho$ and $P$, similarly to what is obtained with other SPH codes}.}
    \label{fig:hydrotestsedovL2err}
\end{figure}

\gl{We perform a Sedov-Taylor blast wave  test \citep{1959Sedov, 1950Taylor.a, 1950Taylor.b} by first setting up a medium of uniform density $\rho=1$ with $u=0$ and $\gamma=5/3$, in a 3D box of dimensions $[-0.6, 0.6]^3$. The particles are arranged locally on a compact hexagonal lattice. The smoothing length is initially converged by iterating a white time step.
Internal energy is then injected in the centre of the box. This energy peak is smoothed by the SPH kernel according to $u_a = W \left( \mathbf{r}, 2 h_0\right)\times E_0 $, where the total amount of energy injected is fixed at $E_{0} = 1$ and $h_0$ is the smoothing length of the particles after relaxation.
For this test, the CFL condition is lowered to $\tilde{C}_{\mathrm{cour}} = \tilde{C}_{\mathrm{force}} = 0.1$ to prevent leapfrog corrector sub-cycling caused by the strong shock. This result in an enhanced energy conservation, with a maximum relative error of $10^{-6}$ observed across all tests.
The simulation is then evolved up to $t=0.1$. Simulations with $N = 26, 52, 105, 210$ particles per direction (see Fig.~\ref{fig:hydrotestsedov}) have been performed on a single A100 GPU of an NVidia DGX workstation. Simulations with $N = 420$ and $N = 840$ were executed on the \textsc{Adastra} supercomputer (see Sect.~\ref{sec:perf}) on 4 and 32 nodes respectively. The highest resolution blast test involves 1.255 Gpart, including ghost particles. The simulation consists in 14979 iterations performed in 14 hours, including setup and dumps, on 32 nodes corresponding to 128 Mi250X or equivalently 256 GCDs (see Sect.~\ref{sec:hardwaretests} for details). The total energy consumed for this test is $1.94\,\textrm{GJ}$, as reported by \textsc{Slurm}. The power consumption per node is $1195\,\textrm{W}$, which equates to slightly over half of the peak consumption of a single node ($2240\,\textrm{W}$).
Numerical results are compared against analytical solutions. Fig.\ref{fig:hydrotestsedov} shows results obtained for the density for $N^{3}$ particles, with $N = 105, 210, 420, 840$. For the latter case, $\mathcal{L}_2$ errors are of order $\sim 10^{-1}$, which is similar to what is obtained with other SPH codes with this particular setup.
Figure \ref{fig:hydrotestsedovL2err} shows that order one convergence is achieved for $v$, $\rho$ and $P$, similarly to what is obtained with other SPH codes.} 

\subsection{Kelvin-Helmholtz instability}
\label{sec:khhydrotest}

\begin{figure*}
    \centering
    \includegraphics[width=1.1\linewidth]{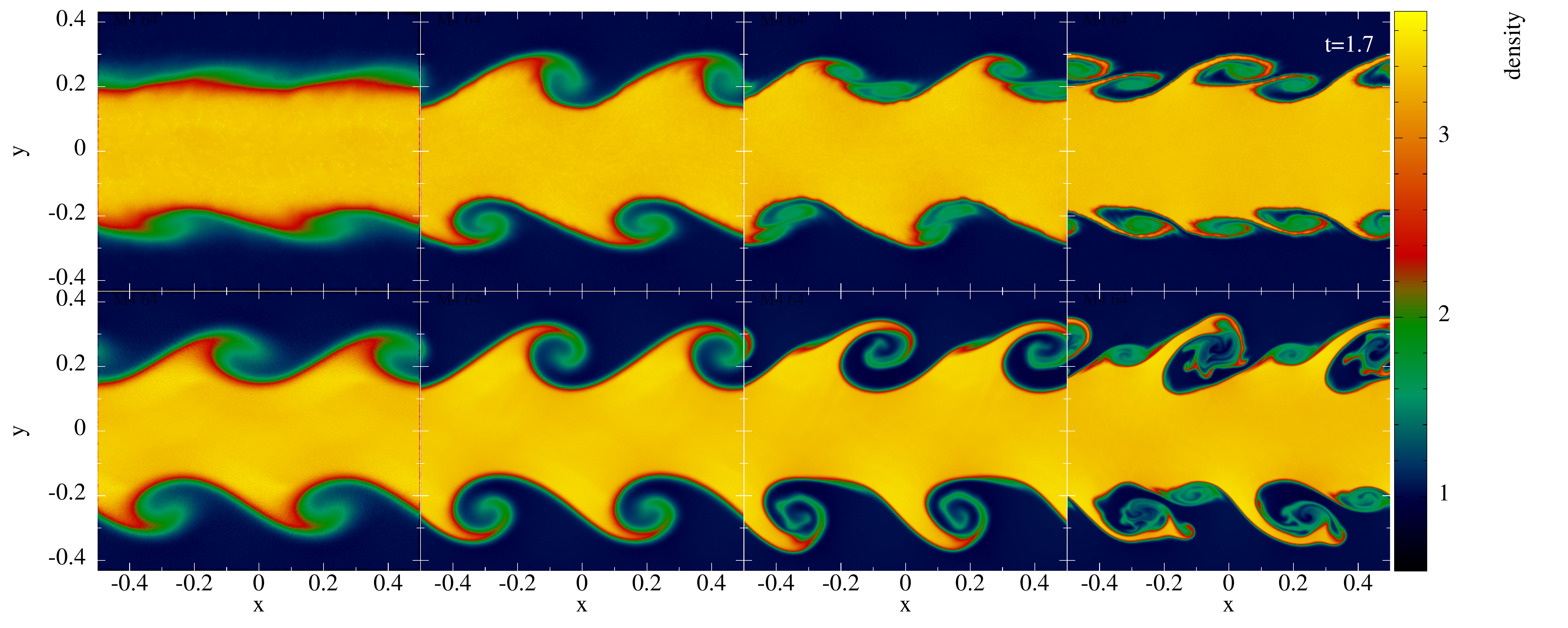}
    \caption{
        \gl{Density profiles obtained in the low-resolution 3D Kelvin-Helmholtz test described in Sect.~~\ref{sec:khhydrotest} at $t = 1.7$ with the $M_4$ kernel (top panel) and the $M_6$ quintic kernel (bottom panel) respectively in code units. The instability is correctly captured with the $M_{6}$ kernel, similarly to the findings of \citet{Tricco2019}. From left to right, the numbers of particles $N_{\rm l}$ and $N_{\rm r}$ used along the $x$ axis for the low-density and the high-density regions are: $ N_{\rm l} = 128$ and $ N_{\rm r} = 192 $, $N_{\rm l} = 256$ and $ N_{\rm r} = 384$, $N_{\rm l} = 512$ and $ N_{\rm r} = 758$, $ N_{\rm l} = 1024$ and $ N_{\rm r} = 1536$, which corresponds to $215\times 10^3$, $860\times 10^3$, $3.4\times 10^6$ and $13.7\times 10^6$ particles respectively.}
    }
    \label{fig:hydrokhdensshamrock}
\end{figure*}

\gl{We test the ability of \SHAMROCK{} to capture instabilities related to discontinuities on internal energy by performing a Kelvin-Helmholtz instability test \citep{2008Price}. We adopt a setup close to the one proposed by \cite{Schaal2015}, that gives rise to secondary instabilities fostering additional turbulence mixing. The initial pressure, density and velocity profiles are initialised according to
\begin{align}
P &= 3.5\\
\rho &= \begin{cases}
        1, \quad & \text{if } \vert y\vert > y_s /4,\\
        (3/2)^3, \quad & \text{if } \vert y\vert \leq y_s /4,\\
    \end{cases}\\
v_x &= \begin{cases}
        \xi/2, \quad & \text{if } \vert y\vert > y_s /4,\\
        -\xi/2, \quad & \text{if } \vert y\vert \leq y_s /4,\\
    \end{cases}\\
v_y &= \varepsilon \sin(4\pi x)\left\lbrace
\exp\left(- \frac{(y-y_s/4)^2}{2 \sigma^2} \right) + \exp\left(- \frac{(y+y_s/4)^2}{2 \sigma^2} \right)
\right\rbrace.
\end{align}
The test is performed in 2.5D, restricting the $z$ axis to a thin layer comprising only 6 SPH particles in the low density region, and 9 particles in the high density region. We opt for a density ratio of $(3/2)^3$ between the two regions to simplify the particle setup process and circumvent unnecessary complexities associated with arranging particles on closed-packed lattices.
We use $\gamma=1.4$. The slip velocity and the perturbation parameters are $\xi = 1$, $\epsilon = 10^{-2}$, $\sigma = 0.05/\sqrt{2}$, similarly to the values used in \cite{Schaal2015}. Simulations are performed on a single A100-40GB GPU. This GPU can accommodate a maximum of approximately $\sim 40\cdot 10^6$ particles for the $M_{4}$ kernel and $\sim 20\cdot 10^6$ particles for the $M_{6}$ kernel.
Fig.~\ref{fig:hydrokhdensshamrock} shows results obtained at increasing resolutions for the $M_4$ kernel (top panel) and the $M_6$ kernel (bottom panel). Similarly to the findings of \cite{Tricco2019}, we first observe that the $M_4$ fails to accurately capture the instability, even at high resolutions, as vortices appear flattened and overly diffused. Conversely, we observe that all these features are effectively captured when employing the $M_6$ kernel.
The further Sect.~\ref{sec:conformance} shows that our results align almost perfectly with those obtained with \textsc{Phantom}. The growth rate observed for the instability matches therefore the findings reported in \citet{Tricco2019}.}

\subsection{Conformance with \textsc{Phantom}}
\label{sec:conformance}

\gl{We aim to benchmark the performance of \SHAMROCK{} against a state-of-the-art,
robust, optimised and extensively tested SPH code running on CPUs. Several SPH codes are in use in the community (e.g. \textsc{Bonsai-SPH} \citealt{BedorfPortegiesZwart2020},\textsc{Gadget} \citealt{gadget}, \textsc{Gasoline} \citealt{WadsleyStadelQuinn2004}, \textsc{Gizmo} \citealt{Hopkins2014}, \textsc{Seren} \citealt{HubberBattyEtAl2011}, \textsc{Swift} \citealt{SchallerGonnetEtAl2018}). We choose
\textsc{Phantom}, since it is optimised for hydrodynamics, well-used by the astrophysical community and extensively tested and documented
\citep{phantom}. Before conducting comparisons, one has to ensure that the two solvers are rigorously identical, up to identified insignificant errors. This is the purpose of the next two tests, that are uncommonly designed to reveal
discrepancies by amplifying errors using lower resolution or less regular
kernels than achievable. For this, we generate the initial conditions with
\textsc{Phantom}, then start an identical simulation from the same dump using a fixed time step.}

\subsubsection{Residuals: Low res Sedov-Taylor blast wave test}

\begin{figure*}
    \centering
    \includegraphics[width=0.75\linewidth]{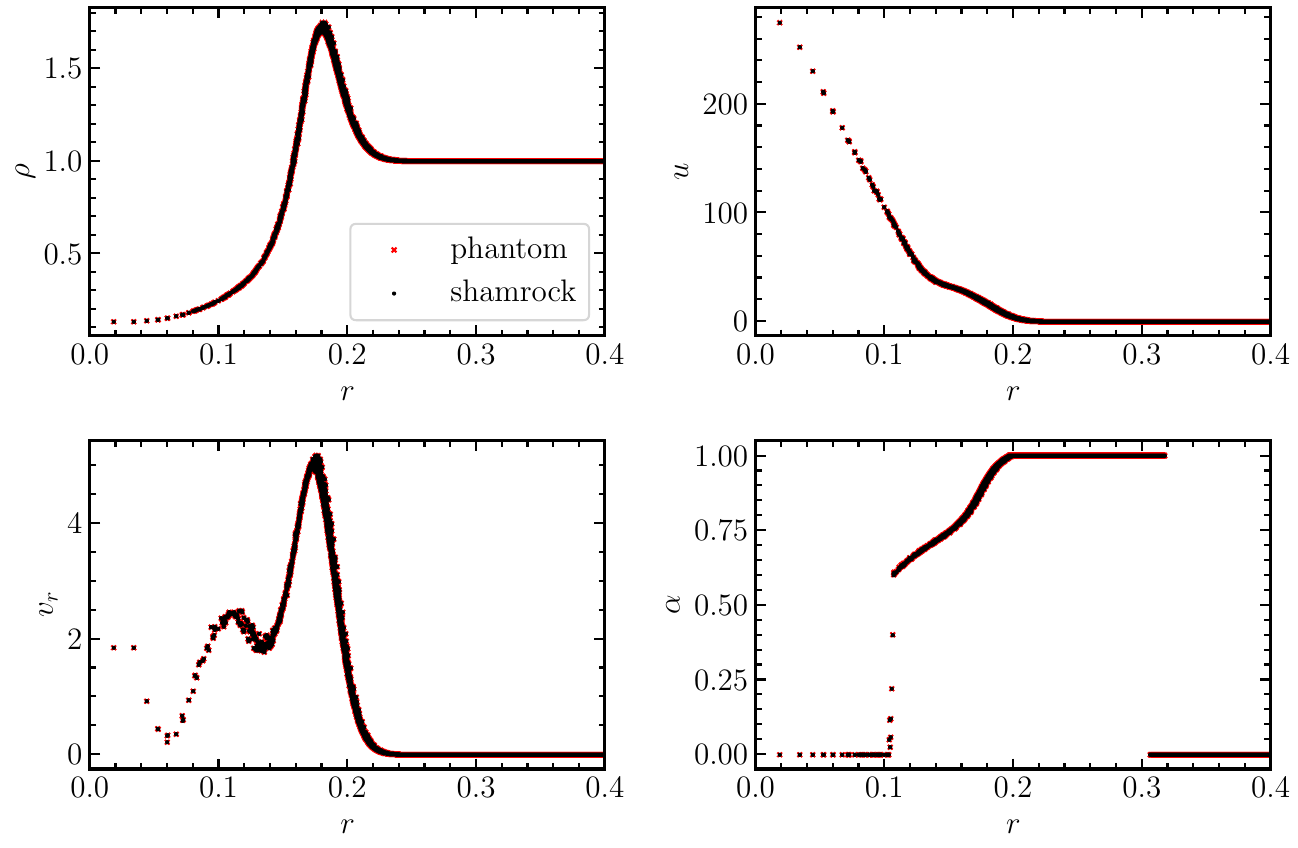}
    \caption{
        \gl{A comparison is made between the densities $\rho$, internal energies $u$, velocities $v_{r}$, and shock detection parameters $\alpha$ obtained at $t = 1$ from two identical low-resolution Sedov-Taylor blast wave tests conducted by \textsc{Phantom} (red dots) and \SHAMROCK{} (black dots). Initially, \textsc{Phantom} is  used to generate the same setup file for the two simulations. Runs are then conducted using a fixed time-step of $dt=10^{-5}$. The dots are indistinguishable by eye (e.g. the $\mathcal{L}_2$ error on the velocity field is of order $5\cdot 10^{-4}$): the implementations of the SPH solver are identical in the two codes.}
    }
    \label{fig:compphantomsedov}
\end{figure*}

\gl{We first measure the residual discrepancies between \SHAMROCK{} and \textsc{Phantom} by comparing results obtained on two identical Sedov-Taylor blast wave tests described in Sect.~\ref{sec:sedovtaylor}, fixing the time step to $dt = 10^{-5}$. This three-dimensional test is highly sensitive to rounding errors, primarily due to the presence of a low-density, zero-energy region surrounding the blast wave. In particular, aligning the behaviours of the shock viscosity parameter $\alpha^{\rm AV}$ proves being particularly challenging. Finally, Fig.~\ref{fig:compphantomsedov} shows that discrepancies between \SHAMROCK{} and \textsc{Phantom} are imperceptible to the naked eye. Quantitatively, the $\mathcal{L}_2$ errors are
\begin{itemize}
   \item relative $\mathcal{L}_2$ distance $r$ : $ 2.0869658802024003e-07 $,
   \item relative $\mathcal{L}_2$ distance $h$ : $ 3.952645327403623e-05 $,
   \item relative $\mathcal{L}_2$ distance $v_r$ : $  0.0005418229957181854 $,
   \item relative $\mathcal{L}_2$ distance $u$ : $ 3.6622341394801246e-05$.
\end{itemize}
Following an in-depth examination, the sole identified distinctions between the two solvers are as follows: in \textsc{Phantom}, the shock parameter $\alpha^{\rm AV}$ and the estimate of $\nabla \cdot \mathbf{v}$ are stored as single-precision floating-point numbers, while in \SHAMROCK{}, they are double-precision.}

\subsubsection{Residuals: Low res Kelvin-Helmholtz instability test}
\label{sec:khshamphantom}

%--ICI---
\begin{figure*}
    \centering
    \includegraphics[width=0.95\linewidth]{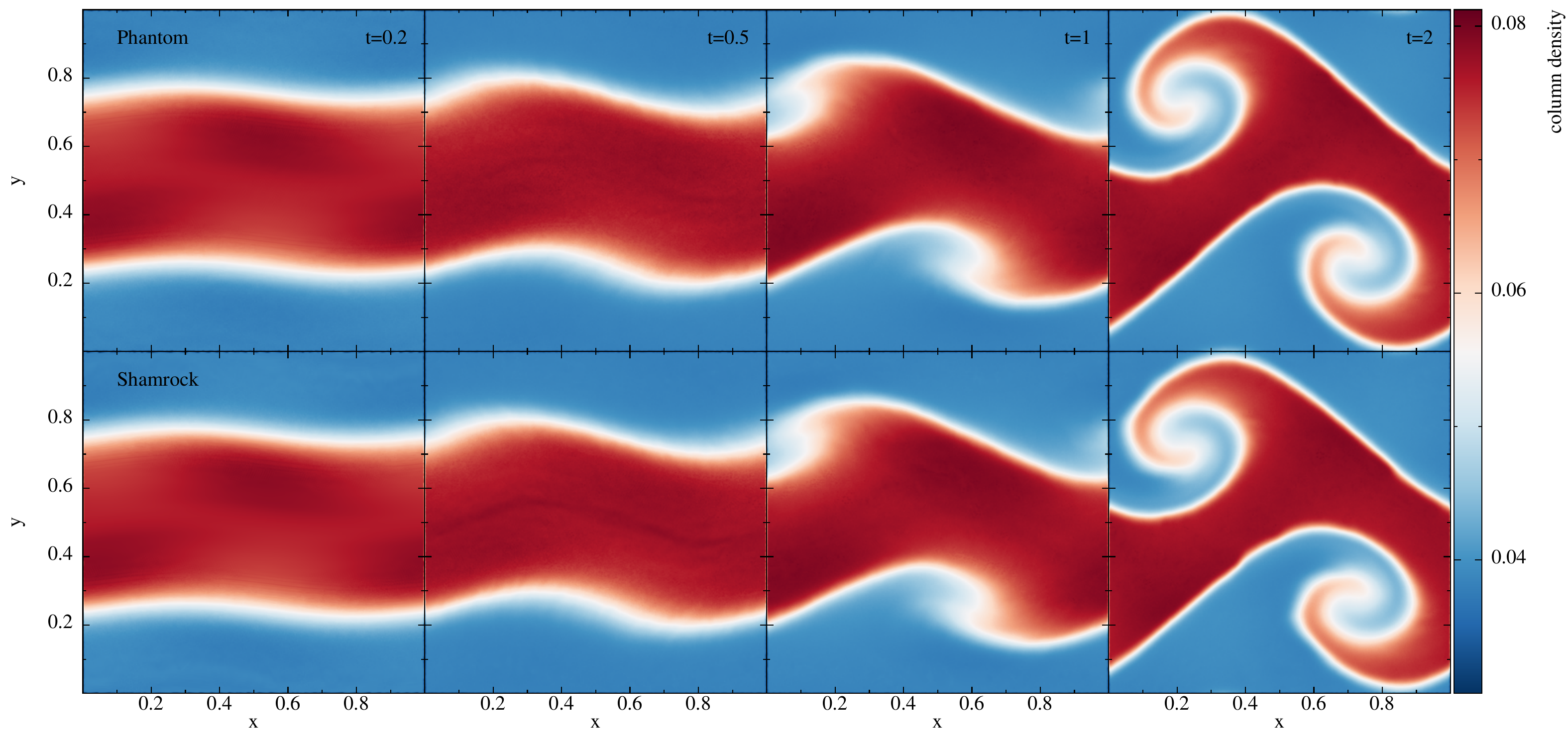}
    \caption{
        \gl{Density profiles obtained at $t = 0.2, 0.5, 1, 2$ on a low resolution Kelvin-Helmholtz test by \textsc{Phantom} (top panel) and \SHAMROCK{} (bottom panel). Results are almost identical. The test is voluntarily performed with an unsuited M4 kernel at low-resolution to accentuate residual discrepancies between the two solvers. Those stem from the use of single precision in one and double precision in the other for shock detection variables.}
    }
    \label{fig:hydrokhdens}
\end{figure*}
\begin{figure*}
    \centering
    \includegraphics[width=0.95\linewidth]{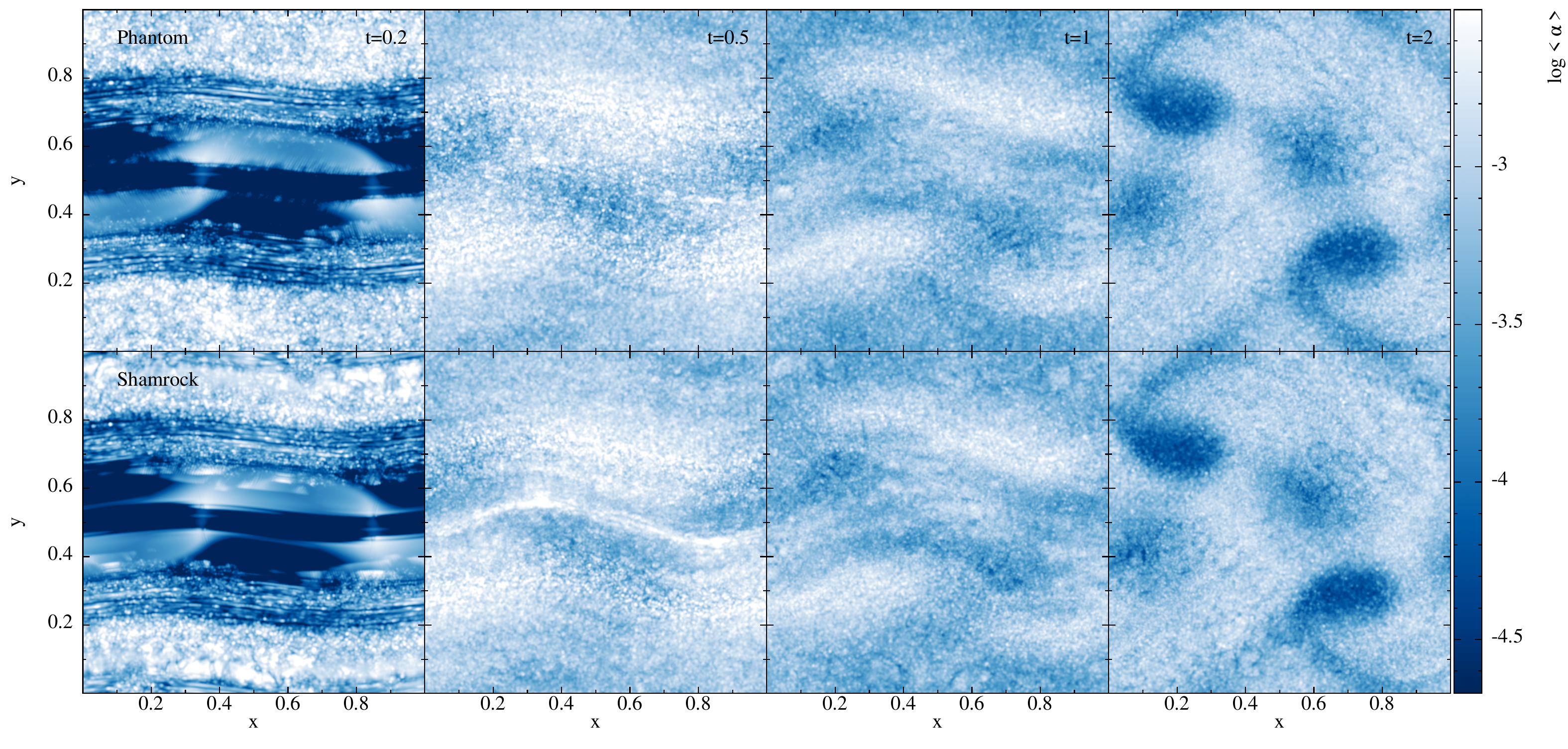}
    \caption{\gl{Same plot as in Fig.\ref{fig:hydrokhdens}, revealing  the amplitude of truncature errors in the shock viscosity
    parameter. To our understanding, these errors represent the sole source of discrepancies between the implementations of SPH in \textsc{Phantom} and \SHAMROCK{}.}}
    \label{fig:hydrokhalpha}
\end{figure*}

We measure the residuals between \SHAMROCK{} and \textsc{Phantom} by performing the Kelvin-Helmholtz instability test implemented in \textsc{Phantom} at commit number \texttt{e01f76c3}, at low resolution. Simulations are evolved to $t=2$, while dumps are produced every $\Delta t=0.1$ to sample the development of the instability. We choose the M4 kernel and a low number of particles to reveal the differences between the codes. Fig.~\ref{fig:hydrokhdens} and Fig.~\ref{fig:hydrokhalpha} show the compared evolutions of the density and of the shock parameter respectively.
At $t=0.2$, no difference is observed in the density field. For shock viscosity, we observe for \textsc{Phantom} a small noise of relative amplitude $\lesssim 0.1\%$ along the line $y=0.5$, which we attribute to $\alpha, h, \nabla \cdot \mathbf{v}$ being stored as single precision fields in \textsc{Phantom}. These fluctuations are not present in \SHAMROCK{}, since these quantities are calculated in double precision.
At $t=0.5$ we can distinguish at this low resolution a small line of higher shock viscosity and density in \SHAMROCK{}, which is not present in \textsc{Phantom}. We attribute these residuals to the fact that the \textsc{Phantom} simulation may have increased numerical viscosity due to single precision errors, while the SPH lattice in the \SHAMROCK{} simulation is still reorganising.
At $t=1$, no significant difference is observed between the two simulations.
Finally, at $t=2$, tiny differences can be observed at the edges of the instability, for the same reasons as at $t = 1$.

\textsc{Phantom} uses mixed precision (single/double), while \SHAMROCK{} uses only double precision. As a result, discrepancies on the order of single precision are expected between the codes.
As a remark, we do not use mixed precision in \SHAMROCK{} because, on GPUs, converting between single and double precision is costly, similar to the latency of a double precision square root. Therefore, any performance gains from mixed precision are offset by the cost of precision conversion.

\subsection{Circumbinary discs}
\label{sec:sim-circumbinary}

\begin{figure*}
    \centering
    %\captionsetup{format=sanslabel}
    \begin{subfigure}{.45\textwidth}
       \centering
       %\fbox{
       \includegraphics[width = 0.99\textwidth]{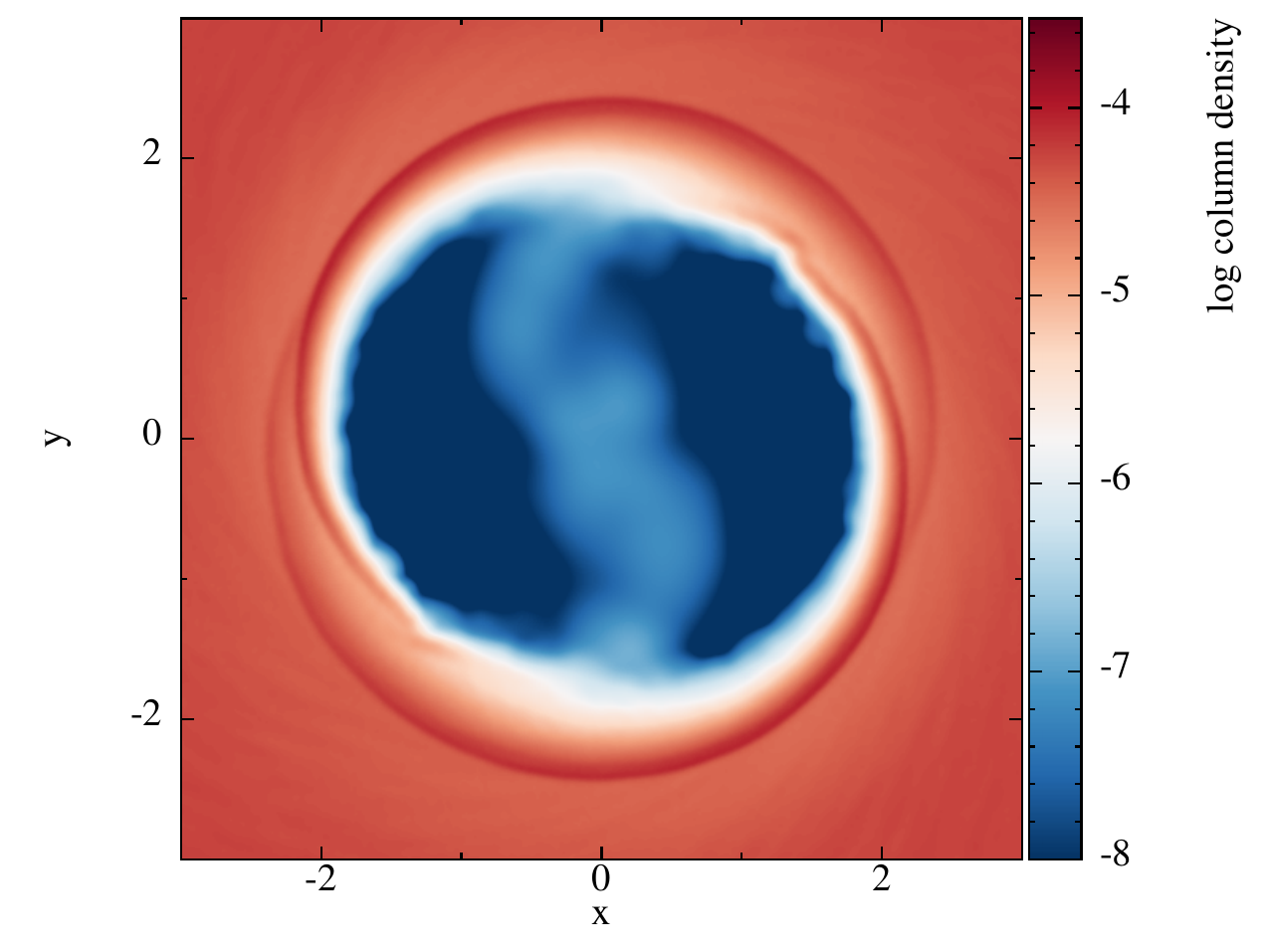}
       \vfill
       %}
       \caption{$10^7$ particles.}
   \end{subfigure}
   \begin{subfigure}{.45\textwidth}
       \centering
       %\fbox{
       \includegraphics[width = 0.99\textwidth]{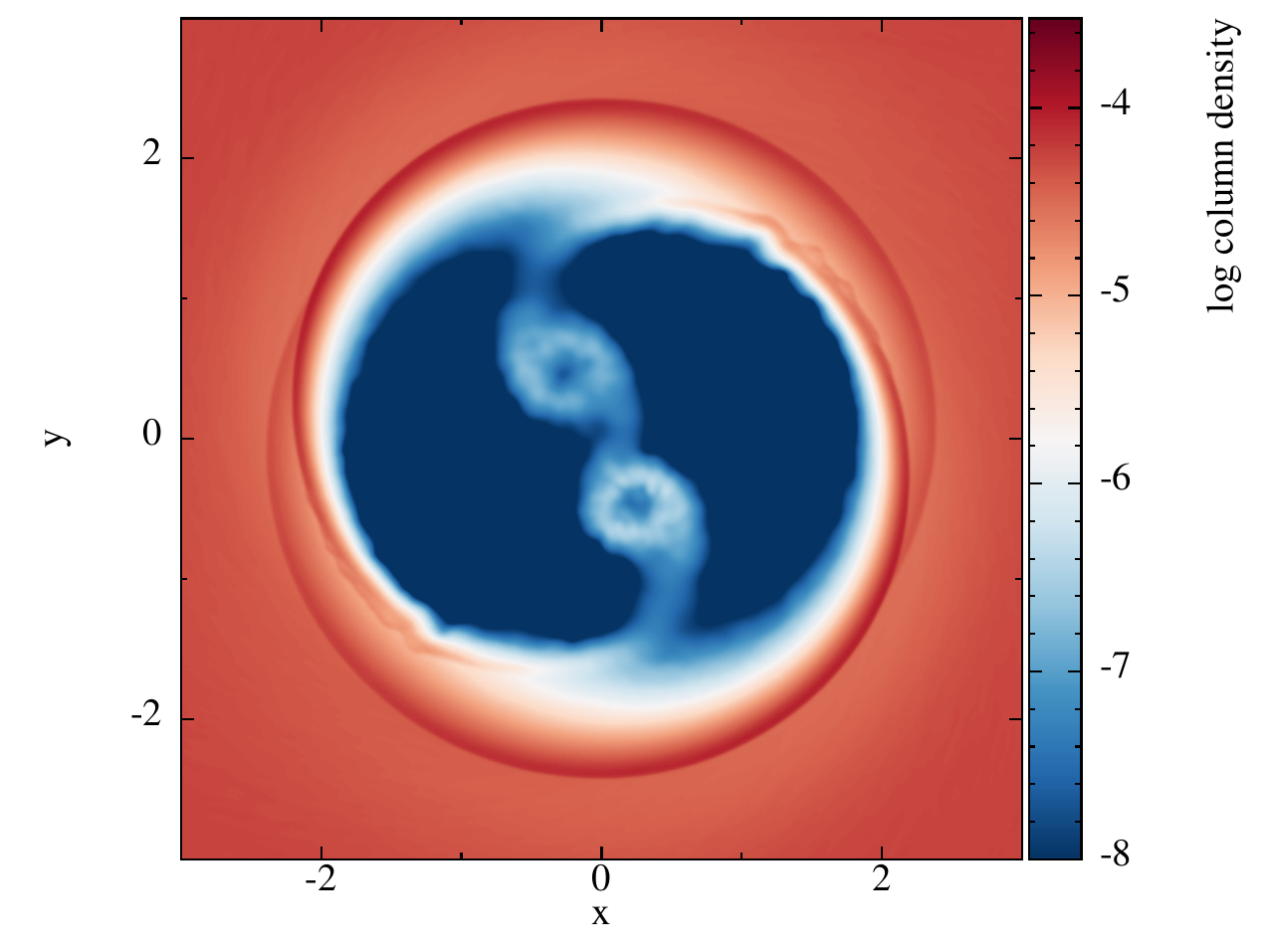}
       \vfill
       %}
       \caption{$10^8$ particles.}   
    \end{subfigure}
    \begin{subfigure}{.45\textwidth}
        \centering
        %\fbox{
        \includegraphics[width = 0.99\textwidth]{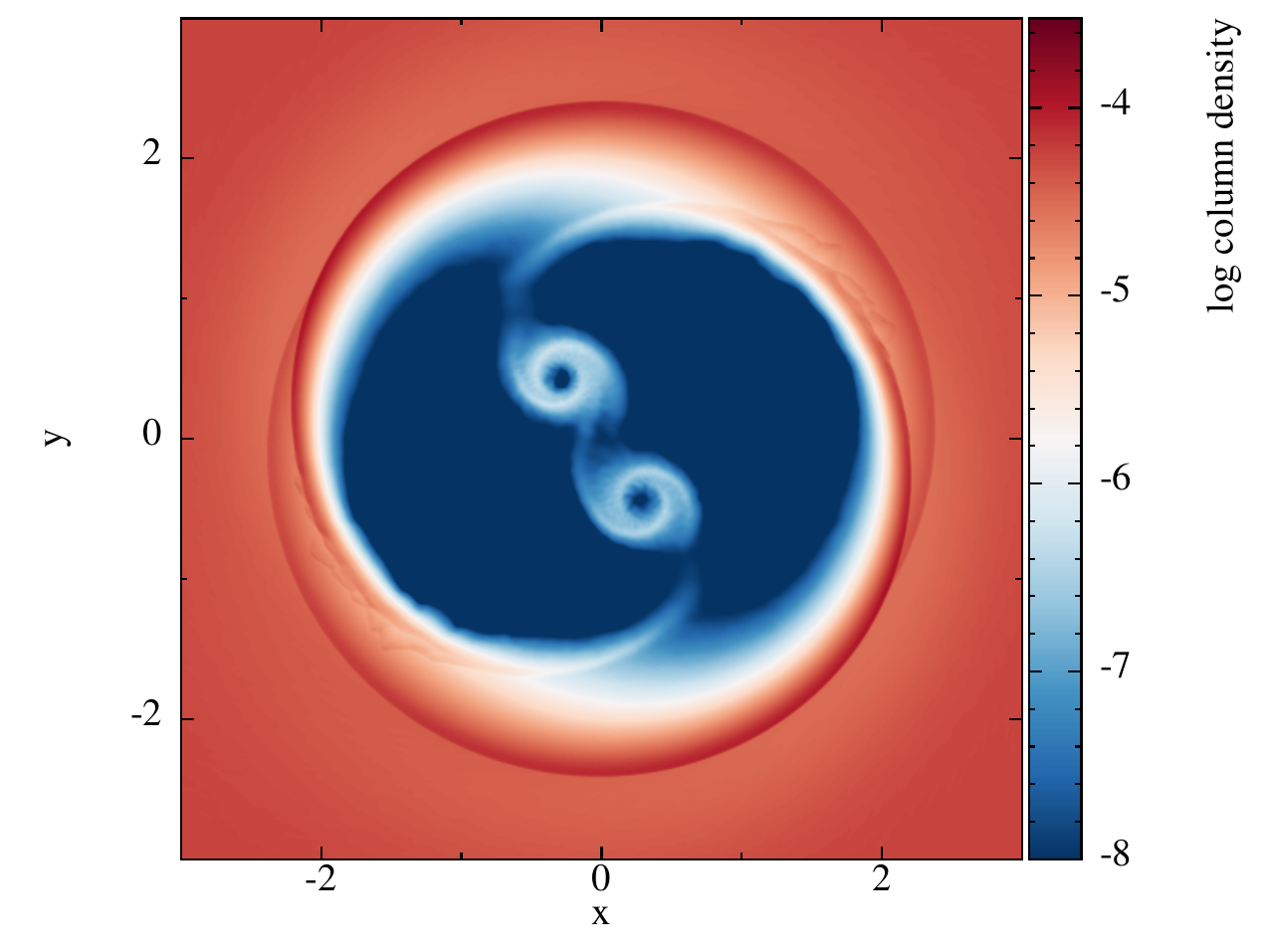}
        \vfill
        %}
        \caption{$10^9$ particles.}
    \end{subfigure}
    \begin{subfigure}{.45\textwidth}
        \centering
        %\fbox{
        \includegraphics[width = 0.80\textwidth]{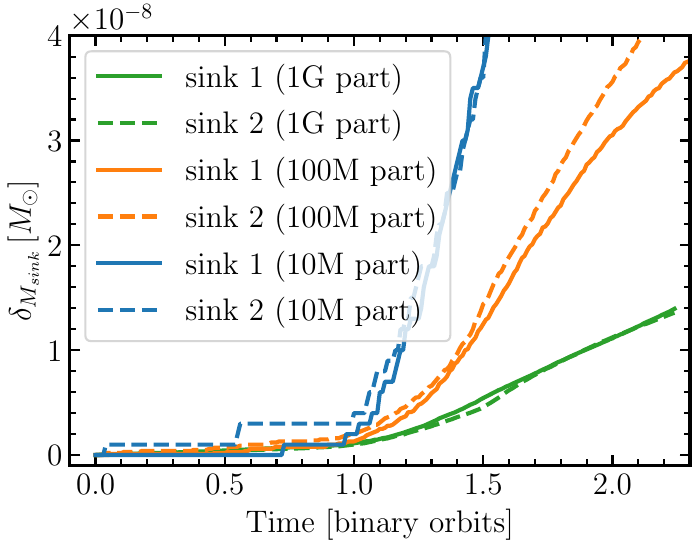}
        \vfill
        %}
        \caption{Evolution of the sinks mass in the simulations.}
    \end{subfigure}
    \caption{
        SPH simulation of a circum-binary discs according to the setup described in Sect.~\ref{sec:sim-circumbinary} performed with \SHAMROCK{} SPH solver. 
        Panels (a), (b) and (c) shows face-on view of the column integrated density after $1.88$ orbits 
        of the discs at different resolutions, respectively $10^7$, $10^8$ and $10^9$ particles.
        The bottom right panel (d) shows the evolution of the mass accreted by the sink particles $\delta M_{\rm sink}$ in the simulations shown in panels (a), (b) and (c) as a function of time.}
    \label{fig:exemple.sim.sham.circum}
\end{figure*}

To evaluate \SHAMROCK{}'s capability for simulating astrophysically relevant systems, we test the code using the setup proposed by \cite{DuffellEtAl2024} for modelling the evolution of a gaseous disc around a binary star. This test is numerically challenging in three dimensions, due to its large resolution requirement. 
The numerical setup consist of two stars of equal masses $M_{1,2} = 0.5 M_{\odot}$ evolving on circular orbits with constant 
binary separation $d=1 \unitau$. Units are set such that 
$G M_{\odot}=1$, for the orbital frequency of the binary to be one in code units, which corresponds to an orbital period of $T_{\rm bin} = 2 \pi$.
A disc is set up around this binary system. Its surface density profile is 
$\Sigma(r)=\Sigma_0\left[\left(1-\delta_0\right) e^{-\left(R_{\mathrm{cav}} / r\right)^{12}}+\delta_0\right]$, 
where the inner cavity radius is set to be $R_{\mathrm{cav}} = 2.5 \unitau$, and the background density to 
$\delta_0 = 10^{-5}$. The simulation is three-dimensional, we adopt a Gaussian vertical density profile 
$\rho(r, z)=\frac{\Sigma(r)}{H} \frac{1}{\sqrt{2 \pi}} \exp \left(-\frac{z^2}{2 H^2}\right)$ where the scale 
height of the discs satisfies $H(r)/r = 0.1$.
Following \cite{DuffellEtAl2024}, we use a locally isothermal equation of state, the sound speed being parametrised by 
 $c_{\rm s}(\mathbf{r}) = ({H \over r}) \sqrt{ - \Phi_1(\mathbf{r}) - \Phi_2(\mathbf{r})}$, 
where $\Phi_{1,2}(\mathbf{r})$ are the gravitational potential of the two stars. The disc density is set up using a Monte Carlo method, as described in \cite{phantom}.
For the velocity profile, we adopt a Keplerian profile modified by a sub-Keplerian correction to account for the radial pressure gradient.
The gas is modelled with the constant $\alpha$-viscosity disc model to have an equivalent $\alpha_{\rm SS} = 10^{-3}$ \citep{LodatoPrice2010}.
Lastly, the binary system is simulated using sink particles with an accretion radius of $r_{\rm acc} = 0.05\unitau$. Particles within this distance are accreted onto the sink, and their mass, linear momentum, and angular momentum are added to the sink particle.
The simulations were performed up to 5 binary orbits, except for the simulations with $10^9$ particles simulations which were stopped after 2.4 orbits. This was due to the remaining hours available in the time allocation on the Adastra supercomputer.
The $10^9$ particle simulation was performed on Adastra supercomputer for 2 days, running at a rate of approximately one orbit per day, using 8 GPU nodes for a total of around 1500 GPU hours.

Fig.~\ref{fig:exemple.sim.sham.circum} shows density profiles obtained after a few orbits for simulations with $10^{7}$, $10^8$ and $10^9$ particles respectively. With $10^7$ particles, 
the inner cavity is under-resolved, resulting in the direct accretion of gas falling into the cavity, which is immediately accreted onto the sink, leading to a high accretion rate. 
With the increased resolution corresponding to $10^8$ particles, the simulation begins to capture the formation of mini-discs of gas around each star. The gas falls onto these mini-discs when passing through the cavity created by the binary, leading to a buffering effect on the accretion, which results in lower accretion rates, as described in \cite{FarrisEtAl2014}.  At this resolution, the accretion rate is significantly reduced compared to the $10^7$ particles case. However, the circumstellar 
discs are still under-resolved since no expected $m = 2$ substructures are observed.
With $10^9$ particles, these substructures are now resolved. The over-diffusivity is reduced, and the accretion rate continues to decrease compared to the previous cases.

Although we observe a significant reduction in the accretion rate due to the resolution of the circumstellar discs in this three-dimensional simulations, even higher resolution would be needed to draw conclusions about the convergence of the results, highlighting the interest of codes like \SHAMROCK{}. Qualitatively, we do not expect the accretion rate to change significantly as the resolution increases, once the circumbinary discs are properly resolved, up to the point where the physics at very small scales governing transport and dissipation may begin to be resolved. This issue must be considered separately from the dependence of the results on the thermal structure of the circumstellar and circumbinary discs, as realistic simulations require incorporating physical radiative transfer. These question are of great interest for astrophysics, but require dedicated numerical resources and a focused study on their own. Nonetheless, these preliminary tests pave the way towards studying resolved accretion rates in double or triple star systems, a topic of active research in the community (e.g. \citealt{CeppiEtAl2022}).

\subsection{Summary}

The hydrodynamic SPH solver implemented in \SHAMROCK{} passes successfully the standard tests (advection, Sod tube, Sedov-Taylor blast, Kelvin-Helmoltz instability). The implementation of the SPH solver in \SHAMROCK{} is identical to that of \textsc{Phantom}. Results obtained with the two codes are almost indistinguishable, the residuals being attributed to the choice of floating-point precision for the quantities $h, \alpha, \nabla \cdot \mathbf{v}$. This sets the basis for rigorous performance comparison. The circumbinary disc test demonstrates the potential of \SHAMROCK{} for simulating astrophysically relevant systems with unprecedented resolution.

\section{Performance}
\label{sec:perf}

\subsection{Characteristics of the benchmarks}

\subsubsection{Hardware specificities}
\label{sec:hardwaretests}

The tests performed to estimate performance with \SHAMROCK{} were conducted on two systems. 
Single GPU and CPU tests were performed on an Nvidia A100-SXM4-40GB GPU of an Nvidia DGX workstation. This workstation is equipped with 4 Nvidia A100 40GB GPUs paired with an Epyc7742 64-core CPU, and are exploited via SIDUS \citep{quemener2013sidus} by the Centre Blaise Pascal at ENS de Lyon . 
CPU tests were carried out on the CPU of the same DGX workstation using AdaptiveCPP \OpenMP{} backend. For those \SHAMROCK{} was compiled using \texttt{-O3 -march=native}.
For single GPU tests of \SHAMROCK{} compilation was performed using the Intel fork of the llvm/clang-19 compiler, also referenced as ONEAPI/DPC++ with optimizations \texttt{-O3 -march=native}. We voluntarily didn't used fast math optimisations as they would not be used in production. We use the CUDA/PTX backend of Intel llvm targetting the CUDA architecture \texttt{sm\char`_s70} corresponding to the compute capability of A100 GPUs. The Intel LLVM CUDA/PTX backend generates code using PTX ISA, the assembly language used to represent \CUDA{} kernels. This result in a program that actually that is first lowered from C++ to PTX, then compiled using Nvidia \texttt{ptxas} tool, which enable the code to be profiled using Nvidia's \CUDA{} tools. The \CUDA{} version used is 12.0.

Multi-GPU and multi node tests were performed on Adastra Supercomputer at CINES in France, using up to 256 compute nodes, each compute node is a HPE Cray EX235a each equipped with 4 Mi250X GPUs paired with a 64 Cores AMD Epyc Trento CPU.
On this platform we used ROCm/HIP backend of Intel llvm targetting the AMD GPU architecture \texttt{gfx90a} corresponding to the compute capability of Mi250X GPUs. 
The Intel llvm compiler was compiled using in the same module environment as \SHAMROCK{}. We used Cray CPE 23.12 with acceleration on \texttt{gfx90a} and Trento on the host, in conjunction with the provided \verb|PrgEnv-intel|. MPI with GPU aware support support was provided by the cray MPIch 8.1.26 module. \ROCM{} support was provided by both \verb|amd-mixed| 5.7.1 and \verb|rocm| 5.7.1.
Although the Mi250X GPU is a single chip, it is made up of two GCDs, which appear as separate instances on the compute node, where one MPI rank is assigned per GCDs.

\subsubsection{Setups}

We present the performances of the SPH hydrodynamical solver of \SHAMROCK{} on the Sedov Taylor blast wave, since it involves contributions of all the different terms in the hydrodynamical solver, and it is neither specific to astrophysics nor SPH. We compare the results with the one obtained with the hydro dynamical solver of \textsc{Phantom} with an almost identical implementation (see Sect.~\ref{sec:conformance}), on a computing units having similar power consumption.\\
We use the $M_4$ kernel and set $h_{fact}=1.2$. To setup the lattice, we first consider a box of size $[-0.6, 0.6]^3$. For a desired number of particle $N$, the volume per particle is $c V / N$, where $c$ is compacity of a close-packing of equal spheres. As such, the spacing $dr$ between particles is $dr = \left( 3 c V / 4 \pi N\right)^{1/3}$. We then adapt the boundaries of the simulation volume to ensure periodicity in all directions for the initial close-packed lattice of particles.\\

\subsection{Single GPU or GPU}

\begin{figure}[t!]
\begin{center}
\includegraphics[width=0.9\linewidth]{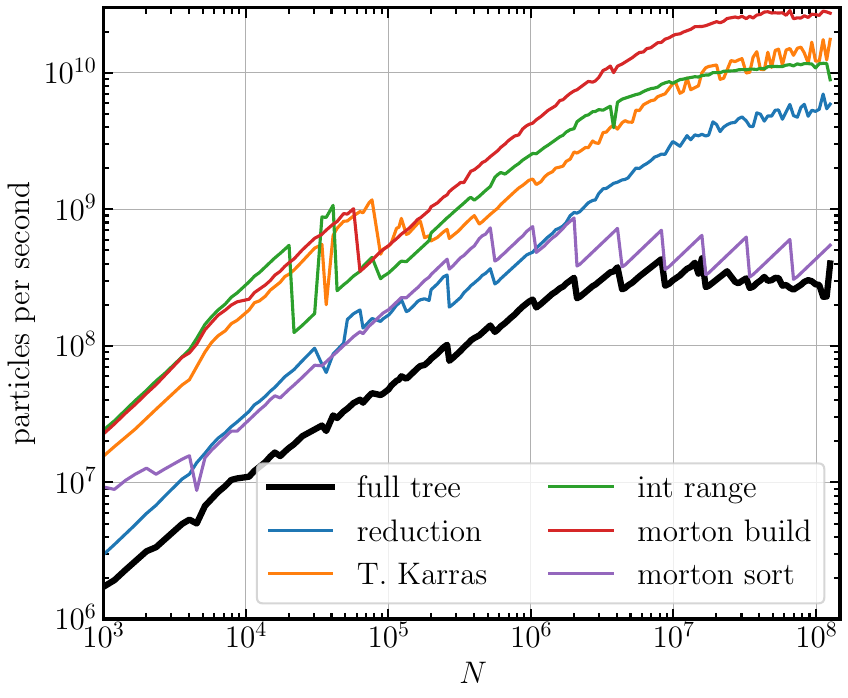}
\caption{
    Benchmark of the performance of the tree building in \SHAMROCK. 
     Each curve represents the number of particles processed per second for various segments of the algorithm.    The thick solid black curve shows the total time to build the tree. The other curves show the performance of the main algorithms involved in the tree building procedure. Those correspond to benchmarks of the isolated algorithms, which 
    break the asynchronous nature of SYCL. As such, the sum of the individual times do not rigorously add up to the exact time of the entire algorithm.
     This benchmark used a dataset of input positions generated from an hexagonal closed packing lattice, with variations in lattice spacing. Varying the initial distribution of particles will not affect total performance of the tree, since overall, the building time is dominated by the bitonic sort.
    In this test, we used single precision Morton codes.}
\label{fig:tree_build_perf}
\end{center}
\end{figure}

\begin{figure*}
\begin{center}
\includegraphics[width=0.45\linewidth]{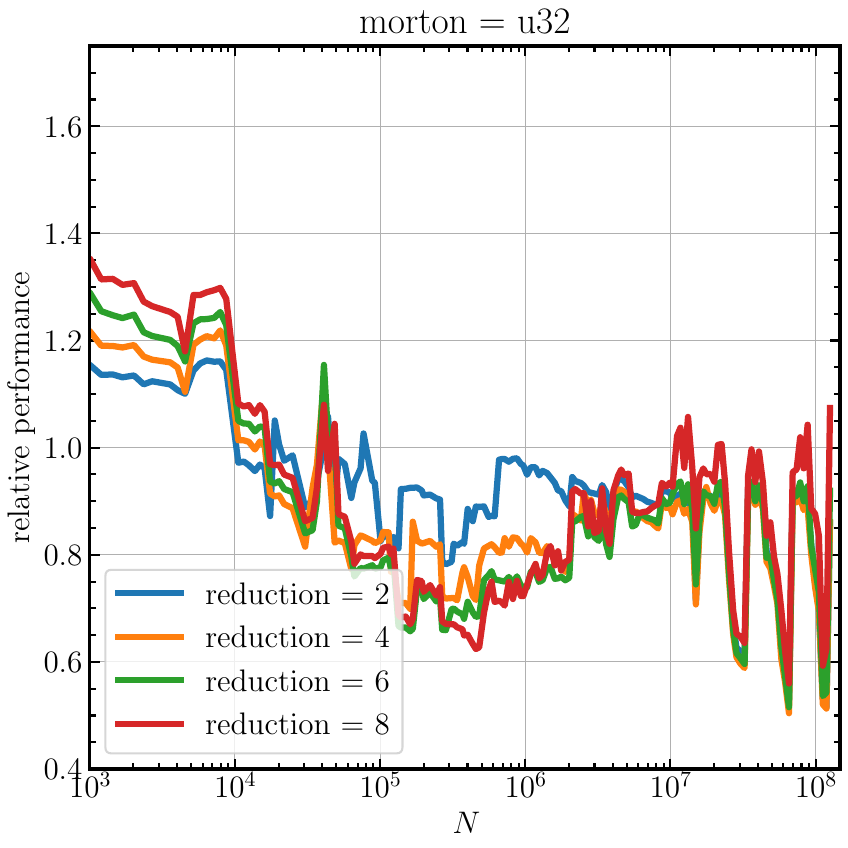}
\includegraphics[width=0.45\linewidth]{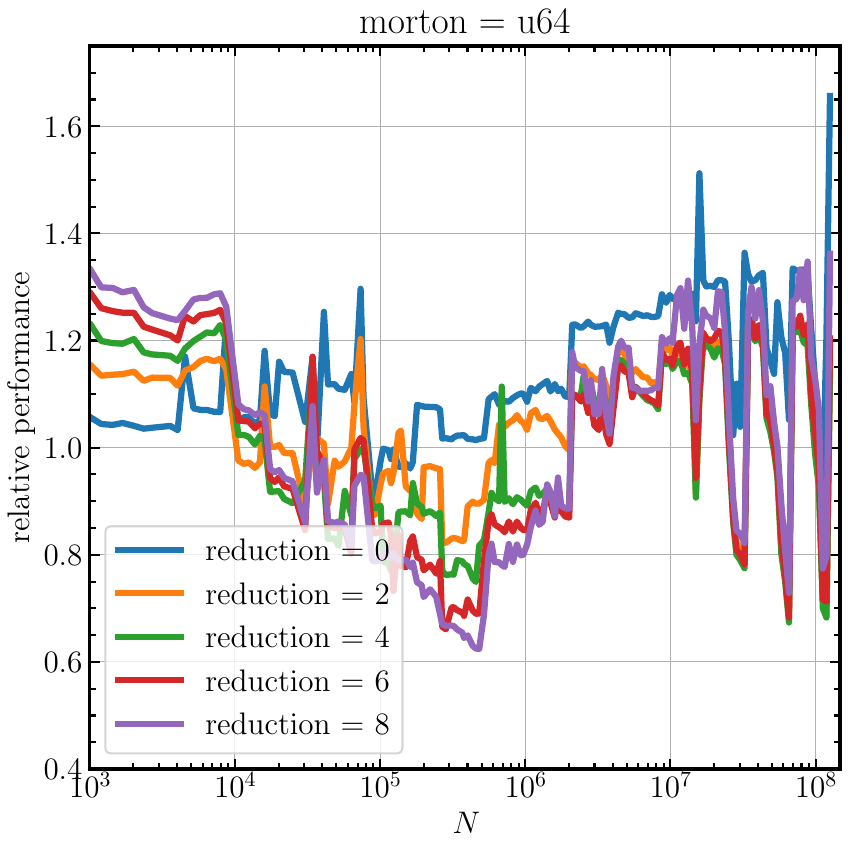}
\caption{Relative performance of the complete tree building procedure for two different types of Morton codes (left: u32, right: u64), for different levels of reduction. The setup for this test is identical to the one presented in Fig.~\ref{fig:tree_build_perf}.
}
\label{fig:tree_build_perf_varprec}
\end{center}
\end{figure*}

\subsubsection{Performance of tree building}
\label{sec:treebuildperf}

Fig.~\ref{fig:tree_build_perf} shows the performance of the \textsc{Shamrock} tree building algorithm described in Sect.~\ref{sec:treebuild}, by presenting results of tests carried out over $10^{3}$ to $10^{8}$ objects distributed on a regular cubic lattice. The results are presented in figures showing the number of object integrated to the tree per second, as a function of the total number of objects. This metric highlights the efficiency threshold of the GPU, where the computation time is shorter than the actual GPU programming overhead. It also highlights any deviation from a linear computation time as a function of the size of the input.

For a small to moderate number of objects $N \lesssim N_{\rm c}$ where $N_{\rm c} \sim 10^{6}$, the overhead of launching a GPU kernel leaves a significant inprint compared to the computational charge. A few million objects per GPU is the typical number of objects above which the algorithm can be used efficiently. For any $N \ge N_{\rm c}$ tested, the tree is built at a typical constant rate of $5 \times 10^{-9}\,\mathrm{s}$ per object. Equivalently, 200 millions of objects per second are processed for Morton codes and the associated positions in double-precision.

For $N \ge N_{\rm c}$, the algorithm achieves an almost constant performance, as long as it could be tested on current hardware. Fluctuations of up to $30\%$ are observed for specific values of $N$. These peaks are consistent across several executions, and therefore probably due to the hardware scheduler on the GPU. Tree construction is dominated by the bitonic sorting algorithm (see Fig.~\ref{fig:tree_build_perf}). Since this algorithm does not depend on the values stored in the buffer, its performance is not affected by the spatial distribution of objects, and regular or randomly arranged points deliver the same performance. The performance of tree building of \SHAMROCK{} is therefore independent of the distribution of objects considered.

Fig.~\ref{fig:tree_build_perf_varprec} shows that performance is almost unaffected by the type (single, double, float or integer) used for the positions ($\sim 5-10\%$, spikes being probably due to the hardware scheduler). Performance is increased by a factor $\sim 30\%$ when the Morton code representation is reduced to single precision.
 
\subsubsection{Performance of neighbour cache building}
\label{sec:cachebuild}

\begin{figure*}
    \centering
\includegraphics[width=0.8\linewidth]{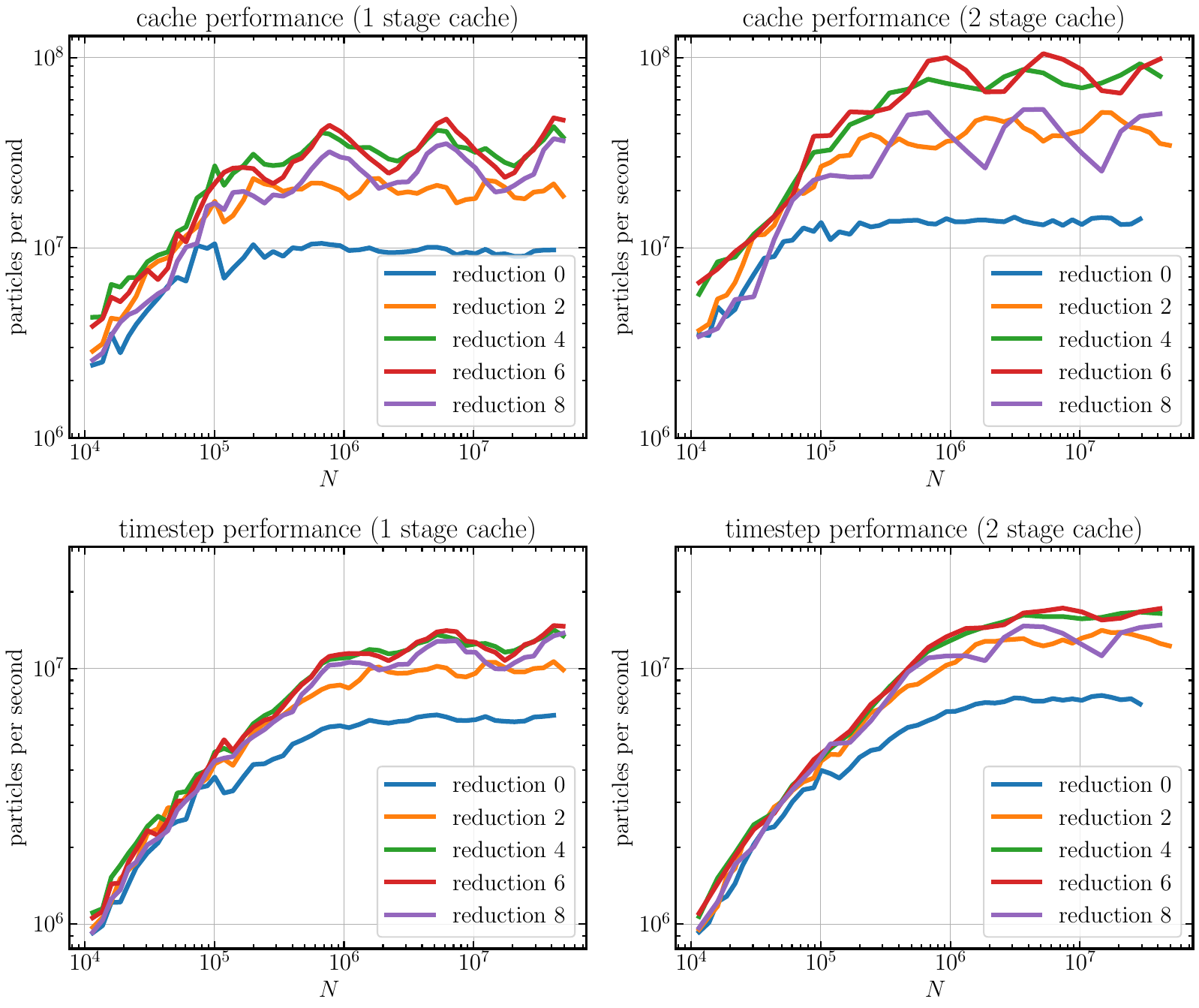}
    \caption{\gl{Performances of cache building and time stepping measured on one time step of the Sedov-Taylor blast wave test presented in Fig.~\ref{fig:hydrotestsedov}.
    Increasing the level of reduction yields improved performance
    overall up to reduction level $6$. Additionally, using two-stages neighbour
    caching improves performance up to a factor of two when used in conjunction with the reduction algorithm. In total, employing both reduction and a two-stage cache enhances cache building performance by a factor of ten, while doubling time-stepping performance.}}
\label{fig:neighcache}
\end{figure*}

To measure performance of cache build and SPH time stepping, we first setup the particles as discussed in Sect.~\ref{sec:treebuildperf}. Additionally, the smoothing length has been converged, resulting in $60$ neighbours for the M4 kernel. After this setup, we perform a single time step. Fig.~\ref{fig:neighcache} reports the time spent during this time step to build the cache and perform the iteration. We compare the results obtained for the single and two stage caches strategies presented in appendices \ref{sec:direct_cache} --\ref{sec:twostageneighcache}, along with different levels of reduction. We find that enhancing the reduction level results in better overall performance, particularly up to reduction level $6$. For higher levels of reduction, performance drops as a consequence of the excessive number of particles per leaf. Optimal configuration corresponds to $\sim 10$ particles per leaf (reduction level of 4), which is similar to the number of particles per leaf in \textsc{Phantom} \citep{phantom}. Additionally, integrating a two-stage neighbour cache alongside the reduction algorithm can double performance. To sum up, the combined use of reduction and a two-stage cache enhances cache building performance by tenfold, while doubling time-stepping performance.

\subsubsection{Performance of time stepping (single GPU \& single CPU)}
\label{sec:single-gpu-cpu-perf}

\begin{figure}
\includegraphics[width=0.91\linewidth]{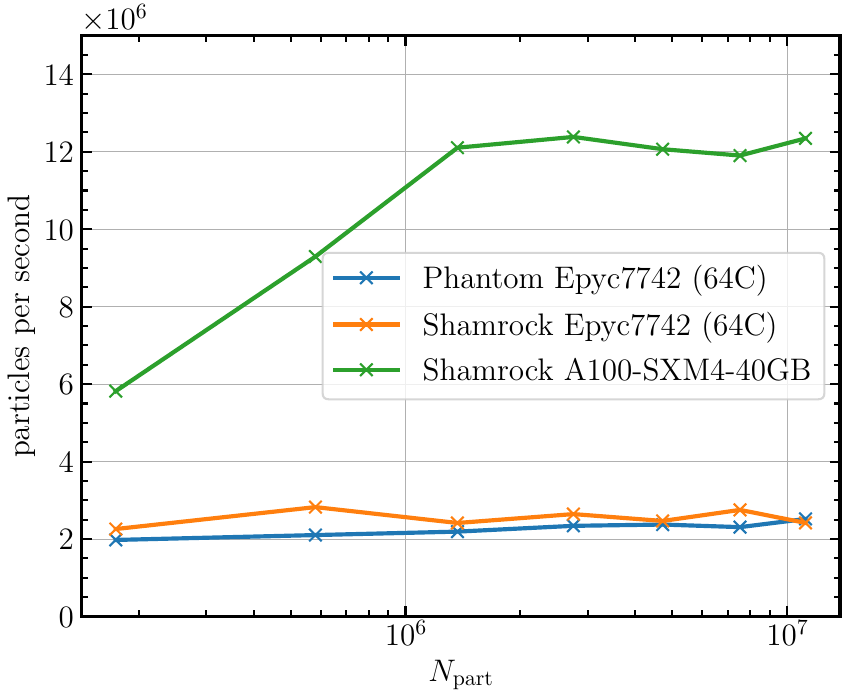}
    \caption{Comparative benchmark of \SHAMROCK{} and \textsc{Phantom} ran on a same restart file of a Sedov-Taylor blast at multiple resolutions produced by \textsc{Phantom}. \SHAMROCK{} does a achieve a slightly higher performance on CPU compared to \textsc{Phantom}, when run a on a single NVIDIA A100-SXM4-40GB GPU the performance is around $5$ times higher for large datasets. Small datasets are not large enough to saturate the GPU explaining the lowered performance on GPU below $10^6$ particles.}
\label{fig:singlegpuperf}
\end{figure}

We evolve setup a Sedov blast using \textsc{Phantom} git pulled at commit number \texttt{e01f76c3}, with compile flags \texttt{IND\char`_TIMESTEPS=no MAXP=50000000} and evolve it with \textsc{Phantom} for a five timesteps and lower both CFL to $10^{-3}$ to avoid leapfrog corrector sub-cycling in both codes and produce a restart file. We then start both \SHAMROCK{} and \textsc{Phantom} on the same restart for 5 iterations, to avoid result being affected by cache warm up. IO has carefully been subtracted from the \textsc{Phantom} measured time.
The performance of \SHAMROCK{} is first tested on a single A100-SXM4-40GB GPU, of total power $275\,\mathrm{W}$. Fig.~\ref{fig:singlegpuperf} shows the number of particles per second iterated as a function as the total number of particles $N_{\rm part}$ in the simulation. As expected, performance increase as the computational pressure on the GPU increases, up to the point where the solver becomes memory-bound ($\sim 10^{6}$ particles). Beyond this threshold, a typical speed of $12\times 10^{6}$ particles per second is achieved.
For a comparison, we perform a similar test with \textsc{Phantom} on an AMD Epyc7742 CPU. On this architecture, \textsc{Phantom} fully exploit its \OpenMP parallelisation across the 64 cores (128 threads). The power consumption is also similar to the one of the A100-SXM4-40GB used for \SHAMROCK{} ($\sim 275$W). For the test described above, one obtains $\gtrsim 2 \times 10^{6}$ particles per second in most cases. 
Note that on this Epyc7742 CPU architecture, \SHAMROCK{} (compiled using AdaptiveCPP \OpenMP backend) achieves slightly higher performance. Despite the limitations inherent in such a comparison, we estimate that \SHAMROCK{} attains approximately a $\sim 5$ factor gain in performance when executed on a single GPU compared to a state-of-the-art SPH CPU code with equivalent power consumption. The performance achieved by \SHAMROCK{} on a single GPU is in line with the increase in bandwidth when running on GPU.

We can not compare directly our results to those of Gadget-4 \cite[Fig.\,61]{gadget}, since it reports the performance in time to simulation without the mention of the total number of iterations in the simulation.
With another SPH kernel, we find that the size of the cache is proportional to the number of neighbours, and the performance, to its inverse.

\subsection{Multi-GPU \& scalability}

\begin{figure*}
\includegraphics[width=1\linewidth]{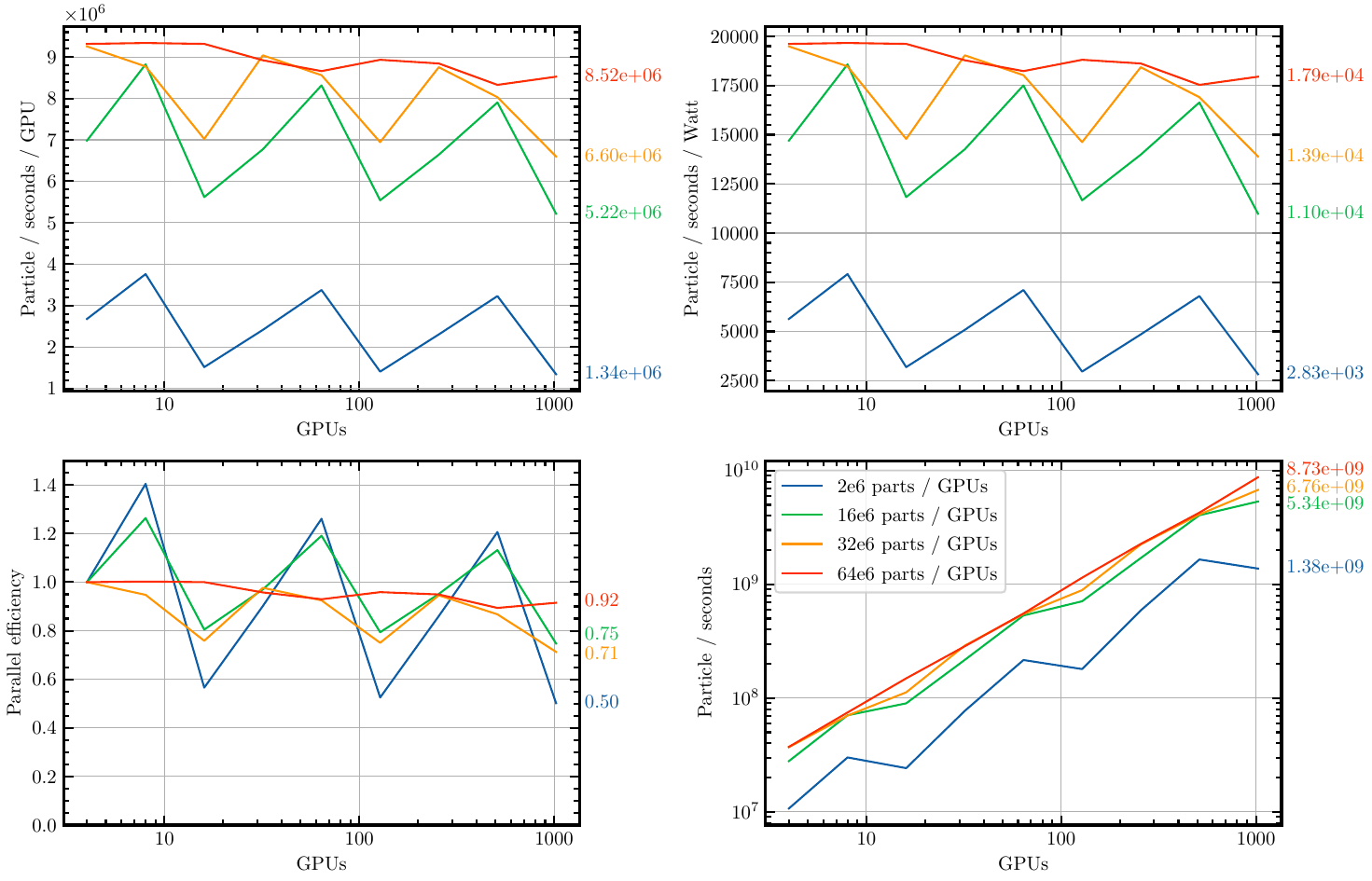}
    \caption{Weak scaling tests conducted on the CINES Adastra supercomputer. These tests are performed for multiple resolutions, from 1 node to 256 nodes, corresponding to using 4 GPUs with 8 \MPI ranks to 1024 GPUs with 2048 \MPI ranks. In these tests, we use the setup of a Sedov-Taylor blast and report the number of particles per GPU. The patch decomposition is set to have at least 8 patches per \MPI ranks. We observe that for large simulations, the scaling test results in 92\% parallel efficiency on $65$ billion particles at 9 billion particle per seconds. Lowering the base resolution reduce the per-GPU performance since GPUs start to be under-utilised. Additionally the variation of the number of particle per patch result in variation in per-GPU performance, resulting in a saw-tooth pattern. We also report the energy efficiency of the tests, were the power consumption used was measured in the single node case and extrapolated as being the product of the number of nodes times the single node consumption.}
\label{fig:perfadastra}
\end{figure*}

We perform the multi-GPU test of \SHAMROCK{} on the \textsc{Adastra} supercomputer of the French CINES, in its early February 2024 configuration. In this multi-GPU test, the split criterion of patches is set at one-sixth of the number of particles per GPU, guaranteeing a minimum of 8 patches per MPI process. We evolve over 5 time steps of the Sedov test and report the performance obtained on the last time step. Unlike in the case of a single GPU, we do not evolve the simulation prior to measurement to limit computational expenses. The scale at which internal energy is injected remains consistent across all tests.

Fig.~\ref{fig:perfadastra} shows that for $65 \times 10^{9}$ particles distributed over 1000 GPUs ($64 \times 10^{6}$ particles per GPU), \SHAMROCK{} achieves $9 \times 10^{9}$ particles iterated per second.  Consequently, iterating one time step over the entire $65 \times 10^{9}$ particles requires 7 seconds on this cluster. These results correspond to around 1.5 times the performance achieved by a single A100 on the same test at the same commit, a value close to what is expected given the hardware specifications. This demonstrates no significant deviation in behaviour attributable to the choice of GPUs. To achieve good load balancing, we find that we need around 10 patches per MPI process. On \textsc{Adastra}, this translates to 20 patches per GPU, amounting to a bit over 1 million particles per Mi250x Graphics Compute Die. Fig.~\ref{fig:singlegpuperf} shows that for small number of particles, the GPU execution units are not loaded efficiently. A correct load corresponds to a typical 2 millions of particles per GPU. Fig.~\ref{fig:perfadastra} also reveals saw tooth shape as the number of particles increases. This feature can be interpreted by noting that every multiple of 8 GPUs, the patches are divided to avoid becoming too large, causing performance to drop below the efficiency threshold. Efficiency then increases again with the number of particles, until another factor of 8 in the resolution is reached, necessitating further patch refinement.
Additionally on Fig.~\ref{fig:perfadastra} we also report the energy efficiency of the weak scaling tests. We measure on every single nodes tests the power consumption related to an iteration of the solver using the hardware counters of the HPE Cray EX235a node. The reported value for multiple nodes is extrapolated from the single node case assuming a total power consumption being the product of the number of nodes times the single node power consumption time the parallel efficiency. Finally we report the power efficiency measured in particles per second per Watt which is also the number of particles processes with a single Joule.
The total power consumption of a node in those tests is not very sensible to the number of particles per GPUs. However the GPU performance can be significantly reduced when GPUs do not have enough particle to process, and having a larger number of particle per GPUs result in the highest efficiency. Maximising the number of particles per GPU maximises efficiency in most cases.

As CPUs are slower than GPUs, communications latencies are of lesser importance than on GPU clusters, and at least a similar level of performance is expected to be achieved on large multi-CPUs clusters.
Memory usage is dominated by the size of the neighbour cache ($\sim 1$ integer/neighbour/particle). On the example of the Adastra super-computer, the limiting factor is found not to be the memory consumption, but the size of the MPI communications (size limited to $2^{31}$ bytes).

In the tests presented in this Section, simulations are limited at small number of particles by latencies, and at large number of particles by the size of MPI communications. Hence, the optimal size of the simulation is between 1 and 64 millions of particles per GPU. We have made different weak scaling tests for different of particles. A strong scaling plot is therefore just a remapping of test presented in Fig. \ref{fig:perfadastra}, and does not contain any new information.

\subsection{Summary}

The larger the simulation, the higher the performance per GPU. Multi-GPU architectures therefore require large simulations to scale and be energy-efficient. With \SHAMROCK{}, this effect can be mitigated by reducing the number of patches, albeit with the potential drawback of rising load balancing issues. Similar benchmarks would be required for further implementations in \SHAMROCK{} of additional physical processes or setups, possibly involving different particle distributions.

\section{Perspectives}
\label{sec:prospects}

\subsection{Multi-physics}

Multi-scale astrophysical problems are often multi-physical. To be consistent, very high-resolution simulations must include realistic physics. The version of \SHAMROCK{} discussed in this article focuses on a purely hydrodynamic SPH solver. Next steps of development consist of implementing local algorithms to address the radiative dynamics of magnetised and dusty fluids. In principle, the modular format of \SHAMROCK{} facilitates the assembly of a new set of known numerical equations into a solver. The biggest challenge is the implementation of gravity, a non-local interaction that requires a new algorithmic layer based on group-group interactions.  Two main algorithms are used to handle gravity by the various astrophysical codes. SPH codes mainly use the method of Fast Multipole Moments (FMM), which takes advantage of the tree structure inherent in particle methods. The technical hurdle lies in achieving numerical efficiency, even for high opening angles involving summation over hundreds or even thousands of neighbours. An alternative approach is to employ a multigrid method, but to our knowledge this has not yet been implemented in an SPH code. These additional physical elements require the implementation of an individual time step to maximise performance (for example using a mask to update only particles of interest. However, it would probably not make sense to execute thinly populated time-steps on the CPU instead of the GPU, since communications between CPU and GPU would likely be a bottleneck for the performance) . It will consequently be also possible to take advantage of \SHAMROCK{}'s efficiency to benchmark individual time stepping against fixed time stepping in simulations that were previously not tractable.

\subsection{Multi-methods}    
    
The \SHAMROCK{} framework relies on the efficient construction and traversal of its tree, irrespective of the numerical object considered. In this study, we have considered particles to develop an SPH solver, but these can be replaced by Eulerian cells in an agnostic way. Since the tree algorithm of \SHAMROCK{} scales almost perfectly to any disordered particle distribution, we can expect similar performance, even on an AMR (Adaptive Mesh Refinement) grid. In principle, various algorithms can be implemented on this grid (such as finite differences or finite volumes), with the moderate cost of incorporating a few specific modules tailored to these solvers (like the accumulation of flows on faces for finite volumes).

The advantages of such a unified framework are twofold. Firstly, to validate an astrophysical model by achieving consistent results with inherently different methods, while the physical model used is rigorously identical (e.g. opacities, cooling rates, resistivities, chemical networks, equations of state). Secondly, to enable rigorous evaluation of numerical methods in terms of accuracy and computational efficiency for specific numerical problems. To our knowledge, no such framework currently exists.

\subsection{Data analysis}
    
The efficiency of the \SHAMROCK{} tree means that data analyses can be carried out very efficiently, whether on particles, cells or both. We plan to develop a native library that will enable these analyses to be performed efficiently on multi-GPU parallel architectures.

\subsection{Optimisation of load balancing}

In this study, a comprehensive analysis of the performance of parallelism were conducted using the Sedov test, where particle distribution is homogeneous in space. The disk test shows that performance is not significantly affected when the distribution exhibits only a large-scale gradient. The goal would be to achieve performance that is only weakly dependent on mass distribution, even when gravity creates highly dense and localised structures. To this end, we have abstracted the load-balancing module and plan to explore a more refined load modelling approach in a future study.

\subsection{Optimisation of latencies}

Finally, the \SHAMROCK{} framework has been designed to optimise performance on multi-node architectures. As discussed in the previous section, specificities of modern hardware imply that performance increases with the computational load demanded of the GPUs. Maybe counter-intuitively, \SHAMROCK{} is therefore not designed by default to run small simulations with a large number of iterations. Since efficient execution of such simulations remains a complementary challenge to that of Exascale, a layer of optimisations regarding remnant latencies remains to be implemented. This would enable users to employ \SHAMROCK{} for both small and large-scale simulations. We will benefit from the knowledge of the work done by the \textsc{Gromacs} team towards optimising latencies in \SYCL{} \citep{alekseenko2024gromacs}.

\section{Conclusion}

We introduced \SHAMROCK{}, a modular and versatile framework designed to run efficiently on multi-GPUs architectures, towards Exascale simulations. The efficiency of \SHAMROCK{} is due to its tree, based on a fully parallel binary logic (Karras algorithm). On a single GPU of an A100, the algorithm builds a tree for 200 million particles in one second. The tree traversal speed, while summing over approximately 60 neighbours, reaches 12 million particles per second per GPU. This property makes it possible to build a Smoothed Particle Hydrodynamics (SPH) solver where neighbours are not stored but recalculated on the fly, reconstructing a tree almost instantaneously.

To exploit the efficiency of this framework, we have implemented and tested an hydrodynamic SPH solver in \SHAMROCK{}. For a Sedov test performed with $10^{6}$ particles on a single A100 GPU, a \SHAMROCK{} simulation is around $\sim$6 times faster than an identical simulation performed with \textsc{Phantom} on an Epic 7742 multicore CPU architecture of equivalent power. The parallelisation of \SHAMROCK{} on several nodes relies on an MPI protocol with hollow communications between the interfaces of a patch system that groups calculations performed on different GPUs. \SHAMROCK{} 's scaling has been tested on the \textsc{Adastra} supercomputer (on 1024 mi250x GPUs). As expected, the higher the  computational load on the GPU, the better the efficiency of the code. For $32 \times 10^{6}$ particles per GPU and $65$ billions of particles in total, we achieve $92\%$ efficiency at low scaling, in a simulation where $9 \times 10^{9}$ particles are iterated per second. Iterating one time step over the $65 \times 10^{9}$ particles takes therefore 7 seconds on this architecture. \SHAMROCK{} is therefore a promising framework that will soon be extended to other grid-based algorithms with adaptive refinement.

\section*{Acknowledgements}
We warmly thank G. Lesur for guidance on the use of multiple GPU systems, and advices on Grid-5000 for prototyping, as well as on the Adastra cluster.
We also thank A. Alpay (\textsc{AdaptiveCPP}) and A. Alekseenko (\textsc{Gromacs}) for guidance on the use of \SYCL{} and related optimizations, 
F. Lovascio, B. Commerçon, L. Sewanou, J. Fensch, A. Durocher and L. Marchal for useful comments and discussions, 
E. Quemener and the Centre Blaise Pascal de Simulation et de Modélisation Numérique for support on the benchmarks, the ENS de Lyon for partly funding the DGX Nvidia Workstation and the electric costs associated to local simulations, 
the CINES for having granted access and support for the \textsc{Adastra} machine during the duration of the Adastra GPU hackaton in Feb. 2024, and in particular E. Malaboeuf, J.-Y. Vet for technical discussions.
A. Charlet, E. Lynch, R. Lenoble for comments on the manuscript.
The \SHAMROCK{} code follows the developments of the WP 5.1 of the Programme et équipements prioritaires de recherche (PEPR) \textit{Origins} (PI: A. Morbidelli). 
We acknowledge funding from the ERC CoG project PODCAST No 864965 as well as funding from the European Research Council (ERC) under the European Union's Horizon Europe research and innovation program (grant agreement No. 101053020, project Dust2Planets).
Lastly, we thank the referees for very insightful reviews.

%%%%%%%%%%%%%%%%%%%%%%%%%%%%%%%%%%%%%%%%%%%%%%%%%%
\section*{Data Availability}

The code and the relevant sources are publicly distributed on Github \faGithub\,under the open source CeCILL v2.1 license (\url{https://github.com/Shamrock-code/Shamrock}).



\bibliographystyle{mnras}
\bibliography{biblio} % if your bibtex file is called example.bib

\begin{thebibliography}{}
\makeatletter
\relax
\def\mn@urlcharsother{\let\do\@makeother \do\$\do\&\do\#\do\^\do\_\do\%\do\~}
\def\mn@doi{\begingroup\mn@urlcharsother \@ifnextchar [ {\mn@doi@}
  {\mn@doi@[]}}
\def\mn@doi@[#1]#2{\def\@tempa{#1}\ifx\@tempa\@empty \href
  {http://dx.doi.org/#2} {doi:#2}\else \href {http://dx.doi.org/#2} {#1}\fi
  \endgroup}
\def\mn@eprint#1#2{\mn@eprint@#1:#2::\@nil}
\def\mn@eprint@arXiv#1{\href {http://arxiv.org/abs/#1} {{\tt arXiv:#1}}}
\def\mn@eprint@dblp#1{\href {http://dblp.uni-trier.de/rec/bibtex/#1.xml}
  {dblp:#1}}
\def\mn@eprint@#1:#2:#3:#4\@nil{\def\@tempa {#1}\def\@tempb {#2}\def\@tempc
  {#3}\ifx \@tempc \@empty \let \@tempc \@tempb \let \@tempb \@tempa \fi \ifx
  \@tempb \@empty \def\@tempb {arXiv}\fi \@ifundefined
  {mn@eprint@\@tempb}{\@tempb:\@tempc}{\expandafter \expandafter \csname
  mn@eprint@\@tempb\endcsname \expandafter{\@tempc}}}

\bibitem[\protect\citeauthoryear{Abraham, Murtola, Schulz, P{\'a}ll, Smith,
  Hess  \& Lindahl}{Abraham et~al.}{2015}]{abraham2015gromacs}
Abraham M.~J.,  Murtola T.,  Schulz R.,  P{\'a}ll S.,  Smith J.~C.,  Hess B.,
  Lindahl E.,  2015, SoftwareX, 1, 19

\bibitem[\protect\citeauthoryear{Adinets \& Merrill}{Adinets \&
  Merrill}{2022}]{adinets2022onesweep}
Adinets A.,  Merrill D.,  2022, arXiv preprint arXiv:2206.01784

\bibitem[\protect\citeauthoryear{Alekseenko \& P\'{a}ll}{Alekseenko \&
  P\'{a}ll}{2023}]{GromacsSycl}
Alekseenko A.,  P\'{a}ll S.,  2023, in Proceedings of the 2023 International
  Workshop on OpenCL. IWOCL '23.
Association for Computing Machinery, New York, NY, USA,
  \mn@doi{10.1145/3585341.3585350}, \url
  {https://doi.org/10.1145/3585341.3585350}

\bibitem[\protect\citeauthoryear{Alekseenko, P{\'a}ll  \& Lindahl}{Alekseenko
  et~al.}{2024}]{alekseenko2024gromacs}
Alekseenko A.,  P{\'a}ll S.,   Lindahl E.,  2024, arXiv preprint
  arXiv:2405.01420

\bibitem[\protect\citeauthoryear{Alpay \& Heuveline}{Alpay \&
  Heuveline}{2020}]{HipSYCL1}
Alpay A.,  Heuveline V.,  2020, in Proceedings of the International Workshop on
  OpenCL. IWOCL '20.
Association for Computing Machinery, New York, NY, USA,
  \mn@doi{10.1145/3388333.3388658}, \url
  {https://doi.org/10.1145/3388333.3388658}

\bibitem[\protect\citeauthoryear{Alpay, Soproni, W\"{u}nsche  \&
  Heuveline}{Alpay et~al.}{2022}]{HipSYCL2}
Alpay A.,  Soproni B.,  W\"{u}nsche H.,   Heuveline V.,  2022, in International
  Workshop on OpenCL. IWOCL'22.
Association for Computing Machinery, New York, NY, USA,
  \mn@doi{10.1145/3529538.3530005}, \url
  {https://doi.org/10.1145/3529538.3530005}

\bibitem[\protect\citeauthoryear{{Arkhipov}, {Wu}, {Li}  \& {Regan}}{{Arkhipov}
  et~al.}{2017}]{2017arXiv170902520A}
{Arkhipov} D.~I.,  {Wu} D.,  {Li} K.,   {Regan} A.~C.,  2017, \mn@doi [arXiv
  e-prints] {10.48550/arXiv.1709.02520}, \href
  {https://ui.adsabs.harvard.edu/abs/2017arXiv170902520A} {p. arXiv:1709.02520}

\bibitem[\protect\citeauthoryear{Batcher}{Batcher}{1968}]{Batcher1968}
Batcher K.~E.,  1968, in Proceedings of the April 30--May 2, 1968, Spring Joint
  Computer Conference. AFIPS '68 (Spring).
Association for Computing Machinery, New York, NY, USA, p. 307–314,
  \mn@doi{10.1145/1468075.1468121}, \url
  {https://doi.org/10.1145/1468075.1468121}

\bibitem[\protect\citeauthoryear{{Bedorf} \& {Portegies Zwart}}{{Bedorf} \&
  {Portegies Zwart}}{2020}]{BedorfPortegiesZwart2020}
{Bedorf} J.,  {Portegies Zwart} S.,  2020, \mn@doi [SciPost Astronomy]
  {10.21468/SciPostAstro.1.1.001}, \href
  {https://ui.adsabs.harvard.edu/abs/2020SciPA...1....1B} {1, 001}

\bibitem[\protect\citeauthoryear{Blelloch}{Blelloch}{1990}]{blelloch1990prefix}
Blelloch G.~E.,  1990, School of Computer Science, Carnegie Mellon University
  Pittsburgh, PA, USA

\bibitem[\protect\citeauthoryear{{Ceppi}, {Cuello}, {Lodato}, {Clarke}, {Toci}
  \& {Price}}{{Ceppi} et~al.}{2022}]{CeppiEtAl2022}
{Ceppi} S.,  {Cuello} N.,  {Lodato} G.,  {Clarke} C.,  {Toci} C.,   {Price}
  D.~J.,  2022, \mn@doi [\mnras] {10.1093/mnras/stac1390}, \href
  {https://ui.adsabs.harvard.edu/abs/2022MNRAS.514..906C} {514, 906}

\bibitem[\protect\citeauthoryear{{Courant}, {Friedrichs}  \& {Lewy}}{{Courant}
  et~al.}{1928}]{CourantFriedrichsLewy1928}
{Courant} R.,  {Friedrichs} K.,   {Lewy} H.,  1928, \mn@doi [Mathematische
  Annalen] {10.1007/BF01448839}, \href
  {https://ui.adsabs.harvard.edu/abs/1928MatAn.100...32C} {100, 32}

\bibitem[\protect\citeauthoryear{Deakin \& McIntosh-Smith}{Deakin \&
  McIntosh-Smith}{2020}]{CloverLeafSYCL}
Deakin T.,  McIntosh-Smith S.,  2020, in Proceedings of the International
  Workshop on OpenCL. IWOCL '20.
Association for Computing Machinery, New York, NY, USA,
  \mn@doi{10.1145/3388333.3388643}, \url
  {https://doi.org/10.1145/3388333.3388643}

\bibitem[\protect\citeauthoryear{{Duffell} et~al.,}{{Duffell}
  et~al.}{2024}]{DuffellEtAl2024}
{Duffell} P.~C.,  et~al., 2024, \mn@doi [\apj] {10.3847/1538-4357/ad5a7e},
  \href {https://ui.adsabs.harvard.edu/abs/2024ApJ...970..156D} {970, 156}

\bibitem[\protect\citeauthoryear{{Durier} \& {Dalla Vecchia}}{{Durier} \&
  {Dalla Vecchia}}{2012}]{DurierDallaVecchia2012}
{Durier} F.,  {Dalla Vecchia} C.,  2012, \mn@doi [\mnras]
  {10.1111/j.1365-2966.2011.19712.x}, \href
  {https://ui.adsabs.harvard.edu/abs/2012MNRAS.419..465D} {419, 465}

\bibitem[\protect\citeauthoryear{{Farris}, {Duffell}, {MacFadyen}  \&
  {Haiman}}{{Farris} et~al.}{2014}]{FarrisEtAl2014}
{Farris} B.~D.,  {Duffell} P.,  {MacFadyen} A.~I.,   {Haiman} Z.,  2014,
  \mn@doi [\apj] {10.1088/0004-637X/783/2/134}, \href
  {https://ui.adsabs.harvard.edu/abs/2014ApJ...783..134F} {783, 134}

\bibitem[\protect\citeauthoryear{{Gafton} \& {Rosswog}}{{Gafton} \&
  {Rosswog}}{2011}]{2011Gafton}
{Gafton} E.,  {Rosswog} S.,  2011, \mn@doi [\mnras]
  {10.1111/j.1365-2966.2011.19528.x}, \href
  {https://ui.adsabs.harvard.edu/abs/2011MNRAS.418..770G} {418, 770}

\bibitem[\protect\citeauthoryear{{Gingold} \& {Monaghan}}{{Gingold} \&
  {Monaghan}}{1977}]{1977Gingold}
{Gingold} R.~A.,  {Monaghan} J.~J.,  1977, \mn@doi [\mnras]
  {10.1093/mnras/181.3.375}, \href
  {https://ui.adsabs.harvard.edu/abs/1977MNRAS.181..375G} {181, 375}

\bibitem[\protect\citeauthoryear{{Grete} et~al.,}{{Grete}
  et~al.}{2022}]{Parthenon}
{Grete} P.,  et~al., 2022, \mn@doi [arXiv e-prints]
  {10.48550/arXiv.2202.12309}, \href
  {https://ui.adsabs.harvard.edu/abs/2022arXiv220212309G} {p. arXiv:2202.12309}

\bibitem[\protect\citeauthoryear{{Hairer}, {Lubich}  \& {Wanner}}{{Hairer}
  et~al.}{2003}]{HairerLubichWanner2003}
{Hairer} E.,  {Lubich} C.,   {Wanner} G.,  2003, \mn@doi [Acta Numerica]
  {10.1017/S0962492902000144}, \href
  {https://ui.adsabs.harvard.edu/abs/2003AcNum..12..399H} {12, 399}

\bibitem[\protect\citeauthoryear{{Hopkins}}{{Hopkins}}{2014}]{Hopkins2014}
{Hopkins} P.~F.,  2014, {GIZMO: Multi-method magneto-hydrodynamics+gravity
  code}, Astrophysics Source Code Library, record ascl:1410.003

\bibitem[\protect\citeauthoryear{{Hopkins}}{{Hopkins}}{2015}]{gizmo}
{Hopkins} P.~F.,  2015, \mn@doi [\mnras] {10.1093/mnras/stv195}, \href
  {https://ui.adsabs.harvard.edu/abs/2015MNRAS.450...53H} {450, 53}

\bibitem[\protect\citeauthoryear{Horn}{Horn}{2005}]{horn2005stream}
Horn D.,  2005, Gpu gems, 2, 573

\bibitem[\protect\citeauthoryear{{Hubber}, {Batty}, {McLeod}  \&
  {Whitworth}}{{Hubber} et~al.}{2011}]{HubberBattyEtAl2011}
{Hubber} D.~A.,  {Batty} C.~P.,  {McLeod} A.,   {Whitworth} A.~P.,  2011,
  \mn@doi [\aap] {10.1051/0004-6361/201014949}, \href
  {https://ui.adsabs.harvard.edu/abs/2011A&A...529A..27H} {529, A27}

\bibitem[\protect\citeauthoryear{Jakob, Rhinelander  \& Moldovan}{Jakob
  et~al.}{2024}]{jakobpybind11}
Jakob W.,  Rhinelander J.,   Moldovan D.,  2024, URL: https://github.
  com/pybind/pybind11

\bibitem[\protect\citeauthoryear{Jin \& Vetter}{Jin \&
  Vetter}{2022}]{PerfSYCLCUDA}
Jin Z.,  Vetter J.~S.,  2022, in Proceedings of the 13th ACM International
  Conference on Bioinformatics, Computational Biology and Health Informatics.
  BCB '22.
Association for Computing Machinery, New York, NY, USA,
  \mn@doi{10.1145/3535508.3545591}, \url
  {https://doi.org/10.1145/3535508.3545591}

\bibitem[\protect\citeauthoryear{Karras}{Karras}{2012}]{karras2012maximizing}
Karras T.,  2012, in Proceedings of the Fourth ACM SIGGRAPH/Eurographics
  conference on High-Performance Graphics. pp 33--37

\bibitem[\protect\citeauthoryear{Lattanzio, Monaghan, Pongracic  \&
  Schwarz}{Lattanzio et~al.}{1986}]{LattanzioMonaghan.et.al.1986}
Lattanzio J.,  Monaghan J.,  Pongracic H.,   Schwarz M.,  1986, SIAM Journal on
  Scientific and Statistical Computing, 7, 591

\bibitem[\protect\citeauthoryear{Lattner \& Adve}{Lattner \&
  Adve}{2004}]{llvmproject}
Lattner C.,  Adve V.,  2004, in International Symposium on Code Generation and
  Optimization, 2004. CGO 2004.. pp 75--86, \mn@doi{10.1109/CGO.2004.1281665}

\bibitem[\protect\citeauthoryear{Lauterbach, Garland, Sengupta, Luebke  \&
  Manocha}{Lauterbach et~al.}{2009}]{Lauterbach2009FastBC}
Lauterbach C.,  Garland M.,  Sengupta S.,  Luebke D.~P.,   Manocha D.,  2009,
  Computer Graphics Forum, 28

\bibitem[\protect\citeauthoryear{{Lesur}, {Baghdadi}, {Wafflard-Fernandez},
  {Mauxion}, {Robert}  \& {Van den Bossche}}{{Lesur} et~al.}{2023}]{Idefix}
{Lesur} G.~R.~J.,  {Baghdadi} S.,  {Wafflard-Fernandez} G.,  {Mauxion} J.,
  {Robert} C.~M.~T.,   {Van den Bossche} M.,  2023, \mn@doi [arXiv e-prints]
  {10.48550/arXiv.2304.13746}, \href
  {https://ui.adsabs.harvard.edu/abs/2023arXiv230413746L} {p. arXiv:2304.13746}

\bibitem[\protect\citeauthoryear{{Lodato} \& {Price}}{{Lodato} \&
  {Price}}{2010}]{LodatoPrice2010}
{Lodato} G.,  {Price} D.~J.,  2010, \mn@doi [\mnras]
  {10.1111/j.1365-2966.2010.16526.x}, \href
  {https://ui.adsabs.harvard.edu/abs/2010MNRAS.405.1212L} {405, 1212}

\bibitem[\protect\citeauthoryear{{Lucy}}{{Lucy}}{1977}]{1977Lucy}
{Lucy} L.~B.,  1977, \mn@doi [\aj] {10.1086/112164}, \href
  {https://ui.adsabs.harvard.edu/abs/1977AJ.....82.1013L} {82, 1013}

\bibitem[\protect\citeauthoryear{Markomanolis et~al.,}{Markomanolis
  et~al.}{2022}]{LumiSYCL}
Markomanolis G.~S.,  et~al., 2022, in Supercomputing Frontiers: 7th Asian
  Conference, SCFA 2022, Singapore, March 1–3, 2022, Proceedings.
  Springer-Verlag, Berlin, Heidelberg, p. 79–101,
  \mn@doi{10.1007/978-3-031-10419-0_6}, \url
  {https://doi.org/10.1007/978-3-031-10419-0_6}

\bibitem[\protect\citeauthoryear{Merrill \& Garland}{Merrill \&
  Garland}{2016}]{merrill2016single}
Merrill D.,  Garland M.,  2016, NVIDIA, Tech. Rep. NVR-2016-002

\bibitem[\protect\citeauthoryear{{Monaghan}}{{Monaghan}}{1997}]{Monaghan1997}
{Monaghan} J.~J.,  1997, \mn@doi [Journal of Computational Physics]
  {10.1006/jcph.1997.5732}, \href
  {https://ui.adsabs.harvard.edu/abs/1997JCoPh.136..298M} {136, 298}

\bibitem[\protect\citeauthoryear{Morton}{Morton}{1966}]{morton1966}
Morton G.~M.,  1966, International Business Machines Company New York

\bibitem[\protect\citeauthoryear{Nassimi \& Sahni}{Nassimi \&
  Sahni}{1979}]{Nassimi1679}
Nassimi Sahni 1979, \mn@doi [IEEE Transactions on Computers]
  {10.1109/TC.1979.1675216}, C-28, 2

\bibitem[\protect\citeauthoryear{{Price}}{{Price}}{2008}]{2008Price}
{Price} D.~J.,  2008, \mn@doi [Journal of Computational Physics]
  {10.1016/j.jcp.2008.08.011}, \href
  {https://ui.adsabs.harvard.edu/abs/2008JCoPh.22710040P} {227, 10040}

\bibitem[\protect\citeauthoryear{{Price}}{{Price}}{2012}]{2012Price}
{Price} D.~J.,  2012, \mn@doi [Journal of Computational Physics]
  {10.1016/j.jcp.2010.12.011}, \href
  {https://ui.adsabs.harvard.edu/abs/2012JCoPh.231..759P} {231, 759}

\bibitem[\protect\citeauthoryear{{Price} et~al.,}{{Price}
  et~al.}{2018}]{phantom}
{Price} D.~J.,  et~al., 2018, \mn@doi [\pasa] {10.1017/pasa.2018.25}, \href
  {https://ui.adsabs.harvard.edu/abs/2018PASA...35...31P} {35, e031}

\bibitem[\protect\citeauthoryear{Quemener \& Corvellec}{Quemener \&
  Corvellec}{2013}]{quemener2013sidus}
Quemener E.,  Corvellec M.,  2013, Linux Journal, 2013, 3

\bibitem[\protect\citeauthoryear{Rubin \& Whitted}{Rubin \&
  Whitted}{1980}]{RubinWhitted1980}
Rubin S.~M.,  Whitted T.,  1980, \mn@doi [SIGGRAPH Comput. Graph.]
  {10.1145/965105.807479}, 14, 110–116

\bibitem[\protect\citeauthoryear{{Saitoh} \& {Makino}}{{Saitoh} \&
  {Makino}}{2009}]{SaitohMakino2009}
{Saitoh} T.~R.,  {Makino} J.,  2009, \mn@doi [\apjl]
  {10.1088/0004-637X/697/2/L99}, \href
  {https://ui.adsabs.harvard.edu/abs/2009ApJ...697L..99S} {697, L99}

\bibitem[\protect\citeauthoryear{Samet}{Samet}{2006}]{samet2006foundations}
Samet H.,  2006, Foundations of multidimensional and metric data structures.
Morgan Kaufmann

\bibitem[\protect\citeauthoryear{{Schaal}, {Bauer}, {Chandrashekar}, {Pakmor},
  {Klingenberg}  \& {Springel}}{{Schaal} et~al.}{2015}]{Schaal2015}
{Schaal} K.,  {Bauer} A.,  {Chandrashekar} P.,  {Pakmor} R.,  {Klingenberg} C.,
    {Springel} V.,  2015, \mn@doi [\mnras] {10.1093/mnras/stv1859}, \href
  {https://ui.adsabs.harvard.edu/abs/2015MNRAS.453.4278S} {453, 4278}

\bibitem[\protect\citeauthoryear{{Schaller}, {Gonnet}, {Draper}, {Chalk},
  {Bower}, {Willis}  \& {Hausammann}}{{Schaller}
  et~al.}{2018}]{SchallerGonnetEtAl2018}
{Schaller} M.,  {Gonnet} P.,  {Draper} P.~W.,  {Chalk} A. B.~G.,  {Bower}
  R.~G.,  {Willis} J.,   {Hausammann} L.,  2018, {SWIFT: SPH With
  Inter-dependent Fine-grained Tasking}, Astrophysics Source Code Library,
  record ascl:1805.020

\bibitem[\protect\citeauthoryear{Schoenberg}{Schoenberg}{1946}]{Schoenberg1946}
Schoenberg I.~J.,  1946, Quarterly of Applied Mathematics, 4, 45

\bibitem[\protect\citeauthoryear{{Sedov}}{{Sedov}}{1959}]{1959Sedov}
{Sedov} L.~I.,  1959, {Similarity and Dimensional Methods in Mechanics}

\bibitem[\protect\citeauthoryear{{Sod}}{{Sod}}{1978}]{1978Sod}
{Sod} G.~A.,  1978, \mn@doi [Journal of Computational Physics]
  {10.1016/0021-9991(78)90023-2}, \href
  {https://ui.adsabs.harvard.edu/abs/1978JCoPh..27....1S} {27, 1}

\bibitem[\protect\citeauthoryear{{Springel}, {Pakmor}, {Zier}  \&
  {Reinecke}}{{Springel} et~al.}{2021}]{gadget}
{Springel} V.,  {Pakmor} R.,  {Zier} O.,   {Reinecke} M.,  2021, \mn@doi
  [\mnras] {10.1093/mnras/stab1855}, \href
  {https://ui.adsabs.harvard.edu/abs/2021MNRAS.506.2871S} {506, 2871}

\bibitem[\protect\citeauthoryear{{Taylor}}{{Taylor}}{1950a}]{1950Taylor.a}
{Taylor} G.,  1950a, \mn@doi [Proceedings of the Royal Society of London Series
  A] {10.1098/rspa.1950.0049}, \href
  {https://ui.adsabs.harvard.edu/abs/1950RSPSA.201..159T} {201, 159}

\bibitem[\protect\citeauthoryear{{Taylor}}{{Taylor}}{1950b}]{1950Taylor.b}
{Taylor} G.,  1950b, \mn@doi [Proceedings of the Royal Society of London Series
  A] {10.1098/rspa.1950.0050}, \href
  {https://ui.adsabs.harvard.edu/abs/1950RSPSA.201..175T} {201, 175}

\bibitem[\protect\citeauthoryear{{Tokuue} \& {Ishiyama}}{{Tokuue} \&
  {Ishiyama}}{2024}]{TokuueIshiyama2024}
{Tokuue} T.,  {Ishiyama} T.,  2024, \mn@doi [\mnras] {10.1093/mnras/stad4001},
  \href {https://ui.adsabs.harvard.edu/abs/2024MNRAS.528..821T} {528, 821}

\bibitem[\protect\citeauthoryear{{Tricco}}{{Tricco}}{2019}]{Tricco2019}
{Tricco} T.~S.,  2019, \mn@doi [\mnras] {10.1093/mnras/stz2042}, \href
  {https://ui.adsabs.harvard.edu/abs/2019MNRAS.488.5210T} {488, 5210}

\bibitem[\protect\citeauthoryear{{Trott} et~al.,}{{Trott}
  et~al.}{2021}]{Kokkos}
{Trott} C.,  et~al., 2021, \mn@doi [Computing in Science and Engineering]
  {10.1109/MCSE.2021.3098509}, \href
  {https://ui.adsabs.harvard.edu/abs/2021CSE....23e..10T} {23, 10}

\bibitem[\protect\citeauthoryear{{Verlet}}{{Verlet}}{1967}]{Verlet1967}
{Verlet} L.,  1967, \mn@doi [Physical Review] {10.1103/PhysRev.159.98}, \href
  {https://ui.adsabs.harvard.edu/abs/1967PhRv..159...98V} {159, 98}

\bibitem[\protect\citeauthoryear{{Wadsley}, {Stadel}  \& {Quinn}}{{Wadsley}
  et~al.}{2004}]{WadsleyStadelQuinn2004}
{Wadsley} J.~W.,  {Stadel} J.,   {Quinn} T.,  2004, \mn@doi [\na]
  {10.1016/j.newast.2003.08.004}, \href
  {https://ui.adsabs.harvard.edu/abs/2004NewA....9..137W} {9, 137}

\bibitem[\protect\citeauthoryear{Warren \& Salmon}{Warren \&
  Salmon}{1993}]{WarrenSalmon1993}
Warren M.~S.,  Salmon J.~K.,  1993, in Proceedings of the 1993 ACM/IEEE
  Conference on Supercomputing. Supercomputing '93.
Association for Computing Machinery, New York, NY, USA, p. 12–21,
  \mn@doi{10.1145/169627.169640}, \url {https://doi.org/10.1145/169627.169640}

\bibitem[\protect\citeauthoryear{Wendland}{Wendland}{1995}]{Wendland1995}
Wendland H.,  1995, Advances in computational Mathematics, 4, 389

\bibitem[\protect\citeauthoryear{{Wibking} \& {Krumholz}}{{Wibking} \&
  {Krumholz}}{2022}]{QUOKKA}
{Wibking} B.~D.,  {Krumholz} M.~R.,  2022, \mn@doi [\mnras]
  {10.1093/mnras/stac439}, \href
  {https://ui.adsabs.harvard.edu/abs/2022MNRAS.512.1430W} {512, 1430}

\makeatother
\end{thebibliography}

%\begin{thebibliography}{99}
%\end{thebibliography}

%%%%%%%%%%%%%%%%%%%%%%%%%%%%%%%%%%%%%%%%%%%%%%%%%%

%%%%%%%%%%%%%%%%% APPENDICES %%%%%%%%%%%%%%%%%%%%%

\appendix

\section{Domain decomposition \& \MPI}
\label{sec:MPI}

\subsection{Simulation box}

The three-dimensional volume on which a numerical simulation is performed can be embedded in a cube, whose edges define axes for Cartesian coordinates.
This cube is often referred as an Aligned Axis Bounding Box, or \AABB{}, particularly within the ray-tracing community. The box is  parametrised by two values, $r_{\min}$ and $r_{\max}$, which are chosen to represent the minimum and maximum possible coordinates inside the cube in all three dimensions. For convenience, we shall refer to this \AABB{} as the \textit{simulation box} of \SHAMROCK{}.
Within this box, coordinates can be mapped to a grid of integers, by subdividing the simulation box coordinates into $N_{\rm g}$ grid points on each axis, where $N_{\rm g}$ is a power of two.
In pratice, we use $N_{\rm g} = 2^{21}$ or $N_{\rm g} = 2^{42}$ (see Sect.~\ref{sec:load-balancing}).

\subsection{Patch decomposition}
\label{sec:abstractpatch}

\subsection{Data Structure}

Each patch in \SHAMROCK{} is associated with two types of information. The first type is the \textit{patch metadata}, which encompasses the current status, location and identifier of the patch. The second, called \textit{patch data}, comprises the data pertaining to the fields processed by the patch.

\subsubsection{Patch metadata}

Within \SHAMROCK{}, metadata is synchronised across all \MPI{} ranks. This synchronisation is made possible by the use of a class of small size (80 bytes when compiled). The metadata of a \SHAMROCK{} patch is represented in the code with the following class
\begin{tcolorbox}
    \begin{Verbatim}[commandchars=\\\{\}]
\PY{k}{template}\PY{o}{\PYZlt{}}\PY{n}{u32}\PY{+w}{ }\PY{n}{dim}\PY{o}{\PYZgt{}}
\PY{k}{struct}\PY{+w}{ }\PY{n+nc}{Patch}\PY{p}{\PYZob{}}
\PY{+w}{    }\PY{n}{u64}\PY{+w}{ }\PY{n}{id\PYZus{}patch}\PY{p}{;}
\PY{+w}{    }\PY{n}{u64}\PY{+w}{ }\PY{n}{pack\PYZus{}node\PYZus{}index}\PY{p}{;}
\PY{+w}{    }\PY{n}{u64}\PY{+w}{ }\PY{n}{load\PYZus{}value}\PY{p}{;}

\PY{+w}{    }\PY{n}{std}\PY{o}{:}\PY{o}{:}\PY{n}{array}\PY{o}{\PYZlt{}}\PY{n}{u64}\PY{p}{,}\PY{n}{dim}\PY{o}{\PYZgt{}}\PY{+w}{ }\PY{n}{coord\PYZus{}min}\PY{p}{;}
\PY{+w}{    }\PY{n}{std}\PY{o}{:}\PY{o}{:}\PY{n}{array}\PY{o}{\PYZlt{}}\PY{n}{u64}\PY{p}{,}\PY{n}{dim}\PY{o}{\PYZgt{}}\PY{+w}{ }\PY{n}{coord\PYZus{}max}\PY{p}{;}

\PY{+w}{    }\PY{n}{u32}\PY{+w}{ }\PY{n}{node\PYZus{}owner\PYZus{}id}\PY{p}{;}
\PY{p}{\PYZcb{}}\PY{p}{;}
\end{Verbatim}
  
\end{tcolorbox}
In this class, \texttt{u32} denotes 32 bits unsigned integers and \texttt{u64} their 64 bits variants, \texttt{id\char`_patch} the patch unique identifier, \texttt{load\char`_value} the estimated load of a patch (see Sect.~\ref{sec:abstractpatch}), \texttt{coord\char`_min} and \texttt{coord\char`_max} represent edges of the AABB patch on the integer grid, \texttt{node\char`_owner} the \MPI{} rank owning the current patch. Finally, \texttt{pack\char`_node\char`_index} is an additional field used to specify that a patch aims to reside in the same \MPI{} ranks as another one (see section \ref{sec:schedstep} for more details). We also provide a dedicated \MPI{} type to facilitate the utilisation of collective operations on patch metadata.%

\subsubsection{Patch data} 

The \textit{patch data} of a patch is a list of fields related to a collection of objects (cells or particles). 
A field can contain one or multiple values per object, as long as the number of values per object is constant. 
The first field, so-called the \textit{main field} in \SHAMROCK{}, must have one value per object and store the positions of every object in the patch. 
Domain decomposition and load balancing are executed based on the positions stored in the main field.
When a patch is moved, split or merged, the corresponding operations are applied to the other fields as well. This ensures that communications are implicitly modified when the layout of the data is changed, eliminating the need for direct user intervention.
For efficient implementation of new physics, the fields stored in the patch data can encompass a wide range of types (scalar, vector, or matrices, with float, double, or integer data), arranged in any order. This versatility is enabled by representing the patch data as a \texttt{std::vector} of \texttt{std::variant} encompassing all possible field types. This aspect is abstracted from the user, as only field identifiers and types are required. One example of such use is 
\begin{tcolorbox}
    \begin{Verbatim}[commandchars=\\\{\},fontsize=\smaller]
\PY{n}{PatchData}\PY{+w}{ }\PY{o}{\PYZam{}}\PY{+w}{ }\PY{n}{pdat}\PY{+w}{ }\PY{o}{=}\PY{+w}{ }\PY{p}{.}\PY{p}{.}\PY{p}{.}
\PY{c+c1}{// get the layout of the patch data}
\PY{n}{PatchDataLayout}\PY{+w}{ }\PY{o}{\PYZam{}}\PY{n}{pdl}\PY{+w}{ }\PY{o}{=}\PY{+w}{ }\PY{n}{pdat}\PY{p}{.}\PY{n}{pdl}\PY{p}{;}
\PY{c+c1}{// get id of the field (name and type specified) }
\PY{c+c1}{// f64\PYZus{}3 is a 3 dimensional double precision vector}
\PY{k}{const}\PY{+w}{ }\PY{n}{u32}\PY{+w}{ }\PY{n}{ivxyz}\PY{+w}{      }\PY{o}{=}\PY{+w}{ }\PY{n}{pdl}\PY{p}{.}\PY{n}{get\PYZus{}field\PYZus{}idx}\PY{o}{\PYZlt{}}\PY{n}{f64\PYZus{}3}\PY{o}{\PYZgt{}}\PY{p}{(}\PY{l+s}{\PYZdq{}}\PY{l+s}{vxyz}\PY{l+s}{\PYZdq{}}\PY{p}{)}\PY{p}{;}
\PY{c+c1}{// get the field at this id}
\PY{n}{PatchDataField}\PY{o}{\PYZlt{}}\PY{n}{f64\PYZus{}3}\PY{o}{\PYZgt{}}\PY{+w}{ }\PY{o}{\PYZam{}}\PY{+w}{ }\PY{n}{vxyz}\PY{+w}{ }\PY{o}{=}\PY{+w}{ }
\PY{+w}{                           }\PY{n}{pdat}\PY{p}{.}\PY{n}{get\PYZus{}field}\PY{o}{\PYZlt{}}\PY{n}{f64\PYZus{}3}\PY{o}{\PYZgt{}}\PY{p}{(}\PY{n}{ivxyz}\PY{p}{)}\PY{p}{;}
\end{Verbatim}
  
\end{tcolorbox}

\subsubsection{Patch scheduler}

In \SHAMROCK{}, a single class is responsible of managing patches, distributing data to \MPI ranks and processing the refinement of the patch grid. This class contains the patch octree, patch metadata, and patch data. It is referred internally as the \texttt{PatchScheduler}.
This class is only controlled by four parameters: the patch data layout, which specifies the list of fields and the corresponding number of variables, the split criterion $\mathcal{W}_{\rm max}$ and the merge criterion $\mathcal{W}_{\rm min}$ that control patch refinement, and the load balancing configuration.
The patch scheduler is designed to operate as a black box for the user. The user calls the \texttt{scheduler\char`_step} function, which triggers the scheduler to execute merge, split, and load balancing operations. The \texttt{scheduler\char`_step} is called at the beginning of every time step in practice.
Multiple `for each' functions are provided in \SHAMROCK{} as abstractions for iterating over patches. An example of such use is
\begin{tcolorbox}
    \begin{Verbatim}[commandchars=\\\{\},fontsize=\smaller]
\PY{n}{PatchScheduler}\PY{+w}{ }\PY{o}{\PYZam{}}\PY{+w}{ }\PY{n}{scheduler}\PY{+w}{ }\PY{o}{=}\PY{+w}{ }\PY{p}{.}\PY{p}{.}\PY{p}{.}

\PY{n}{scheduler}\PY{p}{.}\PY{n}{for\PYZus{}each\PYZus{}patchdata}\PY{p}{(}
\PY{+w}{    }\PY{c+c1}{// the c++ lambda contain the operation}
\PY{+w}{    }\PY{c+c1}{// to perform on the patches}
\PY{+w}{    }\PY{p}{[}\PY{o}{\PYZam{}}\PY{p}{]}\PY{p}{(}\PY{k}{const}\PY{+w}{ }\PY{n}{Patch}\PY{+w}{ }\PY{o}{\PYZam{}}\PY{+w}{ }\PY{n}{p}\PY{p}{,}\PY{+w}{ }\PY{n}{PatchData}\PY{+w}{ }\PY{o}{\PYZam{}}\PY{n}{pdat}\PY{p}{)}\PY{+w}{ }\PY{p}{\PYZob{}}
\PY{+w}{        }\PY{c+c1}{// do someting on the patch}
\PY{+w}{    }\PY{p}{\PYZcb{}}
\PY{p}{)}\PY{p}{;}
\end{Verbatim}
  
\end{tcolorbox}
These abstractions shield the end user from interactions with the \MPI layer. The strategy is as follows: one does not need to be aware of which patches reside on which \MPI ranks. Indeed, operations are conducted solely through `for each' calls to the patches, and the scheduler handles the other tasks.

\subsection{Scheduler step}
\label{sec:schedstep}

\begin{figure}
    \centering
    \includegraphics{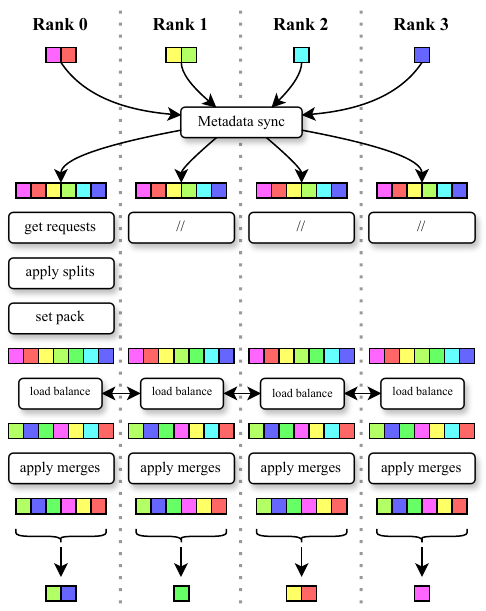}
    \caption{Illustration of a scheduler step. Initially, a synchronisation of the patch metadata occurs across all \MPI ranks, resulting in each rank possessing an identical list of all patch metadata. Subsequently, each \MPI rank generates a list of split and merge requests. Split requests are then executed, followed by setting the packing index. The subsequent operation consists in performing load balancing on all patches. Finally, merge requests are carried out to complete the step.}
    \label{fig:schedstep}
\end{figure}

Fig.~\ref{fig:schedstep} illustrates a single scheduler step in \SHAMROCK{}. During this step, patch data are exclusively processed on their current \MPI{} rank, while patch metadata and the patch tree remain unchanged over all \MPI ranks.

%--ICI---

\subsubsection{Synchronising metadata}

\begin{figure}
    \centering
    \includegraphics{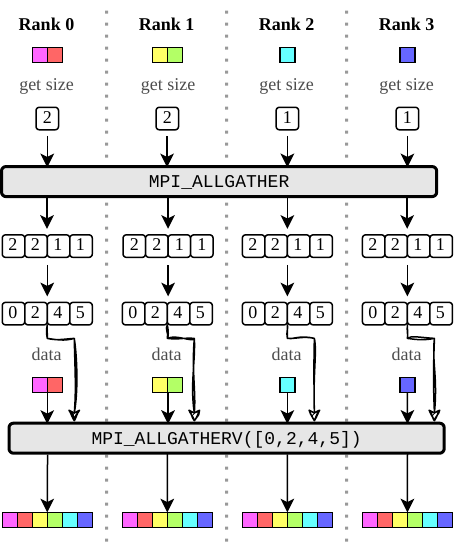}
    \caption{Illustration of the steps performed in a
    \texttt{vector\char`_allgatherv} operation. Firstly, the size of each sent vector is retrieved. Secondly, all \MPI ranks gather the sizes sent by each other \MPI{} ranks. An exclusive scan is performed on this list to obtain the offset at which the data of each \MPI{} rank will be inserted in the final vector. Finally, a \texttt{MPI\char`_ALLGATHER\char`_V} is called using the provided offsets to retrieve the gathered list in each \MPI rank.}
    \label{fig:allgatherv}
 \end{figure}

The initial operation conducted during a scheduler step consists in synchronising the metadata across the \MPI{} ranks. This operation, named \texttt{vector\char`_allgatherv} in \SHAMROCK{}, is implemented as an extension of the \MPI primitive \texttt{MPI\char`_ALLGATHER\char`_V} (see fig.\ref{fig:allgatherv}). Given a \texttt{std::vector} in each \MPI ranks, \texttt{vector\char`_allgatherv} returns on all ranks the same \texttt{std::vector} made by concatenating the input vectors in each ranks. We create an \MPI{} type for the patch metadata, and use \texttt{vector\char`_allgatherv}
to gather all the metadata of all patches on all MPI ranks. This operation returns the list containing the metadata of all patches in the simulation (the step `Metadata sync' in Fig.\ref{fig:schedstep}).

\subsubsection{Listing requests}

The operation `Get requests' depicted in Fig.\ref{fig:schedstep} provides the list of identifiers for patches requiring merging or splitting.
A patch splits when its estimated load exceeds the split criterion (see Sect.\ref{sec:abstractpatch}). If all children of a node in the patch tree meet the merge criterion, they merge, resulting in the parent node being marked for pending child merge and consequently transitioning into a tree leaf.

\subsubsection{Patch splitting}

Subsequent split operations on the metadata and the patch tree are carried out in each \MPI{} rank. If the \MPI{} rank holds the patch data associated with the patch being split into eight new patches, the patch data is then subdivided into eight new patch data objects corresponding to the eight newly formed patches.

\subsubsection{Collecting information on ranks}

The \textit{pack index} is a list containing necessary information indicating whether a given patch \textit{a} must reside in the same \MPI{} rank as another patch \textit{b}.
After having executed patch splitting, we then go through the list of merge operations along the \MPI ranks.
We use an identifier that denotes the parent of the eight merging children patches. 
With the exception of the first child, all the other children patches have their pack index set to the index of the first child in the global metadata list. This signifies that the seven other children patches must be in the same \MPI rank as the first one, which enables the merge operation to be conducted at a later stage. The pack index is used during the subsequent load balancing step.

\subsubsection{Load balancing}
\label{sec:load-balancing}

Performing load balancing consists of grouping patches in chunks, and distributing the chunks appropriately over the \MPI ranks for their computational charge to be as homogeneous as possible. Load balancing is performed in four sub-steps:
\begin{itemize}
\item The load balancing module receives a list of metadata that includes estimated computational loads. Here, a strategy for patch reorganisation of patches is computed (see Sect.~\ref{sec:load-balancing-strat} for the details of the load balancing procedure). The code returns a list specifying the novel \MPI{} rank assignment for each patch. When compared to the current owner of each patch, this list identifies necessary changes, indicating if a patch must move from one \MPI{} rank to another. Additionally, this list incorporates the pack index described above.
\item The patch reorganisation encoded in the list of changes is subsequently implemented. We iterate through the list of changes a first time. If the sender \MPI{} rank matches the \MPI{} rank of the current process, it sends the corresponding patch data using a non-blocking send of the serialized data (see \ref{sec:serialization} for details).
\item We then go through the list of changes for a second time. If the receiver patch matches the \MPI rank of the current process,  we execute an \MPI{} non-blocking receive operation to obtain the corresponding patch data in the new rank. 
\item Finally, we finish by waiting for all \MPI{} operations to complete, thereby concluding the load balancing step.
\end{itemize}
Generating the list of changes accounts for the pack index. As such, patches intended for merging together are in the same \MPI rank after the load balancing operation.

\subsubsection{Patch merging}

Merge operations require for the eight children patches to be in the same \MPI{} rank, which is garanteed by the packing in the load balancing step. Similarly to split operations, merges are executed across both metadata and the patch tree within all \MPI ranks. Merging is also applied to the patch data on the \MPI{} rank that owns the data.
\vfil
\subsection{Load balancing strategies}
\label{sec:load-balancing-strat}
\vfil
The load balancing module generates the list of owner of each patch determined by abstract estimates of its required computational load. Load balancing is processed consistently across all \MPI{} ranks for identical inputs. The load balancing module initially utilises the list of all patch metadata, with the estimated load values as input. Patches are then arranged along a Hilbert curve, which is subsequently segmented into contiguous chunks of adjacent patches. The objective of optimal load balancing is to identify a collection of chunks wherein the workload is distributed as evenly as possible across all \MPI{} ranks.

\vfil
To achieve this, various load balancing strategies are dynamically evaluated in \SHAMROCK{}  (e.g. analytic decomposition, round-robin method), and the one found to be the most effective is selected.  The computational overhead involved in assessing the benefits of different load balancing strategies is minimal, since it relies on simple estimations. This process yields a list specifying the new \MPI{} ranks for each patches. This list is compared against the current distribution of patches to generate the change list when load balancing is applied.
\vfil
%--ICI--
\subsection{Patch interactions}
\subsubsection{Interaction criteria}
\label{sec:patch-interact-crit}
For a given collection of objects (cells or particles), we can establish a condition indicating whether objects $i$ and $j$ interact, and define a Boolean interaction criterion $\gamma_{o/o}(i,j)$ to signify this condition. For example, in the Smoothed Particle Hydrodynamics method, $\gamma_{o/o}$ is defined as true when particle $i$ is within the interaction radius of particle $j$, or vice versa.
\vfil
A first generalisation of this object-object criterion is an object-group criterion, which describes if there is interaction between an object and a group of objects. A necessary condition for such a criterion is
\begin{align*}
    \gamma_{o/g}( i , \{j\}_j) \Leftarrow \bigvee_j \gamma_{o/o}(i,j)
\end{align*}
This condition formally expresses the fact that if the interaction criterion is fulfilled for any object in the group, it must also hold true for the entire group. Failure to meet this condition would imply the possibility of interaction with an element of the group without interaction with the group as a whole, which is incorrect. Both the object-object and group-object criteria are used in the tree traversal step. \ref{sec:treetrav}.
Another extension of the aforementioned criteria is the group-group interaction criterion, which similarly satisfies the following condition
\begin{align*}
    \gamma_{g/g}( \{i\}_i , \{j\}_j) \Leftarrow \bigvee_i \gamma_{o/g}(i,\{j\}_j)
\end{align*}
This latter condition is used to manage ghost zones (see Sect.~\ref{sec:patchgz}) and perform two-stages neighbour search (see Sect.~\ref{sec:twostageneighcache}).
\newpage
\subsubsection{Interaction graph}
\label{sec:patchgz}
\begin{figure}
    \centering
    \includegraphics[width=0.75\linewidth]{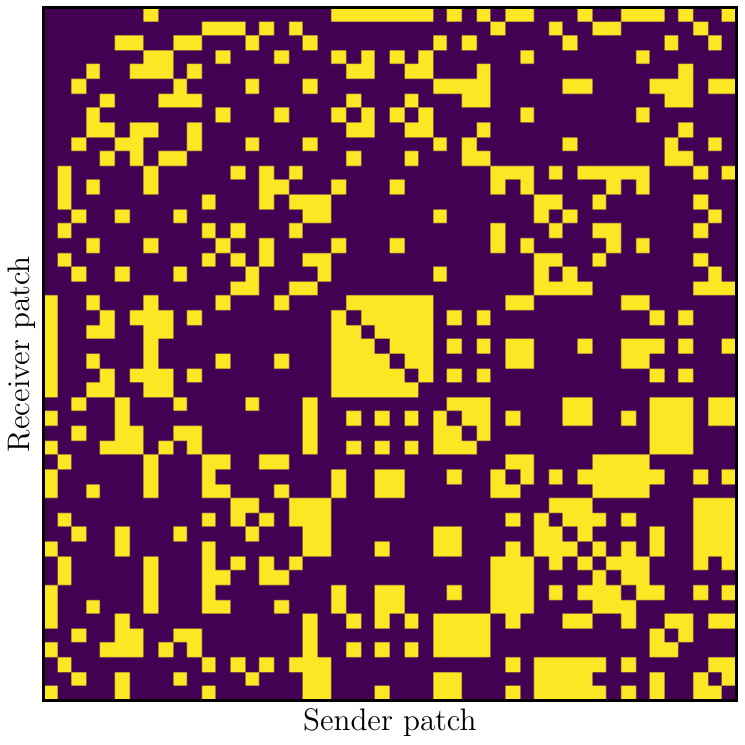}
    \caption{Matrix of the interaction graph between patches extracted from an SPH simulation of a protoplanetary disc, involving several hundred patches. Empty patches have been excluded from the graph as they do not meet any interaction criteria due to their emptiness. The resulting interaction matrix is symmetrical and sparse. This visualisation was generated using a debug tool in \SHAMROCK{}, which creates a dot graph representing the ghost zones of the current time step, which can be rendered in its matrix form here showed.}
    \label{fig:ghostzonegraphexemple}
\end{figure}
Patches themselves are objects that can interact. Their interaction is handled using a group-group interaction criterion $\gamma_{g/g}$. The interaction criterion of an empty patch is always false. After assessing the interaction status between all pairs of patches, we define the \textit{interaction graph} of the patches by considering the list of links such that the group-group interaction criterion $\gamma_{g/g}$ is true. Fig.~\ref{fig:ghostzonegraphexemple} shows an example of such a graph of interactions between patches. 

\subsubsection{Interfaces and ghost zones patch}
We define the \textit{interface between two patches} as the smallest set of individual objects for which the group-group interaction criterion between their parent patches is satisfied. To reduce communications between interacting patches, we communicate not the entire patch content to its neighbour, but only their interfaces. These communicated interfaces consequently manifest as ghost extensions for the neighbouring patch and are therefore called \textit{ghosts zones} of patches. The graph of ghost zones between patches is the same as the interaction graph, with links between vertices representing the ghost zone of one patch being sent to another patch.

\subsection{Serialisation}
\label{sec:serialization}

In \SHAMROCK{}, all communications are serialised, i.e. converted into a stream of bytes to reduce the \MPI overhead by performing less operations, and shield the user from the \MPI layer. To send a patch ghost zone, data are initially packed into a byte buffer. Communication patterns and operations remain therefore unchanged, regardless of the communication content. In particular, the addition of a field simply adds extra data to the serialisation without altering the communication process.

\begin{tcolorbox}
    \begin{Verbatim}[commandchars=\\\{\},fontsize=\smaller]
\PY{c+c1}{// Data to be serialized}
\PY{n}{std}\PY{o}{:}\PY{o}{:}\PY{n}{string}\PY{+w}{ }\PY{n}{str}\PY{+w}{ }\PY{o}{=}\PY{+w}{ }\PY{l+s}{\PYZdq{}}\PY{l+s}{exemple}\PY{l+s}{\PYZdq{}}\PY{p}{;}
\PY{n}{sycl}\PY{o}{:}\PY{o}{:}\PY{n}{buffer}\PY{o}{\PYZlt{}}\PY{n}{f64\PYZus{}3}\PY{o}{\PYZgt{}}\PY{+w}{ }\PY{n}{buffer}\PY{+w}{ }\PY{o}{=}\PY{+w}{ }\PY{p}{.}\PY{p}{.}\PY{p}{.}\PY{p}{;}
\PY{n}{u32}\PY{+w}{ }\PY{n}{buf\PYZus{}size}\PY{+w}{ }\PY{o}{=}\PY{+w}{ }\PY{n}{buffer}\PY{p}{.}\PY{n}{size}\PY{p}{(}\PY{p}{)}\PY{p}{;}

\PY{n}{SerializeHelper}\PY{+w}{ }\PY{n}{ser}\PY{p}{;}

\PY{c+c1}{// Compute byte size of header and content}
\PY{n}{SerializeSize}\PY{+w}{ }\PY{n}{bytelen}\PY{+w}{ }\PY{o}{=}\PY{+w}{ }
\PY{+w}{      }\PY{n}{ser}\PY{p}{.}\PY{n}{serialize\PYZus{}byte\PYZus{}size}\PY{o}{\PYZlt{}}\PY{n}{u32}\PY{o}{\PYZgt{}}\PY{p}{(}\PY{p}{)}\PY{+w}{ }
\PY{+w}{    }\PY{o}{+}\PY{+w}{ }\PY{n}{ser}\PY{p}{.}\PY{n}{serialize\PYZus{}byte\PYZus{}size}\PY{o}{\PYZlt{}}\PY{n}{f64\PYZus{}3}\PY{o}{\PYZgt{}}\PY{p}{(}\PY{n}{buffer}\PY{p}{.}\PY{n}{size}\PY{p}{(}\PY{p}{)}\PY{p}{)}
\PY{+w}{    }\PY{o}{+}\PY{+w}{ }\PY{n}{ser}\PY{p}{.}\PY{n}{serialize\PYZus{}byte\PYZus{}size}\PY{p}{(}\PY{n}{test\PYZus{}str}\PY{p}{)}\PY{p}{;}

\PY{c+c1}{// Allocate memory}
\PY{n}{ser}\PY{p}{.}\PY{n}{allocate}\PY{p}{(}\PY{n}{bytelen}\PY{p}{)}\PY{p}{;}

\PY{c+c1}{// Write data}
\PY{n}{ser}\PY{p}{.}\PY{n}{write}\PY{p}{(}\PY{n}{buf\PYZus{}size}\PY{p}{)}\PY{p}{;}
\PY{n}{ser}\PY{p}{.}\PY{n}{write\PYZus{}buf}\PY{p}{(}\PY{n}{buffer}\PY{p}{,}\PY{+w}{ }\PY{n}{n2}\PY{p}{)}\PY{p}{;}
\PY{n}{ser}\PY{p}{.}\PY{n}{write}\PY{p}{(}\PY{n}{test\PYZus{}str}\PY{p}{)}\PY{p}{;}

\PY{c+c1}{// Recover the result}
\PY{n}{sycl}\PY{o}{:}\PY{o}{:}\PY{n}{buffer}\PY{o}{\PYZlt{}}\PY{n}{u8}\PY{o}{\PYZgt{}}\PY{+w}{ }\PY{n}{res}\PY{+w}{ }\PY{o}{=}\PY{+w}{ }\PY{n}{ser}\PY{p}{.}\PY{n}{finalize}\PY{p}{(}\PY{p}{)}\PY{p}{;}
\end{Verbatim}
  
\end{tcolorbox}

\begin{tcolorbox}
    \begin{Verbatim}[commandchars=\\\{\},fontsize=\smaller]
\PY{c+c1}{// The byte buffer}
\PY{n}{sycl}\PY{o}{:}\PY{o}{:}\PY{n}{buffer}\PY{o}{\PYZlt{}}\PY{n}{u8}\PY{o}{\PYZgt{}}\PY{+w}{ }\PY{n}{res}\PY{+w}{ }\PY{o}{=}\PY{+w}{ }\PY{p}{.}\PY{p}{.}\PY{p}{.}\PY{p}{;}

\PY{c+c1}{// Give the buffer to the helper}
\PY{n}{shamalgs}\PY{o}{:}\PY{o}{:}\PY{n}{SerializeHelper}\PY{+w}{ }\PY{n+nf}{ser}\PY{p}{(}\PY{n}{std}\PY{o}{:}\PY{o}{:}\PY{n}{move}\PY{p}{(}\PY{n}{res}\PY{p}{)}\PY{p}{)}\PY{p}{;}

\PY{c+c1}{// Recover buffer size}
\PY{n}{u32}\PY{+w}{ }\PY{n}{buf\PYZus{}size}\PY{p}{;}
\PY{n}{ser}\PY{p}{.}\PY{n}{load}\PY{p}{(}\PY{n}{buf\PYZus{}size}\PY{p}{)}\PY{p}{;}

\PY{c+c1}{// Allocate buffer and load data}
\PY{n}{sycl}\PY{o}{:}\PY{o}{:}\PY{n}{buffer}\PY{o}{\PYZlt{}}\PY{n}{f64\PYZus{}3}\PY{o}{\PYZgt{}}\PY{+w}{ }\PY{n}{buf}\PY{+w}{ }\PY{p}{(}\PY{n}{buf\PYZus{}size}\PY{p}{)}\PY{p}{;}
\PY{n}{ser}\PY{p}{.}\PY{n}{load\PYZus{}buf}\PY{p}{(}\PY{n}{buf}\PY{p}{,}\PY{+w}{ }\PY{n}{buf\PYZus{}size}\PY{p}{)}\PY{p}{;}

\PY{c+c1}{// Read the string}
\PY{n}{std}\PY{o}{:}\PY{o}{:}\PY{n}{string}\PY{+w}{ }\PY{n}{str}\PY{p}{;}
\PY{n}{ser}\PY{p}{.}\PY{n}{load}\PY{p}{(}\PY{n}{recv\PYZus{}str}\PY{p}{)}\PY{p}{;}
\end{Verbatim}
  
\end{tcolorbox}

Serialisation in \SHAMROCK{} relies on a split header data approach. Individual values are stored in the header on the CPU, while buffer data is stored on the device (CPU or GPU). This organisation ensures that individual value reads incurs minimal latency, thus avoiding high GPU load latency. The entire buffer is only assembled on the device at the end of the serialisation procedure. During deserialisation, the header is initially copied to the CPU. To circumvent constraints imposed by the \CUDA backend, all reads and writes are adjusted to 8-byte length.

\subsection{Sparse \MPI{} communications} 
\label{sec:sparse-mpi-comm}

\begin{figure*}
    \centering
    \includegraphics[width=0.9\linewidth]{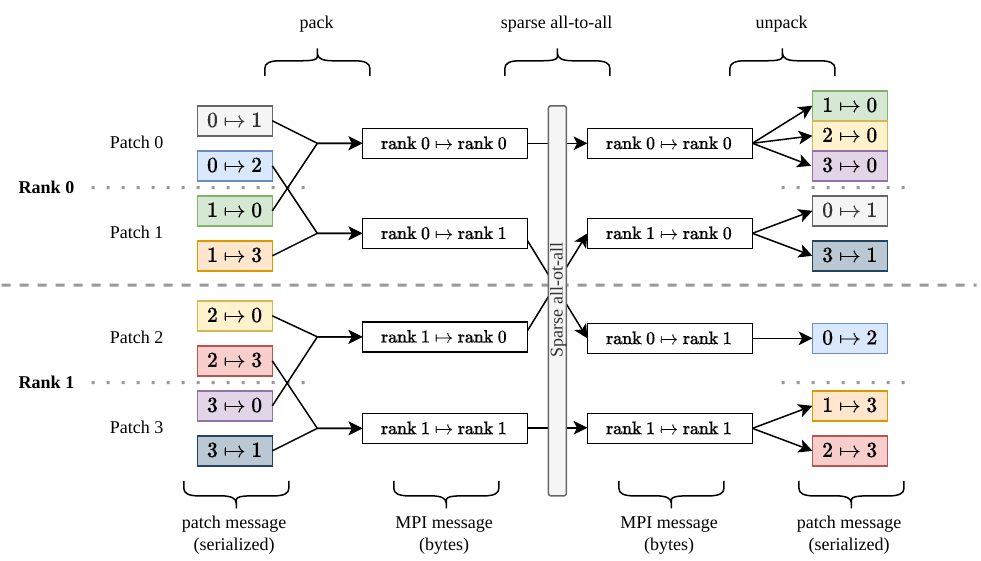}
    \caption{Illustration of the behaviour of a \MPI sparse communication of patches in \SHAMROCK{}. The first step consists of packing communication between a same pair of \MPI ranks together. Subsequently, a sparse all-to-all operation is executed (see Fig.\ref{fig:sparsealltoall}). Finally, the received buffers are unpacked.}
    \label{fig:sparsecommexemple}
 \end{figure*}

 \begin{figure*}
    \centering
    \includegraphics[width=0.9\linewidth]{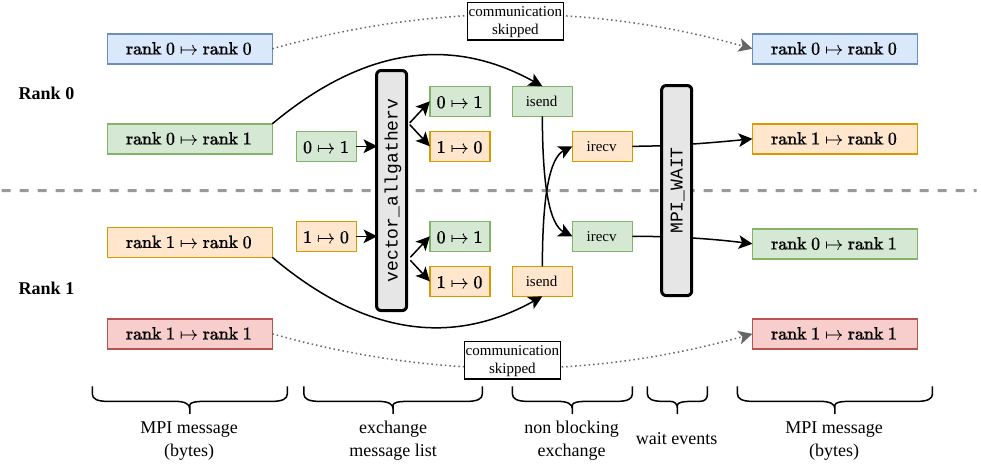}
    \caption{Illustration of the behaviour of a \MPI sparse all-to-all
     communication in \SHAMROCK{}. Firstly, a \texttt{vector\char`_allgatherv} is
     performed on the list of communication. Subsequently, each rank executes a
     non-blocking send of its data. To prepare for receiving, a non-blocking receive is
     launched for every incoming message. The operation is concluded by waiting
     for all non-blocking operation to finish. Communications for
     an \MPI rank to itself are skipped by simply relocating the data within the rank.}
    \label{fig:sparsealltoall}
\end{figure*}

With synchronised metadata, each \MPI ranks holds information of the \MPI rank to which every patch belongs. We therefore group communication between patches involving the same pair of \MPI ranks in a single \textit{patch message} (see Fig.~\ref{fig:sparsecommexemple}).
The graph corresponding to patch messages to be communicated is also sparse ($\text{rank }i \mapsto \text{rank }j$). We therefore apply a \MPI operation that extends \texttt{MPI\char`_Alltoall} to accommodate a sparse graph structure.
The operation, referred to as \textit{sparse all-to-all}, is structured as depicted in Fig.~\ref{fig:sparsealltoall}.  Initially, we compile the list of communications to be executed on each node. Subsequently, on each node, we go through the communication list and execute a non-blocking \MPI send if the sender's rank matches the current \MPI rank. Following this, on each node, we go through the communication list once more and initiate a non-blocking \MPI receive if the recipient's rank aligns with the current \MPI rank. Finally, we conclude the operation by invoking a \MPI wait on all non-blocking communication requests. Exchanges within the patch ghost zone graph are finalised once the sparse all to all operation is completed. If the sender's \MPI rank matches that of the recipient, the communication is disregarded, and an internal memory move is executed instead.
Another approach could involve using an \MPI reduce operation to count the number of messages received, and trigger the corresponding number of non-blocking receives with \texttt{MPI\char`_ANY\char`_SOURCE}. Given the limited number of communications, we observe no practical distinction between the two methods in practice. Moreover, the former approach is easier to debug and optimise, since it eliminates the need for sorting data to ensure determinism in the list of received messages.
So far in \SHAMROCK{}, we have not yet exploited compute/communication overlap, which could help hide communication latencies. This leaves room for potential future optimizations.
Additionally, the strategy of sparse communication ensures that only the actual data communications are sparse. To initiate these communications, each node must have a list of messages to send and receive. Consequently, we use MPI all-to-all operations to communicate this list. This corresponds to $8 \textrm{bytes} \cdot N_{\textrm{patch}}^2 \simeq 33 \, \textrm{GB}$ for 65,000 patches, which takes a few milliseconds given the bandwidth of the Adastra supercomputer. Therefore, this step is relatively small (by a factor of 10) compared to the actual data communication in practice, with the largest test performed in this study involving 8,000 patches.

\section{Morton codes and radix trees}

Specific notations used in this Section are given in Table~\ref{table:symbols}.

\begin{table}
\caption{List of symbols used in Sect.~\ref{sec:num}.}
\label{table:symbols}
\begin{center}
\begin{tabular}{|c|c|c|}
\hline
Symbol & Definition & Meaning \\
\hline
$x,y,z$ & Sect.~\ref{sec:Morton} & particle coordinates\\
$X,Y,Z$ & Sect.~\ref{sec:Morton} & integer particle coordinates\\
$\beta$ & Sect.~\ref{sec:Morton} & bit count\\
$ \mathtt{X}_0 \mathtt{X}_1 \cdots \mathtt{X}_{\beta-1}$ & Sect.~\ref{sec:Morton} & binary representation of $X$\\
$m$ & $\mathtt{X}_0 \mathtt{Y}_0 \mathtt{Z}_0 \cdots $ & generic Morton code\\
$m_1 \equiv 0101$ & Sect.~\ref{sec:morton.prefix} & Morton code example 1\\
$m_2 \equiv 0111$ & Sect.~\ref{sec:morton.prefix} & Morton code example 2\\
$\delta \left(a, b \right)$& eq.\ref{eq:delta_ab} & Karras $\delta$ operator\\
$\texttt{clz}(a)$ & Sect.~\ref{sec:delta} & count leading zeros\\
$ a\hat{~~} b$ & Sect.~\ref{sec:delta} & bitwise XOR operator\\
$a ~\&~ b$ &Sect.~\ref{sec:prefrange} &  bitwise AND operator\\
$a ~\texttt{<<}~ b$ &Sect.~\ref{sec:prefrange} & left bitshift operator\\
$\mathbf{r}_i$ & \, & position of particle $i$ \\
$m_i$ & \, & Morton code of particle $i$\\
$\{\mu_i\}_i$ & $\mu_i = m_{\epsilon_i}$ & sorted Morton codes \\
$\epsilon_i$ & sort : $\epsilon_i \mapsto i$ & sort inverse permutation\\
$\xi_i$ & Sect.~\ref{sec:mortondupremove} & Morton-keep mask\\
$\texttt{id}_i$ & & indexes of kept Morton codes\\
$\mu_{\mathrm{leaf},i}$ & & tree leaf Morton codes\\
\hline
\end{tabular}
\end{center}
\end{table}

\subsection{Morton codes}
\label{sec:Morton}
 
In hydrodynamic simulations, physical fields are represented on a discrete set of elementary numerical elements such as grid cells or interpolation points (or \textit{numerical objects} for a generic terminology). The positions of these objects are represented by coordinates, usually stored as floating point numbers such as $(x,y,z)$ in three dimensions. 
These coordinates are usually sampled on a 3D integer grid, which in turn can be mapped onto a 1D integer fractal curve. The Morton space-filling curve, also called \textit{Morton ordering}, is commonly used for this purpose since it has a natural duality with a tree structure (e.g. \citealt{samet2006foundations}, see below). 
In practice, Morton ordering can be constructed from a list of 3D positions as follows. 
First, the real coordinates in each dimension are remapped over the interval $\left[ 0, 1 \right)^{3}$ (note the exclusion of the value 1) by doing $x \mapsto  ( x - x_{\min} )/ ( x_{\max} - x_{\min}) $, and a similar procedure is applied for $y$ and $z$ respectively. 
This unit cube is then divided into a 3D grid of $(2^{\beta})^3$ elements, where ${\beta}$ is the number of bits used to represent integers. 
Within this grid, the objects possess integer coordinates  $(X,Y,Z) \in [ 0, 2^{\beta} - 1]^{3}$. These integer coordinates are noted in their binary representation $X = \mathtt{X}_0 \mathtt{X}_1 \mathtt{X}_2 \cdots$, where $\mathtt{X}_{i}$ denote the value of the $i^{\rm th}$ bit (the same convention also applies for $Y$ and $Z$). 
The Morton space-filling curve comprises a sequence of integers, called \textit{Morton codes} (or Morton numbers), defined through the following construction in a binary basis: the Morton code $m$ of each object is obtained by interleaving the binary representation of each coordinate $m \equiv \mathtt{X}_0 \mathtt{Y}_0 \mathtt{Z}_0 \mathtt{X}_1 \mathtt{Y}_1 \mathtt{Z}_1 \mathtt{X}_2 \mathtt{Y}_2 \mathtt{Z}_2 \cdots \mathtt{X}_{{\beta} - 1} \mathtt{Y}_{{\beta}-1} \mathtt{Z}_{{\beta} -1} $. 

\begin{figure*}
\center
    \includegraphics[width=0.7\linewidth]{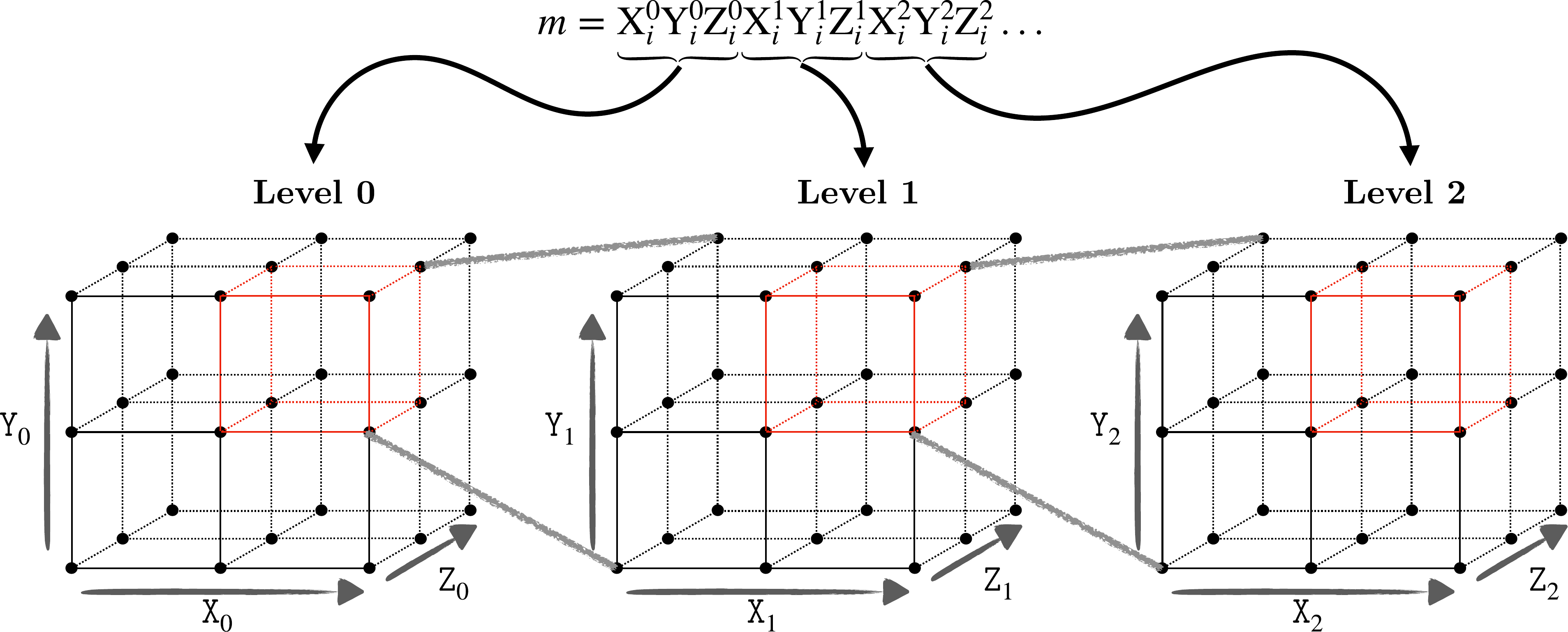}
    \caption{
        Illustration of the duality between Morton codes and the structure of an octree. 3 bits can describe the procedure of dividing a cube into eight smaller cubes. Repeating the procedure with triplets of additional bits produces an octree.}
    \label{fig:octree_morton}
\end{figure*}

By default, and unless specified otherwise, Morton codes are presented in binary notation, while other integers are expressed in decimal hereafter. A Morton code can also be interpreted as an ordered position on an octree with ${\beta} + 1$ levels, or alternatively as a position in a binary tree with $3{\beta} + 1$ levels (Fig.~\ref{fig:octree_morton}). 
To illustrate this duality, let us consider the first bit $\mathtt{X}_0$ of a Morton code. If $\mathtt{X}_0 = 0$, the integer coordinate $X$ belongs to the half space where $X < 2^{{\beta}-1}-1$. If $\mathtt{X}_0 = 1$, the integer coordinate $X$ belongs to the other half space $X \geq 2^{{\beta}-1}-1$. 
The following bits $Y_{0}$ and $Z_{0}$ divide the other dimensions in a similar way. The next sequence of bits $X_{1}, Y_{1}, Z_{1}$ subdivides the subspace characterised by $X_{0}, Y_{0}, Z_{0}$ in a similar manner, and the construction of a tree follows recursively. After going through all bits and reaching $\mathtt{X}_{{\beta} - 1} \mathtt{Y}_{{\beta}-1} \mathtt{Z}_{{\beta} -1} $, one is left with the exact position in the space of integer coordinates.
This tree structure consists of nested volumes where each parent volume encompasses all its children, forming as such a Bounded Volume Hierarchy (BVH). 

\subsection{Prefixes}
\label{sec:morton.prefix}

A \textit{prefix} is the sequence of the first $\gamma \le \beta$ bits of a Morton code. One defines the \textit{longest common prefix} of two Morton codes $a$ and $b$ as the sequence of matching bits starting from $X_{0}$ until two bits differ. As an example, the longest common prefix of $m_1 \equiv \mathbf{01}01$ and $m_2 \equiv \mathbf{01}11$ is $\mathbf{01}$. The longest common prefix of two Morton codes gives the minimal subspace of the integer 3D grid that contains the two Morton codes. The number of bits used to represent the longest common prefix of $a$ and $b$, called the \textit{length of the longest common prefix of $a$ and $b$}, is denoted $\delta\left( a, b\right)$.

\subsection{Bounding boxes}
We use the terminology \textit{prefix class} to refer to a set of Morton codes that have common prefixes. The longest common prefix of any pair of elements in a prefix class is at least of length $\gamma$ (or equivalently, for any pair of Morton code $a,b$ in the prefix class, $\delta(a,b) \geq \gamma$). 

Each prefix class corresponds to an axis aligned bounding box in the space of integer positions, having for generic coordinates $\left[ x_{\rm min}, x_{\max} \right) \times \left[ y_{\rm min}, y_{\max}\right) \times \left[z_{\rm min}, z
_{\max} \right) $ (Fig.~\ref{fig:octree_morton}). We refer to the set of three integers representing the lengths of the edges of this bounding box as \textit{the size of the bounding box}. Mathematically,
\begin{equation}
\mathbf{s}(\gamma) =  \left\lbrace 
2^{\beta - \lfloor \gamma / 3\rfloor},
2^{\beta- \lfloor (\gamma-1) / 3\rfloor},
2^{\beta- \lfloor (\gamma-2) / 3\rfloor}
\right\rbrace ,
\end{equation}
where $\lfloor \cdot \rfloor$ denotes the floor function of a real number. Indeed, for a given $\gamma$, the Morton construction divides the $x$-axis $\lfloor \gamma / 3\rfloor$ times, the $y$-axis $\lfloor (\gamma-1) / 3\rfloor$ times and the $z$-axis, $\lfloor (\gamma-2) / 3\rfloor$ times. The exclusion of the upper bounds in the bounding box ensures that the size on each coordinate axis is a power of 2. Similarly, we define the largest common prefix class between two Morton codes $a,b$ as the prefix class corresponding to the longest common prefix between $a$ and $b$. The size of the corresponding bounding box is then denoted $\mathbf{s}(a,b) = \mathbf{s}\left(\delta(a,b)\right)$.

\subsection{Longest common prefix length}
%\subsection{Calculating $\delta$}
\label{sec:delta}

The length of the longest common prefix of two Morton codes $a$ and $b$ is given by \citep{karras2012maximizing}
\begin{equation}
\delta \left(a, b \right) \equiv \texttt{clz} \left( a\hat{~~} b\right).
\label{eq:delta_ab}
\end{equation}
Eq.~\ref{eq:delta_ab} involves two binary operators. The first one is the bitwise  \texttt{XOR}  $\hat{~~}$ operator (Exclusive  \texttt{OR}), that returns the integer formed in binary by zeros where the bits match and ones when they differ. As an example, 
\begin{equation}
m_{1} \hat{~~} m_{2} = 0010 ,
\label{eq:hat}
\end{equation}
since $m_{1}$ and $m_{2}$ differ only by their third bit. The second operator is Count Leading Zeros. \texttt{clz} operates on a binary integers and returns the numbers of zeros preceding the first $1$ in the binary representation. As an example,
\begin{equation}
\texttt{clz}(0010) = 2 .
\label{eq:ex}
\end{equation}
Following this example, the longest common prefix of $m_1 \equiv \mathbf{01}01$ and $m_2 \equiv \mathbf{01}11$ is $\mathbf{01}$ and is of length 2. Eqs.~\ref{eq:hat}--\ref{eq:ex} allow performance, since instructions \texttt{clz} and \texttt{XOR} use only one CPU or GPU cycle on modern architectures. Getting the length of the longest common prefix take only 2 cycles with such procedure (e.g. a \texttt{xor} followed by \texttt{lzcnt} on Intel Skylake architectures).

\subsection{Finding common prefixes}
\label{sec:prefrange}

To find the longest common prefix between two Morton codes $a$ and $b$, we first construct a mask $c$, which is an integer where the first $p = \delta\left(a , b \right)$ bits are set to 1 while the remaining bites are set to 0. For example, applying this mask to the two Morton codes $m_{1}$ and $m_{2}$ from our previous example yields $1100$. To generate the mask, we take advantage of the bitwise shift-left operator. The bitwise shift-left operator $a ~\texttt{<<}~ i $ returns the binary representation of $a$ where the bits are shifted by $i^{\rm th}$ bits to the left, and zeros are introduced in place of non existing bits. Consider $u$, the integer having only ones in binary representation (i.e $u = 2^\beta - 1$, where $\beta$ is the size of the binary representation). $c$ is obtained with the following binary operation $u ~\texttt{<<}~ (\beta - \delta\left(a ,b \right) ) $. In our previous example, $\beta=4$ and $\beta - \delta \left(m_{1} , m_{2} \right) = 2$ gives $1111 ~\texttt{<<}~ 2 = 1100$. Consider now the bitwise \texttt{AND} operator, denoted by $\&$, that returns the integer formed in binary by ones where the bits match and zeros when they differ ($\&$ is the bitwise negation of the bitwise \texttt{XOR} operator). When applying the bitwise \texttt{AND} between $m_{1}$ or $m_{2}$ and the mask, the result is a binary number where the first bits are the prefix and the subsequent bits are zeros. As an example, applying the bitwise \texttt{AND} between $m_{2}$ and the mask yields $\mathbf{01}11 ~\&~ 1100 = \mathbf{01}00$.

\subsection{Getting coordinates sizes of bounding boxes}
\label{sec:bounding}

Consider the prefix class formed by Morton codes whose longest common prefix with $a$ (or equivalently $b$) is $\delta\left(a, b \right)$. This prefix class is a set of binary numbers whose smaller and larger values, denoted $p_{0}$ and $p_{1}$ respectively, are given by
\begin{align}
p_0(a,b) & \equiv  \left(2^\beta - 1 ~\texttt{<<}~ \beta - \delta (a,b ) \right)  ~\&~ a, \label{eq:lowboundbox} \\
p_1(a,b) & \equiv  \left(2^\beta - 1 ~\texttt{<<}~ \beta - \delta (a,b ) \right)  ~\&~ a + \left(2^{\beta - \delta (a,b )} - 1\right). \label{eq:highboundbox}
\end{align}
These two Morton codes correspond to two integer coordinates, denoted $\mathbf{p}_{0}$ and $\mathbf{p}_{1}$, that are the coordinates of the lower and upper edges of the bounding box, respectively. The size of the bounding box corresponding to this prefix class is
\begin{equation}
\mathbf{s}(a,b) = \mathbf{p}_1(a,b) - \mathbf{p}_0(a,b) + (1,1,1).
\label{eq:sab}
\end{equation}

%
%
%\begin{itemize}
%\item construction
%\item link with geometry
%\item operations on morton codes
%\end{itemize}
%
\subsection{Binary radix tree}
A binary radix tree is a hierarchical representation of the prefixes of a list of bit strings, corresponding here to the binary representation of integers (e.g.  \citealt{Lauterbach2009FastBC,karras2012maximizing}). The tree is defined by a set of hierarchically connected nodes, where nodes without children are called leaf nodes or \textit{leaves} (light orange circles on Fig.~\ref{fig:tree_karras}), and the others ones are called \textit{internal nodes} (blue circles on Fig.~\ref{fig:tree_karras}). The binary radix tree is a complete binary tree: every internal node has exactly two children. As such, a tree having $n$ leaves has exactly $n - 1$ internal nodes. This property allows to know lengths of tables in advance, making it particularly beneficial for GPU programming where dynamic allocation is not feasible. One commonly acknowledged downside of this tree structure is the challenge of efficient hierarchical construction \citep{Lauterbach2009FastBC,karras2012maximizing}. 
The deeper one goes down the tree, the longer the length of the common prefix of the Morton codes. The corresponding bounding boxes for each node in the tree are nested and become progressively smaller as one goes down the tree, while the length of the common prefixes increases. The corresponding radix tree forms a Bounding Volume Hierarchy, since each child in the bounding box is contained within the box of its parents.
\subsection{Karras algorithm}
\label{sec:karras_algo}

\begin{figure}
    \begin{center}
    \includegraphics[width=\linewidth]{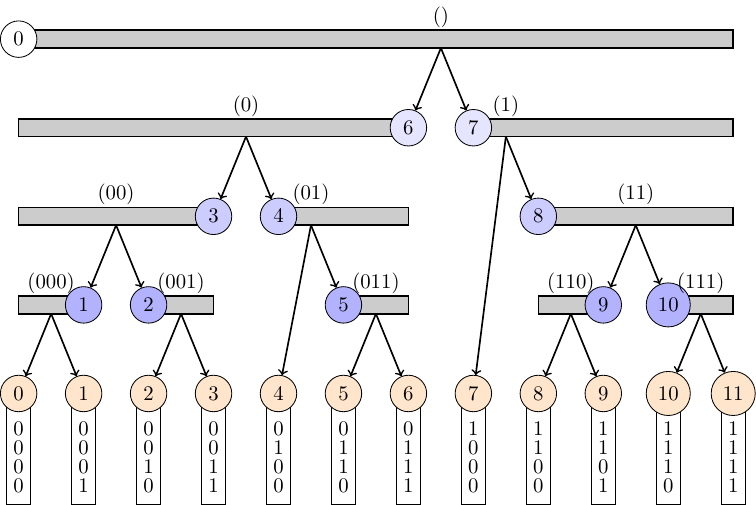}
    \caption{The Karras Algorithm associates a structure of radix tree to a sorted list of unduplicated Morton codes (depicted here within rectangles).  Within this tree, nodes can be either leaves, denoted by integers in light orange circles, or internal nodes, indicated by integers in blue circles. The grey bars represent the ranges of Morton codes covered by each internal node. Common prefixes of these Morton codes are shown in brackets.}
\label{fig:tree_karras}
\end{center}
\end{figure}

%--ICI__

The Karras algorithm overcomes this difficulty with a fully parallel algorithm that constructs a binary radix tree, in which the list of bit strings is a sorted list of Morton codes without any duplicates \cite{karras2012maximizing}.  Fig.~\ref{fig:tree_karras} shows a typical binary radix tree constructed by the Karras algorithm. The integers within the light orange circles represent the indices of the Morton codes in the sorted list, and they also are the indices of the corresponding leaves in the tree. The integers within the blue circles denote the indices of the internal nodes, their value come as a by-product of the algorithm. The grey bars denote intervals of leaf indices corresponding to any Morton code contained in the sub-tree beneath an internal node that shares the same common prefix. In Fig.~\ref{fig:tree_karras}, these prefixes are shown in brackets over the grey bars. The deeper the position in the tree, the longer the prefix. Consider the subset of Morton codes associated  to an internal node. The Karras algorithm divides this list in two sub-lists (arrows Fig.~\ref{fig:tree_karras}), each with a longer common prefix compared to the original. Such a split is unique. Several tables are associated to the construction of Fig.~\ref{fig:tree_karras}. Those are presented on Table~\ref{tabl:arrays_fig_karras}. The split associated to an internal node provides two numbers called the indices of the left and right children respectively, denoted as \texttt{left-child-id} and \texttt{right-child-id} respectively. \textit{A priori}, these indices can correspond to either an internal node or a leaf. This distinction is encoded by the value the integers \texttt{left-child-flag} and \texttt{right-child-flag}, where a 1 means that the corresponding child is a leaf and a 0, an internal node. The grey bar of an internal node has two ends. One corresponds to the index of the node itself, while the other is stored in \texttt{endrange}. Although this value is of no use in the construction of the tree itself, will be important later for calculating the sizes of a bounding box associated to a prefix class and for iterating over objects contained in leafs. The Karras algorithm performs dichotomous searches to compute the values of Table~\ref{tabl:arrays_fig_karras} in parallel, with no prerequisites other than the Morton codes (we refer to the pseudo-code of the algorithm in \citealt{karras2012maximizing} for details). Its efficiency relies firstly on the ability to pre-allocate tables before building the tree, and secondly on the sole use of the $\delta$ operator defined in Sect.~\ref{sec:delta}, which requires just 2 binary operations on dedicated hardware.
 
\begin{table}
    
    \begin{center}
    \resizebox{\hsize}{!}{
       \begin{tabular}{c  | c c c c c c c c c c c}\hline
       Internal cell id  & 0 & 1 & 2 & 3 & 4 & 5 & 6 & 7 & 8 & 9 & 10 \\  \hline\\[-1em]
       \texttt{left-child-id}      & 6 & 0 & 2 & 1 & 4 & 5 & 3 & 7 & 9 & 8 & 10 \\ 
       \texttt{right-child-id}     & 7 & 1 & 3 & 2 & 5 & 6 & 4 & 8 & 10 & 9 & 11 \\ 
       \texttt{left-child-flag}    & 0 & 1 & 1 & 0 & 1 & 1 & 0 & 1 & 0 & 1 & 1 \\ 
       \texttt{right-child-flag}  & 0 & 1 & 1 & 0 & 0 & 1 & 0 & 0 & 0 & 1 & 1 \\ 
       \texttt{endrange}        & 11 & 0 & 3 & 0 & 6 & 6 & 0 & 11 & 11 & 8 & 11 \\ 
       \end{tabular}
    }
    \end{center}
    \caption{Tables corresponding to the tree shown on Fig. \ref{fig:tree_karras} as returned by the Karras algorithm.}
    \label{tabl:arrays_fig_karras}
\end{table}%

\section{A reduction algorithm for radix trees}

\subsection{Removal of duplicated codes}
\label{sec:mortondupremove}

As mentioned in Sect.~\ref{sec:karras_algo}, the Karras algorithm requires a sorted list of Morton codes without duplicates (line 1-5 in alg.\ref{alg:mortonmask}). 
To achieve this, we go through the list of sorted Morton codes and compute a mask to select the Morton codes to retain.  The list of Morton codes without duplicates corresponds to the leaves of the tree obtained after applying the Karras algorithm to construct the radix tree.

\begin{algorithm}
\caption{Removal mask initialisation and reduction algorithm}\label{alg:mortonmask}
\KwData{
	$\{m_i\}_{i \in [0,n)}$ The morton codes.
	}
\KwResult{$\{\xi_i\}_{i \in [0,n)}$ The mask list.}
\BlankLine
\tcp{Flag removal of duplicates}
\For{$i$ \textbf{in parralel}}{
    \uIf{$i == 0$ }{
        $\xi_i \leftarrow \textbf{true}$\;
    }\uElse{
        $\xi_i \leftarrow not (m_i = m_{i-1})$\;
    }
}
\BlankLine
\tcp{Reduction passes}
\For{$n_{red}$ reduction steps}{
    \For{$i$ \textbf{in parralel}}{
    
        \tcp{Get kept morton codes indexes}
        $i_{-1} \leftarrow i - 1$\;
        \While{$(\xi_{i_{-1}} = \textbf{false} \And i_{-1} \geq 0)$}{
            $i_{-1} \leftarrow i_{-1} -1$\;
        }
        $i_{-2} \leftarrow i_{-1} - 1$\;
        \While{$(\xi_{i_{-2}} = \textbf{false} \And i_{-2} \geq 0)$}{
            $i_{-2} \leftarrow i_{-2} -1$\;
        }
        $i_{+1} \leftarrow i + 1$\;
        \While{$(\xi_{i_{+1}} = \textbf{false} \And i_{+1} < N_{morton})$}{
            $i_{+1} \leftarrow i_{+1} +1$\;
        }
        
        \tcp{Reduction criterion}
        $\delta_0 \leftarrow \delta(\mu_i, \mu_{i_{+1}})$\;
        $\delta_{-1} \leftarrow \delta(\mu_{i_{-1}},\mu_i)$\;
        $\delta_{-2} \leftarrow \delta(\mu_{i_{-2}},\mu_{i_{-1}})$\;

        \uIf{$ not(\delta_0 < \delta_{-1} \And \delta_{-2} < \delta_{-1})\And \xi_i = \textbf{true}$ }{
            $\xi_i \leftarrow \textbf{true}$\;
        }\uElse{
            $\xi_i \leftarrow \textbf{false}$\;
        }
    }
}

\end{algorithm}

\begin{algorithm}
\caption{Leaf object iteration}\label{alg:leafiterobjects}
\KwData{$\texttt{id}_i$ The leaf index map.\\$i$ the leaf index we want to unpack}
\BlankLine
\For{$j \in [id_i, id_{i+1})$}{
  $k \leftarrow \epsilon_j$ \tcp{index map of the sort}
  $\mathcal{F}(k)$
}
\end{algorithm}

\subsection{Reduction}
\label{sec:reduc-alg}

In certain situations, an object may interact with a large number of neighbours, resulting in multiple leaves containing these neighbours for the object. One such situation arises frequently in a Smoothed Particle Hydrodynamics solver, where each particle typically interacts with an average of $\sim 60$ neighbours.
One optimisation strategy to speed up the tree traversal consists in reducing the number of leaves containing these 60 neighbours by grouping some leaves at the lower levels of the tree before applying the Karras algorithm. We have integrated a so-called step of \textit{reduction} to achieve this. The resulting tree mirrors the initial one, but with grouped leaves. 

To perform reduction,  we require a criterion determining when two leaves, each containing two Morton codes, can be removed to yield the internal cell positioned just above them. This procedure is carried out using Alg.\ref{alg:mortonmask}: if a Morton code constitutes the second leaf of a shared parent, then it is removable. This property is implemented in the radix tree by verifying when 
$\delta(\mu_{i_{-2}},\mu_{i_{-1}})<  \delta(\mu_{i_{-1}},\mu_i)> \delta(\mu_i, \mu_{i_{+1}})$. When this condition is satisfied, the Morton code $i$ is removable.
The reduction step modifies the Morton tree list associated to the initial tree built. The tree is therefore already reduced when it is built and has never had any additional nodes.

\section{The \SHAMROCK{} tree}
\label{sec:num}

\subsection{Tree building}
\label{sec:treebuild}

\begin{figure}
\begin{center}
\includegraphics[width=0.9\linewidth]{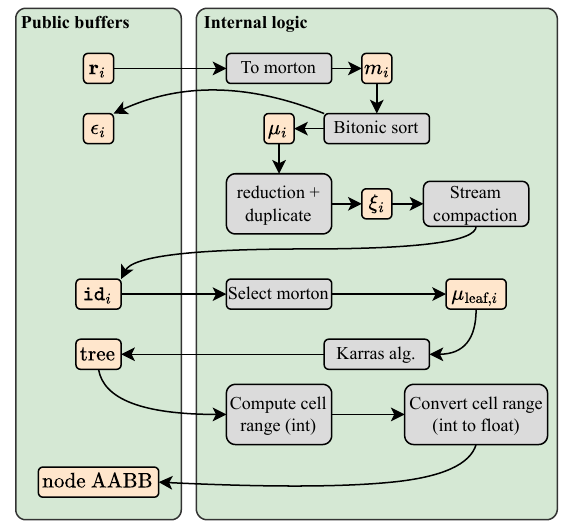}
\caption{
Flowchart illustrating the tree-building procedure, indicating the interdependence between each algorithm (grey boxes) and the related buffers (orange boxes). The internal logic box corresponds to the part of the algorithm inaccessible to the user. Buffers depicted outside this box are structures used in other parts of the code.}
\label{fig:tree_build_flow}
\end{center}
\end{figure}

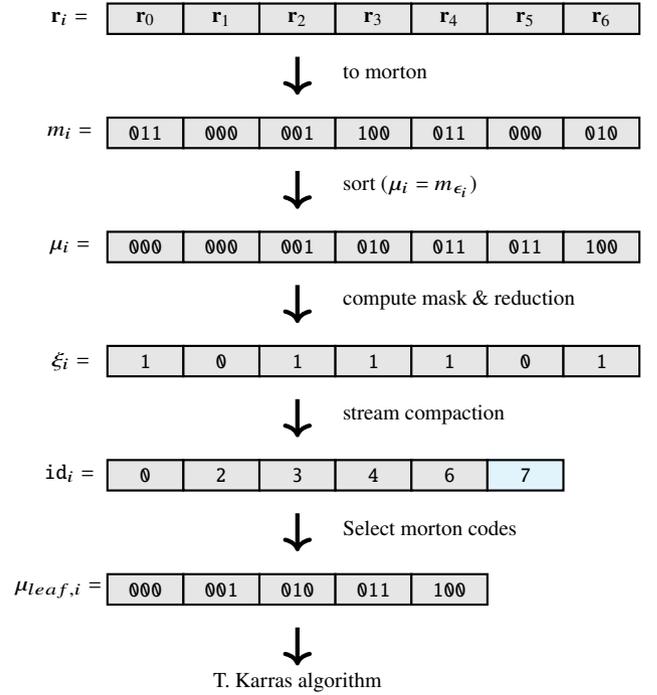
\begin{figure}
    \centering
    \begin{tikzpicture}

    \begin{scope}[shift={(0,3)}]
        \draw ({0.5 -1},0.2) node {$\mathbf{r}_i =$};
        \foreach \px in {0,1,2,3,4,5,6}{
        	\draw[fill=black!10,thick] ({0 + \px},0) rectangle ++(1,0.4);
        }
        
        \draw ({0.5 + 0},0.2) node {$\mathbf{r}_0$};
        \draw ({0.5 + 1},0.2) node {$\mathbf{r}_1$};
        \draw ({0.5 + 2},0.2) node {$\mathbf{r}_2$};
        \draw ({0.5 + 3},0.2) node {$\mathbf{r}_3$};
        \draw ({0.5 + 4},0.2) node {$\mathbf{r}_4$};
        \draw ({0.5 + 5},0.2) node {$\mathbf{r}_5$};
        \draw ({0.5 + 6},0.2) node {$\mathbf{r}_6$};
    
        \draw[->, ultra thick] (2.5, -0.3) --  ++(0,-0.5);
        \draw (3, -0.5) node[anchor=west] {to morton };
    \end{scope}

    \begin{scope}[shift={(0,1.5)}]
        \draw ({0.5 -1},0.2) node {$m_i =$};
        \foreach \px in {0,1,2,3,4,5,6}{
        	\draw[fill=black!10,thick] ({0 + \px},0) rectangle ++(1,0.4);
        }
        \draw ({0.5 + 0},0.2) node {\verb|011|};
        \draw ({0.5 + 1},0.2) node {\verb|000|};
        \draw ({0.5 + 2},0.2) node {\verb|001|};
        \draw ({0.5 + 3},0.2) node {\verb|100|};
        \draw ({0.5 + 4},0.2) node {\verb|011|};
        \draw ({0.5 + 5},0.2) node {\verb|000|};
        \draw ({0.5 + 6},0.2) node {\verb|010|};
    
        \draw[->, ultra thick] (2.5, -0.3) --  ++(0,-0.5);
        \draw (3, -0.5) node[anchor=west] {sort ($\mu_i = m_{\epsilon_i}$)};
    \end{scope}
    
    \draw ({0.5 -1},0.2) node {$\mu_i =$};
    \foreach \px in {0,1,2,3,4,5,6}{
    	\draw[fill=black!10,thick] ({0 + \px},0) rectangle ++(1,0.4);
    }
        \draw ({0.5 + 0},0.2) node {\verb|000|};
        \draw ({0.5 + 1},0.2) node {\verb|000|};
        \draw ({0.5 + 2},0.2) node {\verb|001|};
        \draw ({0.5 + 3},0.2) node {\verb|010|};
        \draw ({0.5 + 4},0.2) node {\verb|011|};
        \draw ({0.5 + 5},0.2) node {\verb|011|};
        \draw ({0.5 + 6},0.2) node {\verb|100|};

    \draw[->, ultra thick] (2.5, -0.3) --  ++(0,-0.5);
        \draw (3, -0.5) node[anchor=west] {compute mask \& reduction};

    \begin{scope}[shift={(0,-1.5)}]
        \draw ({0.5 -1},0.2) node {$\xi_i =$};
        \foreach \px in {0,1,2,3,4,5,6}{
        	\draw[fill=black!10,thick] ({0 + \px},0) rectangle ++(1,0.4);
        }
        
        \draw ({0.5 + 0},0.2) node {\verb|1|};
        \draw ({0.5 + 1},0.2) node {\verb|0|};
        \draw ({0.5 + 2},0.2) node {\verb|1|};
        \draw ({0.5 + 3},0.2) node {\verb|1|};
        \draw ({0.5 + 4},0.2) node {\verb|1|};
        \draw ({0.5 + 5},0.2) node {\verb|0|};
        \draw ({0.5 + 6},0.2) node {\verb|1|};
    
        \draw[->, ultra thick] (2.5, -0.3) --  ++(0,-0.5);
        \draw (3, -0.5) node[anchor=west] {stream compaction};
    \end{scope}

    \begin{scope}[shift={(0,-3)}]
        \draw ({0.5 -1},0.2) node {$\verb|id|_i =$};
    
        \draw[fill=black!10,thick] ({0 + 0},0) rectangle ++(1,0.4);
        \draw[fill=black!10,thick] ({0 + 1},0) rectangle ++(1,0.4);
        \draw[fill=black!10,thick] ({0 + 2},0) rectangle ++(1,0.4);
        \draw[fill=black!10,thick] ({0 + 3},0) rectangle ++(1,0.4);
        \draw[fill=black!10,thick] ({0 + 4},0) rectangle ++(1,0.4);
        \draw[fill=cyan!10,thick] ({0 + 5},0) rectangle ++(1,0.4);
        
        \draw ({0.5 + 0},0.2) node {\verb|0|};
        \draw ({0.5 + 1},0.2) node {\verb|2|};
        \draw ({0.5 + 2},0.2) node {\verb|3|};
        \draw ({0.5 + 3},0.2) node {\verb|4|};
        \draw ({0.5 + 4},0.2) node {\verb|6|};
        \draw ({0.5 + 5},0.2) node {\verb|7|};
        
        \draw[->, ultra thick] (2.5, -0.3) --  ++(0,-0.5);
        \draw (3, -0.5) node[anchor=west] {Select morton codes};
    
    \end{scope}

    \begin{scope}[shift={(0,-4.5)}]
        \draw ({0.35 -1},0.2) node {$\mu_{leaf,i} =$};
    
        \draw[fill=black!10,thick] ({0 + 0},0) rectangle ++(1,0.4);
        \draw[fill=black!10,thick] ({0 + 1},0) rectangle ++(1,0.4);
        \draw[fill=black!10,thick] ({0 + 2},0) rectangle ++(1,0.4);
        \draw[fill=black!10,thick] ({0 + 3},0) rectangle ++(1,0.4);
        \draw[fill=black!10,thick] ({0 + 4},0) rectangle ++(1,0.4);
        
        \draw ({0.5 + 0},0.2) node {\verb|000|};
        \draw ({0.5 + 1},0.2) node {\verb|001|};
        \draw ({0.5 + 2},0.2) node {\verb|010|};
        \draw ({0.5 + 3},0.2) node {\verb|011|};
        \draw ({0.5 + 4},0.2) node {\verb|100|};
        
        \draw[->, ultra thick] (2.5, -0.3) --  ++(0,-0.5);
        \draw (2.5, -0.8) node[anchor=north] {T. Karras algorithm};
    
    \end{scope}
    
    \end{tikzpicture}
    \caption{The cyan slot in the $\texttt{id}_i$ row is the total lenght of the input array. $\epsilon_i$ is the resulting permutation applied by the sort algorithm.}
    \label{fig:mortonlistbuild}
\end{figure}

Fig.~\ref{fig:tree_build_flow} outlines the tree building algorithm of \textsc{Shamrock}. initially, Morton codes are generated from coordinates and efficiently sorted while eliminating duplicates. Morton tables are then prepared and pre-processed (a summary of these steps is sketched in Fig.~\ref{fig:mortonlistbuild}) before filling the values characterising the tree as in Table~\ref{tabl:arrays_fig_karras}. The lengths associated to the coordinates of the cells are finally calculated. The algorithms described in this section are implemented using \texttt{C++} metaprogramming, enabling versatile use of any kind of spatial coordinates in practice.
\subsubsection*{Compute Morton codes}
Morton codes are calculated entirely in parallel (step \texttt{"To Morton"} in Fig.\ref{fig:tree_build_flow}). Initially, a buffer storing the positions of the elementary numerical elements is allocated. These positions are mapped to an integer grid following the procedure described in Sect.~\ref{sec:Morton}. The construction is tested by appropriate sanity checks. The resulting integer coordinates are converted to Morton codes in a Morton code buffer ($m_i$ in Fig.\ref{fig:tree_build_flow}).

\subsubsection*{Sort by Key}
Initially, the list of Morton codes corresponding to the positions of elementary numerical elements is unsorted. A key-value pair sorting algorithm is therefore used to sort the Morton codes while keeping track of the original index of the object within the list. For this task, we use a GPU Bitonic sorting algorithm that we have re-implemented using \texttt{Sycl}. The Bitonic algorithm is simple and its performance is not heavily reliant on the hardware used (step \texttt{"Bitonic sort"} in Fig.\ref{fig:tree_build_flow}, see e.g. \citealt{Batcher1968,Nassimi1679}). While more efficient alternatives have been suggested in the literature, our observation is that they are more difficult to implement and are not as portable across architectures (e.g. \citealt{2017arXiv170902520A,adinets2022onesweep}).

\subsubsection*{Reduction}

From the sorted list of Morton codes, we remove duplicates and apply reduction with a procedure in two steps. In the first step, we generate a buffer of integers where each value is 1 if the Morton code is retained at a given index and 0 otherwise. This information is stored in a buffer called Keep Morton flag buffer ($\xi_i$ in Fig.\ref{fig:tree_build_flow}). In the second step, we use this buffer to perform a stream compaction algorithm (e.g. \citealt{blelloch1990prefix,horn2005stream}, see example in Fig.~\ref{fig:mortonlistbuild}) to construct simultaneously two lists: a list of Morton codes without duplicates, and the list of the indices of the preserved Morton code prior stream compaction. The stream compaction algorithm heavily depends on an internal exclusive scan algorithm. This algorithm, when applied to the array $\{a_i\}_{i\in [0,n]}$ returns the array $\{\sum _{j=0}^{i-1} a_j\}_{i\in [1,n]}$ and $0$ when $i = 0$. In our case, we implemented the single-pass prefix sum with decoupled look-back algorithm \citep{merrill2016single}.

\subsubsection*{Compute tree tables}

At this point, we have a set of Morton codes sorted without duplicates. We then apply the Karras algorithm described in Sect.~\ref{sec:karras_algo} to generate in parallel the tables from which the properties of the tree can be reconstructed (listed on Table~\ref{tabl:arrays_fig_karras}).

%--ICI----

\subsubsection*{Compute tree cell sizes}

We define a \textit{tree cell} as the bounding box that corresponds to the Morton codes of the leaves under a given node. This node can either be an internal node or a leaf. Tree cells are therefore the geometric representation of the tree, and needs to be computed for neighbour finding (see Sect.~\ref{sec:treetrav}). In practice, it is sufficient to compute the boundaries of the edges of the cell $\left[ x_{\rm min}, x_{\max} \right) \times \left[ y_{\rm min}, y_{\max}\right) \times \left[z_{\rm min}, z
_{\max} \right)$  (Sect.~\ref{sec:bounding}).

\begin{algorithm}
\caption{Compute tree cell sizes}\label{alg:computemortonranges}
\KwData{$\texttt{morton}$ The morton code buffer.}
\KwResult{$\texttt{bmin}$, $\texttt{bmax}$, the bounds of the cells.}

$m_1 = \texttt{morton}[i]$

$m_2 = \texttt{morton}[\texttt{endrange}[i]]$

$\sigma = \delta_{\rm karras} ( m_1, m_2 )$

$\textbf{f}_0 = \mathbf{s}(\sigma)$

$\textbf{f}_1 = \mathbf{s}(\sigma + 1)$

$\texttt{mask} = \texttt{maxint} << (\texttt{bitlen} - \sigma)$

$\mathbf{p}_0 = (\text{morton} \rightarrow \text{real space} ) \, (\texttt{m}[i] \, \& \, \texttt{mask})$

$\texttt{bmin} [i] =  \mathbf{p}_0$

$\texttt{bmax} [i] =  \mathbf{p}_0 + \textbf{f}_0$

\If{ $\text{left child flag}[i]$ }{

$\texttt{bmin} [\texttt{rid}[i] + N_{\text{internal}}] = \mathbf{p}_0$

$\texttt{bmax} [\texttt{lid}[i] + N_{\text{internal}}] = \mathbf{p}_0 + \textbf{f}_1$

}

\If{ $\text{right child flag}[i]$ }{

$\texttt{tmp} = \textbf{f}_0 - \textbf{f}_1$ \label{offsetrchild}

$\texttt{bmin} [\texttt{rid}[i] + N_{\text{internal}}] =  
    \mathbf{p}_0 + \texttt{tmp}$

$\texttt{bmax} [\texttt{lid}[i] + N_{\text{internal}}] =  
    \mathbf{p}_0 + \texttt{tmp} + \textbf{f}_1$

}

\end{algorithm}

Alg.~\ref{alg:computemortonranges} provides the procedure to compute the size of tree cells, using the vector position $\mathbf{s}$ defined by Eq.~\ref{eq:sab} and the quantities $p_{0}$ and $p_{1}$ defined by Eqs.~\ref{eq:lowboundbox} -- \ref{eq:highboundbox}. For internal cells that have leaves as children, the boundary of the edges can be calculated by incrementing the value of $\delta(a,b)$ by one unity and using the new value in Eqs.~\ref{eq:lowboundbox} -- \ref{eq:highboundbox}. This gives the expected result for a left child, an extra shift being added for the right child (cf. line \ref{offsetrchild} of Alg.\ref{alg:computemortonranges}).

\subsection{Tree traversal}
\label{sec:treetrav}

\begin{algorithm}
\caption{Tree traversal}\label{alg:treetrav}
\KwData{\\
    $depth$ : The maximal tree depth,
    $N_{inode}$ : The number of internal nodes in the tree,
    $\{lchild_{id,j}\}_{j\in [0, N_{inode})}$,
    $\{rchild_{id,j}\}_{j\in [0, N_{inode})}$,
    $\{lchild_{flag,j}\}_{j\in [0, N_{inode})}$,
    $\{rchild_{flag,j}\}_{j\in [0, N_{inode})}$
    }
\BlankLine
\tcp{Setup index stack}
$i \leftarrow depth - 1$\;
$s \leftarrow \{ err \}_{i \in [0,depth)}$\;
\tcp{Enqueue the root node}
$s_i \leftarrow 0$\;
\Do{$i < depth$}{
    \tcp{Pop top of the stack}
    $j = s_i$\;
    $s_i = err$\;
    $i \leftarrow i + 1$\;
    \tcp{Check if interaction}
    $\alpha \leftarrow \gamma_{o/g}(\ldots,j)$\;
    \uIf{$\alpha$}{
        \tcp{If the current node is a leaf}
        \uIf{$j \geq N_{inode}$}{
            \tcp{Iterate on objects in leaf}
            $\text{leaf object iteration} (j)$\;
        }\uElse{
            \tcp{Push node childs on the stack}
            $lid \leftarrow lchild_{id,j} + (N_{inode}) * lchild_{flag,j}$\;
            $rid \leftarrow rchild_{id,j} + (N_{inode}) * rchild_{flag,j}$\;
            $i \leftarrow i - 1$\;
            $s_i = rid$\;
            $i \leftarrow i - 1$\;
            $s_i = lid$\;
        }
    }\uElse{
        \tcp{Gravity}
        $\text{leaf exclude case} (j)$\;
    }
}

\end{algorithm}

Each cell, leaf or internal of the tree constructed by the procedure described above consists of an axis-aligned bounding boxes and containing several numerical objects. Searching for the neighbours of an object $a$ therefore requires checking the existence of an interaction between a cell of the tree $c$ and the object $a$, using the object-group interaction criterion $\gamma_{o/g}(a,c)$. Per construction, if the criterion is true for a child cell, it is also true for its parent.
Neighbour finding requires therefore starting from the root node and going down the tree, checking at each step whether the interaction criterion is still verified or not. The result is a set of retained tree leaves, that are likely to contain neighbours. The set of neighbours of a given object is then obtained by verifying the object-object interaction criterion on each object in each of the targeted leaves.
The algorithmic procedure for these steps is detailed in Alg.~\ref{alg:treetrav}. It is based on the property that the depth of the tree is shorter than the length of the Morton code representation. This allows a stack of known size to be used to traverse the tree, which can be added at compile time and run on the GPU since there is no dynamic memory allocation. The first step in the algorithm is to push the root node onto the stack. In each subsequent step, we pop the node on top of the stack, and we check whether or not it interacts with the object. If it does, and if it is an internal node, we push its children onto the top of the stack and move on to the next step. Otherwise, if it is a leaf, we iterate through the objects contained in the leaf (Alg.~\ref{alg:leafiterobjects}), and check the object-object interaction criterion for each object in the given leaf. In the source code of \SHAMROCK{}, we abstract Alg.~\ref{alg:treetrav} under the \texttt{rtree\char`_for}. It can be called from within a kernel on the device and can be associated with any interaction criteria. It will then provide an abstract loop over the objects found using the criteria.

\subsection{Direct neighbour cache}
\label{sec:direct_cache}

Using neighbour search directly is technically feasible, but conducting it repeatedly would result in substantial costs due to its intricate logic.  Moreover, executing computations within the core of a device kernel with extensive branching would negatively impact performance. To circumvent these issues, we instead build a neighbour cache when traversing the tree, and then reuse this cache for subsequent computations on the particles.
The benefits are twofold: firstly, it increases performance for the reasons outlined above, and secondly, it decouples neighbour finding from calculations carried out on the particles, enabling optimisation efforts to be better targeted.
Conversely, using such an approach means that we store an integer index for each pair of neighbours, which in SPH is roughly 60 times the number of particles. The memory footprint therefore increases significantly. Taking everything into account, we opt for the neighbour caching strategy due to  its better performance and extensibility.

\begin{algorithm}
\caption{Neighbour caching}\label{alg:neighcachedirect}
\KwData{
    $N$ : The number of objects to build cache for,
    $\gamma_{o/g}$ : the object-group interaction criterion, $\gamma_{o/o}$ the object object interaction criterion.
    }
\KwResult{$\{\xi_i\}_{i \in [0,N+1)}$ The offset map. $\{\Xi_i\}_{i \in [0,N_{neigh})}$ The neighbour id map.}
\BlankLine

$\{c_i \leftarrow 0\}_{i \in [0,N+1)}$\;

\tcp{First pass to count neighbours}
\For{$i \in [0,N)$ \textbf{in parallel}}{
    $c \leftarrow 0$\;
    \For{ $j \leftarrow \texttt{rtree\char`_for}[\gamma_{o/g}(i,\ldots)] $}{
        \If{$\gamma_{o/o}(i,j)$}{
            $c \leftarrow c +1$\;
        }
    }
    $c_i \leftarrow c$\;
}
\tcp{$c_i$ contain the neighbours counts}
$\{\xi_i\}_{i \in [0,N+1)} \leftarrow \texttt{exclusive scan} (\{c_i\}_{i \in [0,N+1)})$\;

\tcp{$\xi_i$ contain the neighbour map offset}
$N_{neigh} \leftarrow c_N$\;

$\{\Xi_i \leftarrow 0\}_{i \in [0,N_{neigh})}$\;

\tcp{Second pass to get neighbours ids}
\For{$i \in [0,N)$ \textbf{in parallel}}{
    $off \leftarrow \xi_i$\;

    \For{ $j \leftarrow \texttt{rtree\char`_for}[\gamma_{o/g}(i,\ldots)] $}{
        \If{$\gamma_{o/o}(i,j)$}{
            $\Xi_{off} \leftarrow j$\;
            $off \leftarrow off +1$\;
        }
    }
}

\end{algorithm}

We start by allocating a buffer to store the neighbour count for each object. We perform an initial loop over all the objects and do a tree traversal for each of them to obtain the neighbour counts. We then perform an exclusive scan, which gives the offset used to write in the neighbour index map from our neighbour count buffer. The neighbour count buffer has an extra element that is set to zero at its end, this allows us to obtain the total number of neighbours in this slot after the exclusive scan. A final loop writes the indexes of the neighbours to the neighbour index map. Details of this procedure are given in Alg.~\ref{alg:neighcachedirect}.
\begin{algorithm}
\caption{Neighbour cache usage}\label{alg:neighcacheiter}
\KwData{$\{\xi_i \leftarrow 0\}_{i \in [0,N+1)}$ the offset map, $\{\Xi_i\}_{i \in [0,N_{neigh})}$ the neighbour cache}
\BlankLine
\For{$j \in [\xi_i, \xi_{i+1})$}{
  $k \leftarrow \Xi_j$ \tcp{index of neighbour}
  $\mathcal{F}(k)$
}
\end{algorithm}
We can use a procedure similar to the one used for the tree leafs in Alg.~\ref{alg:leafiterobjects} to iterate over the neighbours stored in the neighbour cache, as depicted in Alg.~\ref{alg:neighcacheiter}.

\subsection{Two-stages neighbour cache}
\label{sec:twostageneighcache}

The procedure described in Sect.~\ref{sec:direct_cache} consists of a direct neighbour cache, in the sense that for each object we search directly for its neighbours. A more sophisticated approach, likely to improve performance in most cases, involves splitting the direct case into two stages. In the first step, we search for the neighbours of each tree leaf using the group-group interaction criterion and the group-object criterion. In the second step, we first determine in which leaf the object is, then use the leaf neighbour cache to find the neighbour of the object. The first step only searches for neighbours within the leaves of the tree, while the second step produces the same result as in the direct case.
In a two-stages neighbour search, tree traversal is performed once per tree leaf, instead of once per object. When combined with tree reduction, this approach can decrease the number of tree traversals performed by a factor of ten.
On the flip side, adopting a two-stage neighbour caching approach increases the number of kernels to be executed on the device and the allocation pressure (temporarily, as the first step is discarded at the end, the memory footprint is unchanged compared to the direct case, but temporary allocation can introduce additional latency).
Overall, we observe that two-stages neighbour caching generally improves computational efficiency, and when combined with tree reduction, this strategy ultimately yields the best performance.

\section{SPH interaction criterions}
\label{sec:sph-interact-crit}

Eq.~\ref{eq:sphvel} shows that two SPH particles interact when their relative distance is inferior to the maximum of their interaction radius. Formally, the object-object interaction criterion between two particles $a$ and $b$ is
\begin{align}
\gamma_{o/o} (a,b) \equiv \left\{ \vert \mathbf{r}_a - \mathbf{r}_b \vert < \max(l_a, l_b)\right\} .
\end{align}
Consider now a group of SPH particles, and let us embed them in an axis-aligned bounding box (AABB). Consider another SPH particle. A necessary condition for the latter particle to interact with the AABB is: it resides within the volume formed by extending the AABB in all directions by the maximum of all interaction radii of particles inside the AABB, or, a ball centered on the particle, with a radius equal to its interaction radius, intersects the AABB. Formally, the interaction criterion between the particle and the AABB of particles is therefore
\begin{align}
\gamma^1_{o/g} (a,\{b\}_{b \in AABB}) \equiv \nonumber &
\left(r_a \in AABB \oplus l_{AABB,b}\right) \\& \quad\quad \vee \Big(B(\mathbf{r}_a, l_a) \cap AABB \neq \emptyset\Big), \label{eq:objectobjectcritsph}
\end{align}
where $B(\mathbf{r}_a, l_a)$ is a ball centred on $\mathbf{r}_a$ and of diameter $2 l_a$, $l_{AABB,b}$ is the maximum interaction radius of the particles in the AABB, $\max_{b\in AABB}(l_b)$, $AABB \oplus l$ is the operation that extends the AABB in every direction by a distance $l$ and $\vee$ is the boolean \texttt{or} operator. 
Consider now the ball centred on $\mathbf{r}_a$ with a diameter of $2 l_a$. We denote $AABB(\mathbf{r}_a, l_a)$ the square AABB with a side length of $2 l_a$, centred at $\mathbf{r}_a$, ensuring that it encompasses the ball. Replacing the original ball by this AABB in Eq.~\ref{eq:objectobjectcritsph} yields the following group-object criterion
\begin{align}
\gamma^2_{o/g} (a,\{b\}_{b \in AABB}) \equiv \nonumber &
\left(r_a \in AABB \oplus l_{AABB,b}\right) \\& \quad \vee \Big(AABB(\mathbf{r}_a, l_a) \cap AABB \neq \emptyset\Big) .
\label{eq:objectgroupinteractsphpretheo}
\end{align}
Though less stringent than that of  Eq.~\ref{eq:objectobjectcritsph}, this criterion is easier to handle in practice. Indeed, one can show that (App.~\ref{app:AABBpermutationtheo})
\begin{align}
    AABB_1 \oplus h \cap AABB_2 \neq \emptyset \Leftrightarrow AABB_1 \cap AABB_2\oplus h  \neq \emptyset .
\label{eq:AABB_theo}    
\end{align}
Let AABB$_{1 \mathrm{e}}$ and  AABB$_{2\mathrm{e}}$ denote the extended version of AABB$_{1}$ and AABB$_{2}$, extended by the distance $h$ in all three directions respectively. Eq.~\ref{eq:AABB_theo} asserts that if AABB$_{1\mathrm{e}}$ intersects AABB$_{1}$, it is equivalent for AABB$_{1}$ to intersect AABB$_{2\mathrm{e}}$.
Applied on Eq.~\ref{eq:objectgroupinteractsphpretheo}, Eq.~\ref{eq:AABB_theo} guarantees that the object-group interaction criterion can be rewritten by moving the contribution of the interaction radius of the particle $a$ to the term corresponding to the AABB in the second brackets, as follows
\begin{align}
\gamma^2_{o/g} (a,\{b\}_{b \in AABB}) \equiv  &
\left(r_a \in AABB \oplus l_{AABB,b}\right) \vee\\& \quad \vee \Big(AABB(\mathbf{r}_a, 0) \cap AABB  \oplus l_a \neq \emptyset\Big), \nonumber\\
\equiv  & \left(r_a \in AABB \oplus l_{AABB,b}\right)\\
        & \quad  \vee \left(r_a \in AABB \oplus l_a\right) \nonumber
\\
\equiv  &
\left[r_a \in AABB \oplus \max\left(l_{AABB,b}, l_a\right)\right] .
\end{align}
The three criteria discussed above satisfy the hierarchy
\begin{align}
\gamma^2_{o/g} (a,\{b\}_{b \in AABB}) &\Leftarrow \gamma^1_{o/g} (a,\{b\}_{b \in AABB}) \nonumber \\&\Leftarrow \bigvee_b \gamma_{o/o}(a,b) .
\end{align}
Finally, one can extend the first form of $\gamma^2_{o/g}$ to the following group-group interaction criterion 
\begin{align}
\gamma_{g/g} (AABB_1, & AABB_2 ) \equiv  \Big( [AABB_1 \oplus l_{AABB_1,a}] \cap AABB_2 \neq \emptyset \Big) \nonumber \\ &\vee \Big( AABB_1 \cap [AABB_2 \oplus l_{AABB_2,b}] \neq \emptyset \Big) .
\end{align}
Using Eq.~\ref{eq:AABB_theo} similarly as for Eq.\ref{eq:objectgroupinteractsphpretheo} we obtain the form of the group-group interaction criterion used in \SHAMROCK{}, 
\begin{align}
\gamma_{g/g} (&AABB_1, AABB_2) \equiv \nonumber\\ & \Big( AABB_1 \cap [AABB_2 \oplus \max(l_{AABB_1,a}, l_{AABB_2,a})] \neq \emptyset \Big) .
\end{align}
In summary, the interaction criteria used for SPH in \SHAMROCK{} are:
\begin{itemize}
    \item Object-object criterion : 
    \begin{align*}
\gamma_{o/o} &(a,b) = \vert \mathbf{r}_a - \mathbf{r}_b \vert < R_{\rm kern} \max(h_a, h_b) 
    \end{align*}
    \item Object-group criterion : 
    \begin{align*}
\gamma^2_{o/g} &(a,\{b\}_{b \in AABB}) = \left[r_a \in AABB \oplus R_{\rm kern}\max\left(h_{AABB,b}, h_a\right)\right]
    \end{align*}
    \item Group-group criterion : 
    \begin{align*}
\gamma_{g/g} &(AABB_1, AABB_2) =  \Big( AABB_1 \cap [AABB_2 \\ &\quad\quad\quad\quad\quad\oplus R_{\rm kern} \max(h_{AABB_1,a}, h_{AABB_2,a})] \neq \emptyset \Big)
    \end{align*}
\end{itemize}

\section{SPH adaptive smoothing length implementation}
\label{sec:adaptsmoothinglenght}

In astrophysics, a typical choice consists in choosing $h_{a}$ in a way that the resolution follows the density
\begin{equation}
    \rho (h) = m \left({h_{\rm fact} \over h}\right)^3 \label{eq:rhoh} ,
\end{equation}
where $h_{\rm fact}$ is a tabulated dimensionless constant that depends on the kernel (e.g. $h_{\rm fact} = 1.2$ for the $M_{4}$ cubic kernel). This specific form also implies that the averaged number of neighbours within the compact support of a given SPH particle is roughly constant throughout the simulation. Eq.~\ref{eq:rhoh} must itself be consistent with the definition of density Eq.~\ref{eq:rhosphsum}, since $h$ depends on $\rho$ and vice versa. Achieving this requires for density and smoothing length to be calculated simultaneously, by minimising the function
\begin{equation}
    \delta \rho = \rho_a - \rho(h_a) .
\end{equation}
 This approach allows an accurate use of $\rho(h_a)$ in the algorithms rather than calculating the SPH sum. In practice, the iterative procedure is conducted with a Newton-Raphson algorithm. The steps outlined in Alg.~\ref{alg:smoothinglenghtupdate} describe the iterative procedure used to update the smoothing length.
\begin{algorithm}
\caption{Smoothing length update}\label{alg:smoothinglenghtupdate}
\KwData{
	$h_a^n$ The smoothing lengths at timestep $n$,
    $\chi$ The ghost zone size tolerance.
	}
\KwResult{$h_a^{n+1}$ The smoothing lengths at timestep $n+1$.}
\BlankLine
$\{\epsilon_a \leftarrow -1\}_a$\;
\tcp{Use a copy of $h_a^n$ to do iterations}
$\{h_a \leftarrow h_a^n\}_a$\;
\tcp{Outer loop for ghost exchange}
\While{$\min_a(\epsilon_a) = -1$}{
    \BlankLine
    \ldots exchange ghosts positions with tolerance $\chi$ \ldots\;
    \BlankLine
    \tcp{Inner loop for Newton-Rahpson}
\While{$\max_a(\epsilon_a) > \epsilon_c$}{
\BlankLine
\For{$a$ \textbf{in parallel}}{
\tcp{Compute the SPH sum}
$\rho_a \leftarrow \sum_b m_b W_{ab}(h_a)$\;
\BlankLine
\tcp{Newton-Rahpson}
$\delta \rho  \leftarrow \rho_a - \rho(h_a)$\;
$d\delta \rho \leftarrow \sum_b m_b {\partial W_{ab} (h_a) \over \partial h_a} + { 3 \rho_a \over h_a}$\;
$h_a^{n+1} \leftarrow h_a - \delta \rho/d\delta \rho$\;
\BlankLine
\tcp{Avoid over/under-shooting }
\uIf{$h_a^{n+1} > h_a \lambda$ }{
    $h_a^{n+1} \leftarrow h_a \lambda$\;
}\uElseIf{$h_a^{n+1}< h_a / \lambda$}{
    $h_a^{n+1} \leftarrow h_a / \lambda$\;
}
$\epsilon_a \leftarrow \vert h_a^{n+1} - h_a \vert / h_a^n$\;
\tcp{Exceed ghost size}
\uIf{$h_a^{n+1} > h_a^{n} \chi$ }{
    $h_a^{n+1} \leftarrow h_a^{n} \chi$\;
    $\epsilon_a \leftarrow -1$\;
}
}
}
}
\end{algorithm}
 A technicality related to ghost zones arises during this procedure. The size $\gamma_{12}$ of the ghost zone separating two adjacent patches, $P_{1}$ and $P_{2}$, is determined by the group-group interaction criterion between these patches
\begin{equation}
   \gamma_{12} = \max \left( \max_{\{a\}} h_{a},  \max_{\{b\}} h_{b} \right).
\end{equation}
where $a$ and $b$ stem for indices of particles in $P_{1}$ and $P_{2}$ respectively. In \SHAMROCK{}, the size $\gamma_{12}$ is increased by a safety factor $\chi$, termed as the \textit{ghost zone size tolerance}. This factor acknowledges that ghost zone structures should withstand fluctuations in smoothing lengths throughout the iterative process. With this tolerance, the smoothing length can fluctuate by a factor of $\chi$ during density iterations without necessitating \SHAMROCK{} to regenerate the ghost zones.
In practice, we first exchange the ghost zones using a tolerance $\chi = 1.1$, then iterate until all particles converge to the consistent smoothing length or exceed the ghost zone size tolerance. If the latter occurs, we restart the process from the beginning with the updated smoothing length. We find that this almost rarely arises, except during the initial time step when the smoothing length is converged for the first time.
Alg.~\ref{alg:smoothinglenghtupdate} shows that in \SHAMROCK{}, we use an additional safety factor, denoted as $\lambda$, to prevent over- and undershooting throughout the iterations. Without this correction, the iterative procedure may yield unstable negative smoothing lengths. In practice, we use $\lambda = 1.2$.

\section{SPH time stepping}
\label{sec:sph-timestep}

\begin{figure*}
    \centering
    \includegraphics[width=0.95\linewidth]{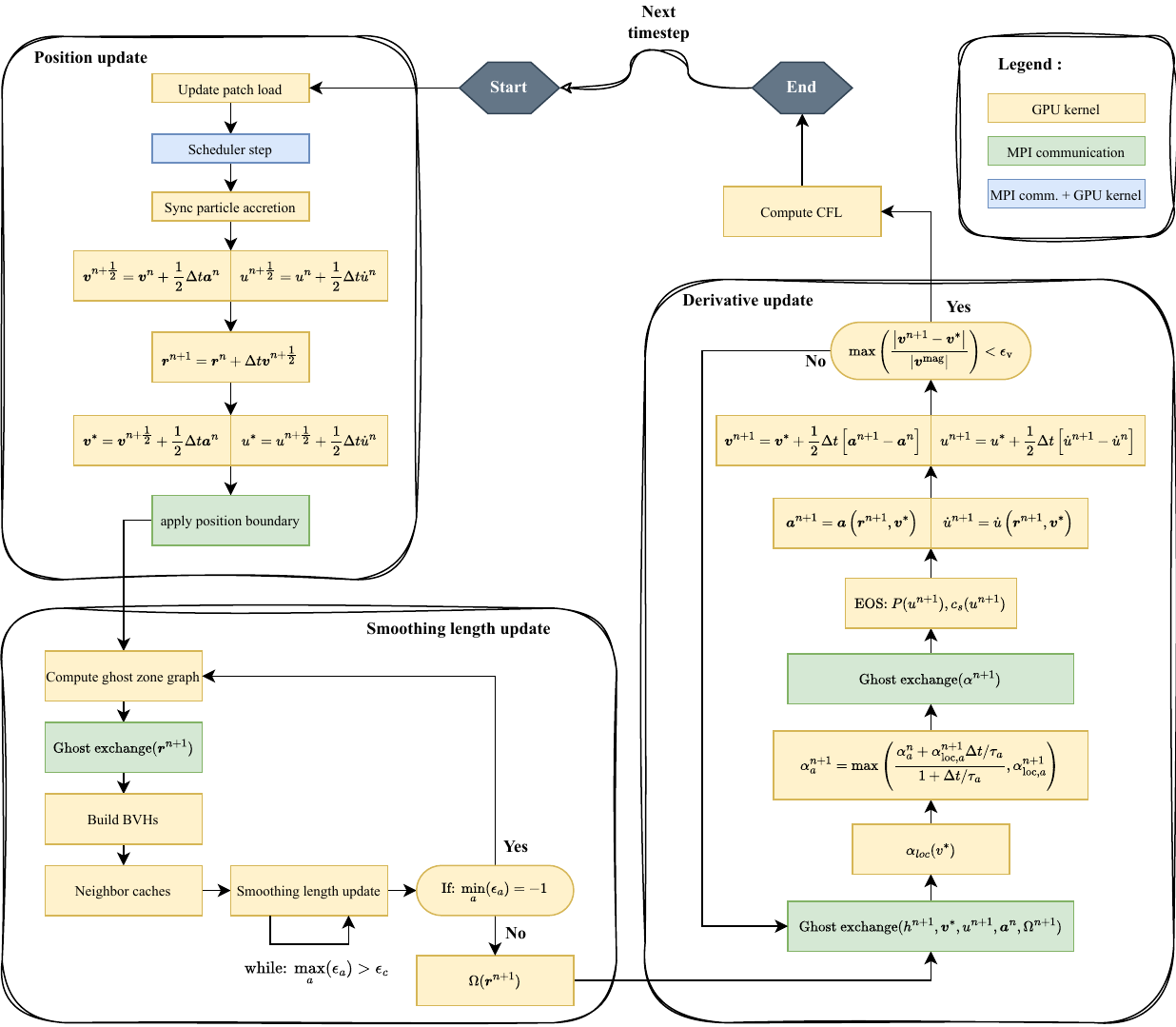}
    \caption{Illustration of an SPH time step through an organisational diagram representing one time step of the SPH scheme, the process being divided into three main sub-steps. Firstly, position updates (scheduling step for patch decomposition, leapfrog prediction, and application of position boundaries if necessary). Secondly, smoothing length updates (generation of ghost zone graph, construction of BVHs, creation of neighbour caches, smoothing length adjustment, computation of $\Omega$). Thirdly, derivative updates (field exchange, viscosity and derivative updates, application of leapfrog corrector). The step concludes with updating the CFL condition. The corresponding equations showed on this flowchart corresponds for the position and derivative update to the equations showed Sect.~\ref{sec:leapfrog} and \ref{sec:shocktimestep}. For the smoothing length the corresponding equations are detailed Sect.~\ref{sec:adaptsmoothinglenght}.}
    \label{fig:timesteppingscheme}
 \end{figure*}

\subsection{Leapfrog integration}
\label{sec:leapfrog}
By construction, standard SPH is conservative and achieves second-order accuracy in space in smooth flows. To ensure consistency, time integration in \SHAMROCK{} employs a symplectic second-order leapfrog integrator, or `Kick-drift-kick' (e.g. \citealt{Verlet1967,HairerLubichWanner2003}):
\begin{align}
\boldsymbol{v}^{n+\frac{1}{2}} & =\boldsymbol{v}^n+\frac{1}{2} \Delta t \boldsymbol{a}^n , \\
\boldsymbol{r}^{n+1} & =\boldsymbol{r}^n+\Delta t \boldsymbol{v}^{n+\frac{1}{2}} , \\
\boldsymbol{v}^* & =\boldsymbol{v}^{n+\frac{1}{2}}+\frac{1}{2} \Delta t \boldsymbol{a}^n \label{eq:velpredict} , \\
\boldsymbol{a}^{n+1} & =\boldsymbol{a}\left(\boldsymbol{r}^{n+1}, \boldsymbol{v}^*\right) , \\
\boldsymbol{v}^{n+1} & =\boldsymbol{v}^*+\frac{1}{2} \Delta t\left[\boldsymbol{a}^{n+1}-\boldsymbol{a}^n\right] ,
\end{align}
where $\mathbf{r}^{n}$, $\mathbf{v}^{n}$ and $\mathbf{a}^{n}$ denote positions, velocities and acceleration at the $n^{\rm -th}$ time step $\Delta t$. In the scheme presented in \citealt{phantom}, a combined iteration is used to calculate the acceleration $\boldsymbol{a}^{n+1}$ and update the smoothing length at the same time. 
To minimise the amount of data communicated, we separate the acceleration and the smoothing length update. In \SHAMROCK{}, the smoothing length is calculated after applying Eq.~\ref{eq:velpredict}. 
Only positions are required for the smoothing length iteration. Once these iterations are complete, we first calculate $\Omega_a$ using Eq.~\ref{eq:sphomega}, then exchange the ghost zones with the required fields, including $\Omega_a$ subsequently used in derivative computations.
Similar to the approach used in \textsc{Phantom}, we use the correction applied to the velocity, calculated during the correction step of the leapfrog, as a reference to check that the resulting solution is reversible over time. The correction applied at the end of the leapfrog scheme is as follows
\begin{align}
\Delta \mathbf{v}_i = \frac{1}{2} \Delta t\left[\boldsymbol{a}^{n+1}_i-\boldsymbol{a}^n_i\right].
\label{eq:deltavi}
\end{align}
We use the result of Eq.~\ref{eq:deltavi} to verify that the maximum correction does not exceed a fraction $\epsilon_{\mathrm{v}}$ of the mean square correction
\begin{align}
\max_{i} \left( \vert \Delta \mathbf{v}_i \vert / \sqrt{{1 \over N} \sum_j \vert \Delta \mathbf{v}_j \vert^2 \vert} \right) <\varepsilon_{\mathrm{v}}. \label{eq:cvleapfrog}
\end{align}
In practice, we set the value $\varepsilon_{\mathrm{v}} = 10^{-2}$.
If any particles fail to meet this criterion, we recalculate the acceleration and apply the correction step again with $\boldsymbol{v}^* \leftarrow \boldsymbol{v}^{n+1}$ instead.

\subsection{Choice of the timestep}

The value of the explicit time step is governed by the Courant-Friedrich-Levy stability condition \citep{CourantFriedrichsLewy1928}. Following 
\citet{phantom} from \cite{LattanzioMonaghan.et.al.1986, Monaghan1997},
\begin{align}
    \Delta t \equiv \min( C_{\mathrm{cour}} \frac{h_a}{v_{\mathrm{sig}, a}^{\mathrm{dt}}},  C_{\text {force }} \sqrt{\frac{h_a}{\left|\boldsymbol{a}_a\right|}}).
\end{align}
The first term allows to correctly capture the propagation of the hydrodynamic characteristic waves in the fluid at a given resolution. Similarly, the second term ensures correct treatment of the action of external forces on the fluid. The safety coefficients are set to the following values $C_{\mathrm{cour}} = 0.3$ and $C_{\text {force }} = 0.25$.

\subsection{CFL multiplier}

To minimize the cost associated with executing the correction cycles of the leapfrog scheme, we reduce the time step for a few iterations when Eq.~\ref{eq:cvleapfrog} is not satisfied, similar to the approach taken in \textsc{Phantom}. To do this in \SHAMROCK{}, we introduce a so-called \textit{CFL multiplier} $\lambda_{\rm CFL}$, which consists of an additional variable factor applied to the CFL condition. Therefore,  
the effective $C_{\mathrm{cour}}$ and $C_{\text{force}}$ used in \SHAMROCK{} SPH solver are
\begin{align}
    C_{\mathrm{cour}} = \lambda_{\rm CFL} \tilde{C}_{\mathrm{cour}}, \quad C_{\mathrm{force}} = \lambda_{\rm CFL} \tilde{C}_{\mathrm{force}},
\end{align}
where $\tilde{C}_{\mathrm{cour}}$ and $\tilde{C}_{\mathrm{force}}$ are the safety coefficients chosen by default by the user. 
If Eq.~\ref{eq:cvleapfrog} is not satisfied, we divide $\lambda_{\rm CFL}$ by a factor of 2. Otherwise, at each time step,
\begin{align}
    \lambda_{\rm CFL}^{n+1} = \frac{1+\lambda_{\rm stiff} \lambda_{\rm CFL}^{n}}{ 1+\lambda_{\rm stiff}},
\end{align}
where $\lambda_{\rm stiff}$ is a coefficient that parameters the stiffness of the evolution of the CFL multiplier. This numerical strategy allows to handle shocks in the simulation, automatically cycling leapfrog iterations over the CFL condition, thereby reducing the time step to enhance energy conservation. This
procedure is particularly effective during the first time steps of the
Sedov-Taylor blast problem.

Our CFL limiter is just a fall back mechanism, that is triggered when the corrector step needs to be re-run. This however does not replace the use of a better CFL criterion (e.g.\,\citealt{DurierDallaVecchia2012, SaitohMakino2009}). It might never be activated in practice.

\subsection{Shock detection}
\label{sec:shocktimestep}

The value of the shock viscosity parameter $\alpha_a$ is evolved using
\begin{align}
    \frac{\mathrm{d} \alpha_a}{\mathrm{~d} t}=-\frac{\left(\alpha_a-\alpha_{\mathrm{loc}, a}\right)}{\tau_a}, \label{eq:evolalphaanalytic}
\end{align}
where the details can be found in \cite[Sect.2.2.9]{phantom}.
The shock viscosity parameter $\alpha$ is evolved according to Eq.\ref{eq:evolalphaanalytic}. After the leafprog prediction step, an implicit time step is used for this integration 
\begin{align}
    \alpha_{\mathrm{loc}, a}^{n+1} =& \alpha_{\mathrm{loc}, a}(\boldsymbol{v}^*, \nabla \boldsymbol{v}^*, \nabla \boldsymbol{a}^n) , \\
    \alpha_a^{n+1}=& \max \left(\frac{\alpha_a^n+\alpha_{\mathrm{loc}, a}^{n+1} \Delta t / \tau_a}{1+\Delta t / \tau_a}, \alpha_{\mathrm{loc}, a}^{n+1} \right) .
\end{align}

\subsection{Summary}

We have implemented in \SHAMROCK{} an SPH hydrodynamical solver with self-consistent smoothing length that handles shock though the combined used of shock viscosity and conductivity with state-of-the-art shock detector. Fig.~\ref{fig:timesteppingscheme} shows a comprehensive overview of one SPH time step in \SHAMROCK{}.

\section{Software design}
\label{sec:sd}

\subsection{Development}

\subsubsection{Codebase organisation}

The \SHAMROCK{} project aims to be fully modular, in the sense that it is made up of several cmake projects which are connected using standardised interfaces. For example, the algorithmic library of \SHAMROCK{} is a cmake sub-project that depends on the backend library. This allows the shamrock sub-projects to be as independent as possible, avoiding merge conflicts and enabling development efforts to be better focused. To date, the project comprises 12 sub-projects. This number is very likely to change in the future, with future additions and refactoring.

\subsubsection{Git}

The \SHAMROCK{} project is hosted on GitHub. We adopt a methodology akin to the one employed by the \textsc{LLVM} project \citep{llvmproject}, where the main branch is protected and can only be modified by pull requests from the feature/fix branches from contributors forks of the project. Releases are performed by branching from the main branch, facilitating the implementation of fixes to existing versions of the code.
The CI test pipeline is routinely executed on GitHub, assessing both the main branch and all incoming pull requests. Successful completion of all tests is mandatory for changes to be merged into the main branch.

\subsection{Testing}

Numerous unit testing and validation options are available for \CPP. However, none of the standard solutions available match our specific requirements, the main one being that tests are integrated with \MPI. Because of this constraint, we have developed our own in-house test library, designed to provide the main features of \textsc{gtest}, while retaining the ability to specify the number of \MPI ranks for a particular test. The current test library is capable of performing unittest, validation tests, and benchmarks. On GitHub, we use self-hosted runners to perform the tests with multiple configurations of compilers, targets and versions.

\subsection{Environment scripts}

    Compiling \SHAMROCK{} on different machines entails dealing with a wide range of diversity.
    Typical technical aspects involve setting up \textsc{LLVM}, \textsc{MPI} and \textsc{SYCL}, which may involve numerous steps on a machine with missing libraries or having complex configuration.    
    To ensure consistency in \SHAMROCK{} configuration across machines, we have designed environment scripts. These scripts aim to produce a build directory with all the requirements for building the code, as well as to provide an `activate' script in this folder, which configures the environment variable and loads the correct modules by sourcing them. In addition, these scripts offer utility functions such as 
    \begin{itemize}
        \item setupcompiler: Setup the SYCL compiler
        \item updatecompiler: Update the environment
        \item shamconfigure: Configure \SHAMROCK{}
        \item shammake: Build \SHAMROCK{}
    \end{itemize}
    This functionality is provided by a `new-env' script that configures the build directory with all requirements, including the compiler \textsc{SYCL}, automatically.
    In summary, only 5 commands are needed to build a working version of \SHAMROCK{}, an example would be
\begin{tcolorbox}
    \begin{Verbatim}[commandchars=\\\{\},fontsize=\smaller]
\PY{c+c1}{\PYZsh{} Setup the environment}
./env/new\PYZhy{}env \PY{l+s+se}{\PYZbs{}}
    \PYZhy{}\PYZhy{}builddir build \PY{l+s+se}{\PYZbs{}}
    \PYZhy{}\PYZhy{}machine debian\PYZhy{}generic.acpp \PY{l+s+se}{\PYZbs{}}
    \PYZhy{}\PYZhy{} \PY{l+s+se}{\PYZbs{}}
    \PYZhy{}\PYZhy{}backend cuda \PY{l+s+se}{\PYZbs{} } 
    \PYZhy{}\PYZhy{}arch sm\PYZus{}70

\PY{c+c1}{\PYZsh{} Now move in the build directory }
\PY{n+nb}{cd} build
\PY{c+c1}{\PYZsh{} Activate the workspace, which will }
\PY{c+c1}{\PYZsh{} define some utility functions}
\PY{n+nb}{source} activate
\PY{c+c1}{\PYZsh{} Configure Shamrock}
shamconfigure
\PY{c+c1}{\PYZsh{} Build Shamrock}
shammake
\end{Verbatim}
  
\end{tcolorbox}

\subsection{Runscripts}

\begin{figure}
\begin{tcolorbox}
    \input{figures/sources/exemple_runscript-py.tex}  
\end{tcolorbox}
\caption{Example of a simplified \SHAMROCK{} runscript}
\label{fig:runscript}
\end{figure}

In \SHAMROCK{}, our aim is to handle setup files and configuration files that would allow great versatility in the use of the code, as well as on-the-fly analysis. Handling such a complexity through configuration files alone is both difficult and non-standard. Moreover, a user should not be required to know \CPP{} to be able to use the code. Using a \Python{} frontend offers a suitable solution to ensure both code versatility and ease of use.
To do this, we use pybind11 \citep{jakobpybind11}, which allows to map \CPP{} functions or classes from the \CPP{} source code to a `\texttt{shamrock}' python library. In the current version of \SHAMROCK{}, two uses are possible. The first is to use \SHAMROCK{} as a python interpreter that will go through and execute the content of a runscript (the script of a \SHAMROCK{} run), which can include, if desired, configuration, simulation and post-processing in a single run and script (see Fig.~\ref{fig:runscript} for an example of a runscript).

The other use is to compile \SHAMROCK{} as a Python library and install it through pip, enabling the code to be used in Jupyter notebooks. Using \SHAMROCK{} as a Python library is ideal for local machine prototyping, while on a cluster, employing \SHAMROCK{} as a Python interpreter is highly recommended.

\subsection{Units}

In \SHAMROCK{} we have chosen to use code units which are a rescaling of base SI units, where the factor is chosen at runtime in the runscript.

\section{AABB extension/intersection permutation}
\label{app:AABBpermutationtheo}

We prove the following theorem:
\begin{align*}
    AABB_1 \oplus h \cap AABB_2 \neq \emptyset \Leftrightarrow AABB_1 \cap AABB_2\oplus h  \neq \emptyset ,
\end{align*}
where $AABB \oplus l$ is the operation that extends the AABB in every direction by a distance $l$.
One initial observation is that an AABB is equivalent to a ball defined using the infinity norm $\vert\vert \cdot \vert\vert_{\infty}$.  Consequently, the intersection of two AABBs is the result of intersecting along each axis independently. Formally, define a first AABB 1 as the Cartesian product of three intervals $AABB_1 = I_{1,x} \times I_{1,y} \times I_{1,z} $, and a second AABB as $AABB_2 = I_{2,x} \times I_{2,y} \times I_{2,z} $. Their intersection is $AABB_1 \cap AABB_2 = (I_{1,x}\cap I_{2,x}) \times (I_{1,y} \cap I_{2,y}) \times ( I_{1,z} \cap I_{2,z})$. Hence, proving the theorem in one dimensions directly extends to three dimensions.
Consider now two one-dimensional intervals $I_1 = [\alpha_1, A_1], I_2 = [\beta_1, B_1]$. With $d(a,b)$ the distance in one dimension between a point $a$ and $b$, and $B(r,h)$ being a ball in one dimension of position $r$ and radius $h$, we have
\begin{align*}
    &\emptyset \neq I_1 \oplus h \cap I_2 \\
    \Leftrightarrow ~& \emptyset \neq [\alpha_1 - h, A_1 + h] \cap[\beta_1, B_1] \\
    \Leftrightarrow ~& \emptyset \neq B \left(\frac{A_1 + \alpha_1}{2}, \frac{A_1 - \alpha_1}{2} + h\right)  \cap B\left(\frac{B_1 + \beta_1}{2}, \frac{B_1 - \beta_1}{2}\right)  \\
    \Leftrightarrow ~& d\left(\frac{A_1 + \alpha_1}{2}, \frac{B_1 + \beta_1}{2}\right) \leq \frac{A_1 - \alpha_1}{2} + h + \frac{B_1 - \beta_1}{2}\\
    \Leftrightarrow ~& \emptyset \neq B\left(\frac{A_1 + \alpha_1}{2}, \frac{A_1 - \alpha_1}{2} \right)  \cap B\left(\frac{B_1 + \beta_1}{2}, \frac{B_1 - \beta_1}{2}+ h\right)  \\
    \Leftrightarrow ~& \emptyset \neq [\alpha_1, A_1 ] \cap[\beta_1 - h, B_1+ h] \\
    \Leftrightarrow ~&\emptyset \neq I_1\cap I_2  \oplus h ,
\end{align*}
which completes the proof.

%%%%%%%%%%%%%%%%%%%%%%%%%%%%%%%%%%%%%%%%%%%%%%%%%%

% Don't change these lines
\bsp	% typesetting comment
\label{lastpage}
\end{document}